\documentclass[twocolumn,aps,prb,showpacs,longbibliography,10pt]{revtex4-2} 

\usepackage{bm}
\usepackage{oplotsymbl}
\usepackage[normalem]{ulem}
\usepackage{amsfonts,amssymb,amsmath,graphicx,color}
\usepackage{comment}
\usepackage{bbold}
\usepackage{array,multirow}
\usepackage{tabularx}
\usepackage{mathtools}
\usepackage{enumitem}
\usepackage{braket}
\usepackage{physics}
\usepackage[version=4]{mhchem}
\usepackage{float}
\usepackage[colorlinks=true,linkcolor=blue,citecolor=blue,urlcolor=blue]{hyperref}
\usepackage{accents}
\newcommand{\dbtilde}[1]{\accentset{\approx}{#1}}

\newcommand\commentout[1]{}

\newcolumntype{P}[1]{>{\centering\arraybackslash}p{#1}}
\newcommand{\e}{\epsilon}

\definecolor{greenPR}{rgb}{0.00, 0.6, 0.00} 

\begin{document}
	 
        \title{Massless multifold Hopf semimetals}
	\author{Ansgar Graf}
	\author{Fr\'ed\'eric Pi\'echon}
	\email{frederic.piechon@universite-paris-saclay.fr}
	\affiliation{%
		Universit\'e Paris-Saclay, CNRS, Laboratoire de Physique des Solides, 91405, Orsay, France\\
	}%
	\date{\today}
	
	\begin{abstract}
		Three-dimensional { massless} topological semimetals exhibit linear energy band crossing points that act as monopoles of Berry curvature. 
		Here, an alternative class of { massless} semimetals is introduced, featuring linear $N$-fold crossing points each of which acts as a source 
		of a \emph{Berry dipole}. We construct continuum and lattice models for such \emph{{ massless} multifold Hopf semimetals (MMHSs)} with $N=3,4,5$ 
		bands and study nontrivial effects of a Berry dipole crossing: (i) A Landau level spectrum that is strongly tunable 
		by the orientation of the magnetic field relative to the dipole axis. (ii) An anomalous Hall conductivity that is an odd function of the Fermi level.
		(iii) Weak-field dissipative magnetoconductivities that resemble the chiral anomaly, chiral magnetic and magnetochiral effects familiar from a pair of coupled Weyl nodes, 
		but that are even functions of the Fermi level. By gapping out MMHSs, multiband Hopf insulators with Hopf numbers as high as $\mathcal{N}_\text{Hopf}=10$ are obtained,
		providing a fertile playground to explore delicate topology.
	\end{abstract}
	
	\maketitle

\section{Introduction}

\begin{figure}
	\centering
	\includegraphics[width=\columnwidth]{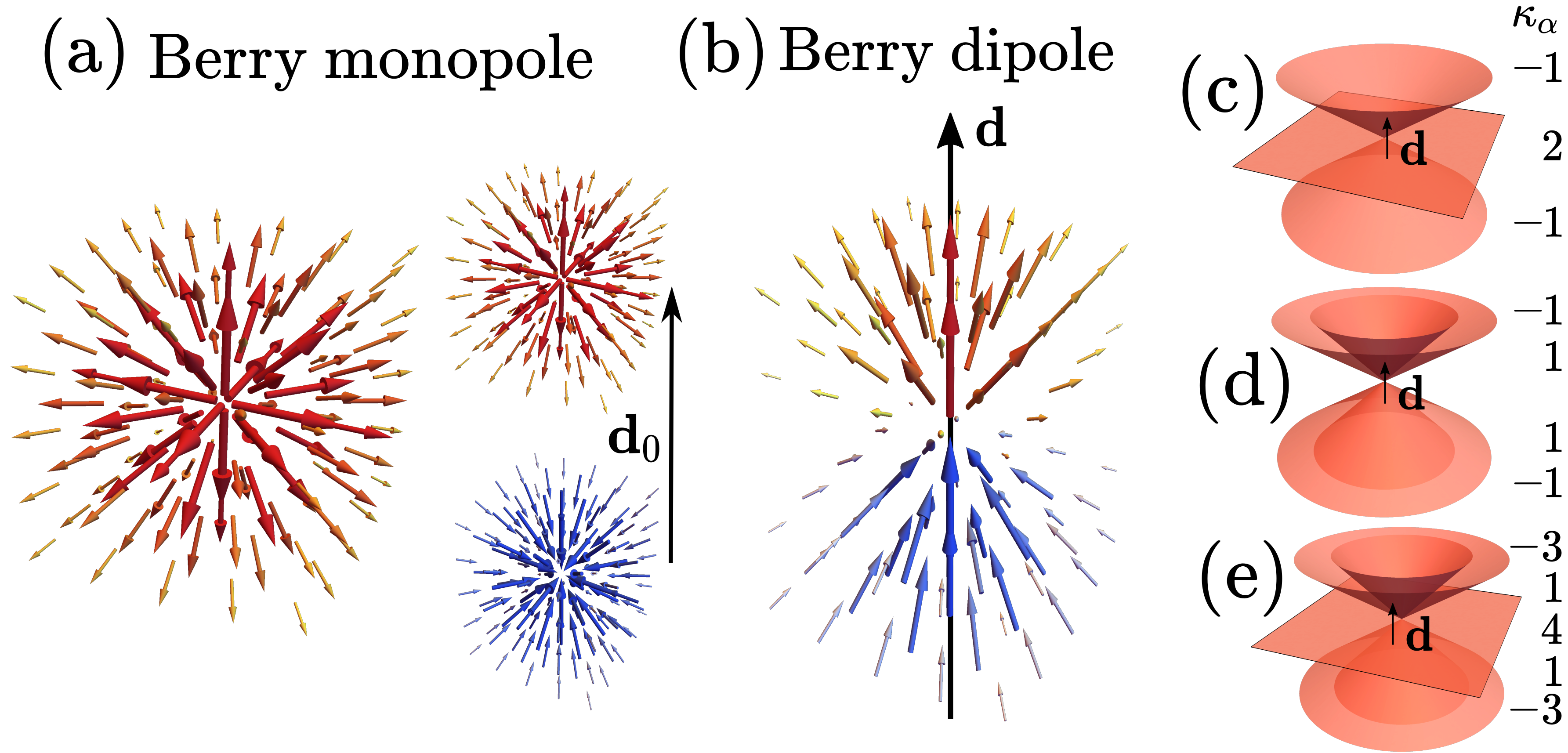}
	\caption{(a) Known topological semimetals are based on linear band crossings each of which is a Berry monopole (left). Those appear in monopole-antimonopole pairs (right). (b) { Massless} multifold Hopf semimetals are based on linear band crossings each of which is a Berry dipole. (c)--(e) Energy spectrum and Berry dipole charges of the continuum models (\ref{Berrydip3})--(\ref{Berrydip5}).}
	\label{fig:multipoles}
\end{figure}

 Three-dimensional (3D) { massless} topological semimetals \cite{Armitage_2018,Lv_2021} are materials with energy band crossing points that act as sources 
 or sinks of Berry curvature  \cite{Berry_1984,Berry_1985a,Volovik_1987,Fang_2003}, so-called Berry monopoles [Fig. \ref{fig:multipoles}(a)].
 
 The simplest example is a Weyl semimetal, with linear two-band crossings described by a Hamiltonian $H_\text{W}(\mathbf{q})=\gamma\,\mathbf{q}\cdot\boldsymbol{\sigma}$, 
 where $\gamma=\pm$ is the chirality, $\mathbf{q}=(q_x,q_y,q_z)$ 
 the momentum measured from the crossing point, and $\boldsymbol{\sigma}$ a vector of Pauli matrices. 
 A Berry monopole of a Weyl semimetal is characterized by a Berry curvature of the form
\begin{equation}
\boldsymbol{\Omega}_\alpha(\mathbf{q})=C_\alpha\frac{\mathbf{q}}{2|\mathbf{q}|^3},
\label{berrymon}
\end{equation}
where $\boldsymbol{\Omega}_\alpha=(\Omega_{\alpha,yz},\Omega_{\alpha,zx},\Omega_{\alpha,xy})$ 
is to be understood as a pseudovector formed from the three inequivalent components 
of the Berry curvature tensor in momentum space; $\alpha=\pm$ labels the two bands involved in the crossing,
and $C_\alpha=-\gamma\alpha$ is the Chern number measuring the quantized flux of Berry curvature (monopole charge).

Berry monopoles always come in pairs [Fig. \ref{fig:multipoles}(a)], which is a manifestation of the Nielsen-Ninomiya theorem \cite{Nielsen_1981}. 
Each pair can be viewed as forming a dipole $\mathbf{d}_0$ in the Brillouin zone; this dipole lies at the heart of various exotic phenomena such as Fermi arcs, 
anomalous Hall effect, and chiral anomaly \cite{Armitage_2018,Lv_2021}.

Any linear two-band ($N=2$) crossing in 3D is described by a Hamiltonian of the Weyl form, $H_\text{W}(\mathbf{q})$, and thus represents a Berry monopole. 
A common belief is that any linear multiband ($N>2$) crossing also represents a Berry monopole. Indeed, Berry monopoles (\ref{berrymon}), with high Chern numbers $C_\alpha$, 
are known to arise from linear multiband crossings that are governed by a generalized Weyl Hamiltonian, 
that is, a pseudospin Hamiltonian $H_\text{s}(\mathbf{q})=\gamma\,\mathbf{q}\cdot\mathbf{S}$ \cite{Bradlyn_2016,Ezawa_2017b}; h
ere $\mathbf{S}$ is the $(2s+1)$-dimensional matrix representation of a pseudospin and $\alpha=-2s,...,2s$. The Weyl Hamiltonian is recovered in the special case $s=1/2$.

The purpose of this paper is to draw attention to the existence of other linear multiband crossings (e.g. { massless multifold semimetals}), 
which are not of the Berry monopole type. In particular, we propose \emph{{ massless} multifold Hopf semimetals (MMHSs)}, a class of semimetals with linear multifold crossing points [Fig. \ref{fig:multipoles}(c)--(e)] each of which carries a \emph{dipolar} Berry curvature
\begin{equation}
	\boldsymbol{\Omega}_\alpha(\mathbf{q})=\kappa_\alpha(\mathbf{d}\cdot\mathbf{q})\frac{\mathbf{q}}{|\mathbf{q}|^4}.
	\label{berryungapped}
\end{equation}
 In contrast to the extended dipole $\mathbf{d}_0$ of a Weyl semimetal, Eq. (\ref{berryungapped}) describes a \emph{point-like} (or \textit{singular}) Berry dipole $\mathbf{d}$ that resides at a single band crossing point [Fig. \ref{fig:multipoles}(b)], with $\mathbf{d}$ representing an anisotropy axis but not a distance in momentum space.
 
 Note that singular Berry dipoles have previously been encountered in the literature, namely (i) when merging a pair of Weyl nodes ($\mathbf{d}_0\rightarrow0$) 
 and (ii) at topological phase transitions between two-band Hopf insulators \cite{Alexandradinata_2021,Nelson_2022}. 
 However, both scenarios fundamentally differ from the MMHSs introduced here in two aspects. First, for MMHSs the Berry dipoles emerge from \emph{linear band crossings}, 
 while they involve quadratic band touchings in both known cases (i) and (ii). Second, the dipole charge $\kappa_\alpha$ in Eq. (\ref{berryungapped}), 
 which is an integer for each band $\alpha$, is \emph{symmetric} with respect to zero energy for the MMHSs, $\kappa_\alpha=\kappa_{-\alpha}$, 
 and thus the Berry curvature has the important property $\boldsymbol{\Omega}_\alpha=\boldsymbol{\Omega}_{-\alpha}$.
 In stark contrast, since the Berry curvature of all bands must sum to zero, one has $\boldsymbol{\Omega}_\alpha=-\boldsymbol{\Omega}_{-\alpha}$ 
 for both known cases (i) and (ii), and more generally for any two-band system or any Berry monopole crossing (\ref{berrymon}).

Our work addresses three main points. First, it shows that linear multiband crossings can carry peculiar quantum geometric structures beyond Berry monopoles. 
We will only discuss the Berry dipole case (\ref{berryungapped}) in detail, but even Berry quadrupoles or Berry octupoles are possible if enough bands cross simultaneously (see Appendix \ref{Appmulti} for examples).

Second, due to the Berry dipoles, MMHSs have physical properties very different from those of known topological semimetals, despite \emph{the same} low-energy spectrum. 
These differences are thus purely rooted in the quantum geometric structure of the Bloch states. In particular, we show that Landau levels, anomalous Hall effect, 
and magnetoconductivity -- all of which have been extensively studied in Weyl semimetals -- exhibit distinct signatures of the Berry dipole (\ref{berryungapped}), 
and it is clear that this extends to a host of other observables.

  Third, we justify the choice of the term ``Hopf semimetals". Indeed, we demonstrate that our semimetals are closely related to the family of Hopf insulators, 
  a peculiar class of band insulators with a so-called \emph{delicate} topology \footnote{See Refs. \cite{Moore_2008,Deng_2013} 
  for an introduction to the well-known two-band Hopf insulator; see Ref. \cite{Nelson_2021a} for a comparison of stable, fragile and delicate topology}. 
  To establish this link, we propose simple lattice models for \emph{multiband Hopf insulators (MHIs)} \cite{Lapierre_2021} and show that MMHSs 
  appear at their topological phase transitions. This extends various recent results  \cite{Alexandradinata_2021,Nelson_2022} regarding the phase transitions of Hopf insulators 
  to the $N>2$ case, and provides { an alternative} class of simple lattice models for studying the physics of delicate topological insulators.
  
  The remainder of the paper is organized as follows. In Section \ref{Seccont} we first propose minimal continuum models for MMHSs and describe their symmetries. 
  We then discuss the peculiar properties of the MMHS models under a magnetic field, in particular focusing on Landau levels, anomalous Hall effect and semiclassical magnetotransport.
   In Section \ref{Seclat} we propose various simple tight-binding models for MMHSs that recover the continuum models in the vicinity of multifold crossing points. 
   These lattice models may feature one or several Berry dipole crossings in the Brillouin zone. We also comment on how the physics of the continuum models extends 
   to the lattice scenario. The relation between (multiband) Hopf semimetals and Hopf insulators will be explained in Section \ref{insulsec}, 
   followed by conclusions and a discussion of possible perspectives in Section \ref{conclusion}.
   
   A number of appendices provide supplemental material. Appendices \ref{Appmulti} and \ref{Apptune} contain examples for extensions of the MMHS continuum models. 
   Appendices \ref{AppLL} and \ref{AppSemi} address the Landau levels of MMHSs, from both an exact quantum approach and an original semiclassical approach. 
   Appendices \ref{Appboltzi} and \ref{Apptransp} provide details on the Boltzmann theory of magnetotransport and its application to MMHS systems. 
   Finally, Appendices \ref{Apphopf} and \ref{AppHaldane} address properties of the multiband Hopf insulators.

\section{Continuum models for massless multifold Hopf semimetals}
\label{Seccont}

\subsection{Description of the continuum models}

 We start by introducing continuum models $H_N^\xi(\mathbf{q})$ for massless Hopf semimetals, with $\xi=\pm$ representing a valley index.  
 They are constructed to have a linear isotropic energy spectrum
 \begin{equation}
E_\alpha(\mathbf{q})=c_\alpha|\mathbf{q}|
\label{spec}
 \end{equation}
that consists of cones and flat bands [Fig. \ref{fig:multipoles}(c)--(e)] with band velocities $c_\alpha$. 
We emphasize that this energy spectrum is \emph{identical} to that of a pseudospin system $H_\text{s}(\mathbf{q})$, 
however the models are constructed to possess a quantum geometry governed by Eq. (\ref{berryungapped}) instead of Eq. (\ref{berrymon}). 
 
 For a three-, four-, and fivefold Hopf semimetal, we consider the models
\begin{subequations}\label{mods}
	\begin{align}
		H_3^\xi(\mathbf{q})&=\begin{pmatrix}
		0 & Q_3^\xi \\
		(Q_3^\xi)^\dagger & 0_2
	\end{pmatrix}, && Q_3^\xi=\begin{pmatrix}
	q_-^\xi & -i q_z
\end{pmatrix},
\label{Berrydip3}\\
	H_4^\xi(\mathbf{q})&=\begin{pmatrix}
	0_2 & Q_4^\xi \\
	(Q_4^\xi)^\dagger & 0_2
\end{pmatrix}, && Q_4^\xi=\begin{pmatrix}
	aq_-^\xi & ia q_z\\
	ibq_z & bq_+^\xi
\end{pmatrix},\label{Berrydip4}\\
H_5^\xi(\mathbf{q})&=\begin{pmatrix}
0_3 & Q_5^\xi \\
(Q_5^\xi)^\dagger & 0_2
\end{pmatrix}, && Q_5^\xi=\begin{pmatrix}
0 & i\sqrt{2}q_z\\
iq_z&  q_+^\xi\\
\sqrt{2}q_+^\xi & 0
\end{pmatrix},\label{Berrydip5}
	\end{align}
\end{subequations}
respectively, where $q_\pm^\xi\equiv\xi q_x\pm iq_y$, and $a,b$ are real parameters such that $a> b>0$. By computing the energy spectrum the band velocities are easily obtained,
as summarized in Table \ref{tab0}.
\begin{table}
	\centering
	\begin{tabularx}{\columnwidth}{c|ccc}
		\hline
		\hline
		& Threefold HS (\ref{Berrydip3}) & Fourfold HS (\ref{Berrydip4}) & Fivefold HS (\ref{Berrydip5}) \\
		\hline
		$c_\alpha$ & $-1,0,1$ & $-a,-b,b,a$ & $-\sqrt{2},-1,0,1,\sqrt{2}$\\
		$\kappa_\alpha$ & $-1,2,-1$ &  $-1,1,1,-1$ & $-3,1,4,1,-3$\\
		$\omega_\alpha$ & $1,0,-1$ & $a,-b,b,-a$ & $\sqrt{2},3,0,-3,-\sqrt{2}$\\
		\hline
		\hline
	\end{tabularx}
	\caption{Coefficients $c_\alpha$, $\kappa_\alpha$ and $\omega_\alpha$ determining the energy spectrum, Berry curvature  
	and orbital magnetic moment of the MMHS continuum models, respectively. Each coefficient is listed from the lowest to the highest band.}
	\label{tab0}
\end{table}

 Note that the terms $\sim q_\pm^\xi$ are familiar from graphene and Weyl semimetals, and the terms $\sim q_z$ provide a third direction fixing the dipole axis as $\mathbf{d}=(0,0,\xi)$. While we will only consider systems with this fixed dipole axis in the remainder of this paper, we emphasize that it is also possible to construct models with a tunable $\mathbf{d}$ vector, see Appendix \ref{Apptune}. 

The Berry curvature of the multiband systems (\ref{mods}) is conveniently computed using eigenprojectors \cite{Graf_2021}, and we find it to be of the form (\ref{berryungapped}) with dipole charges $\kappa_\alpha$
as summarized in Table \ref{tab0} and visualized in Fig. \ref{fig:multipoles}(c)--(e). Interestingly, as a consequence of the symmetry property $\boldsymbol{\Omega}_\alpha=\boldsymbol{\Omega}_{-\alpha}$, or equivalently $\kappa_\alpha=\kappa_{-\alpha}$, large dipole charges are carried by the flat bands.

The models (\ref{mods}) have two important symmetries, linked to the dipole charges and the dipole axis, respectively. Namely, first, a chiral symmetry 
\begin{equation}
\mathcal{S}^{-1}H_N^\xi(\mathbf{q})\mathcal{S}=-H_N^\xi(\mathbf{q})
\label{chirsym}
\end{equation}
with a diagonal matrix $\mathcal{S}$ and $\mathcal{S}^2=\mathbb{1}$. Second, an axial rotation symmetry
\begin{equation}
 [L_d+\Sigma_d,H_N^\xi(\mathbf{q})]=0,
 \label{axsym}
\end{equation}
  with $L_d=\mathbf{d}\cdot\mathbf{L}$ the projection of the angular momentum operator $\mathbf{L}=-i (\mathbf{q}\times\boldsymbol{\nabla_\mathbf{q}})\mathbb{1}$ onto the Berry dipole axis, and with $\Sigma_d$ a diagonal matrix acting as an effective spin projection \footnote{More precisely, we have $\Sigma_d=\frac{1}{3}\text{diag}(1,-2,1)$, $\Sigma_d=\frac{1}{2}\text{diag}(1,-1,-1,1)$, and $\Sigma_d=\frac{1}{5}\text{diag}(4,-1,-6,-1,4)$ for Eqs. (\ref{Berrydip3})--(\ref{Berrydip5}), respectively.}.
  
 These two symmetries determine general properties of the physical responses studied in the following. They are expected to be very different from the responses of a pseudospin system $H_\text{s}(\mathbf{q})$ with Berry monopole, as the latter has a charge-conjugation parity (CP) symmetry $\mathcal{C}^{-1}H_\text{s}(\mathbf{q})\mathcal{C}=-H^*_\text{s}(\mathbf{q})$ with $\mathcal{C}=\text{exp}(i\pi S_y)$ instead of a chiral symmetry, and a full rotation symmetry $\comm{\mathbf{L}+\mathbf{S}}{H_\text{s}(\mathbf{q})}=0$ instead of an axial one.

\subsection{Physical properties of the continuum models}
\label{physcon}

 We now illustrate the impact of the Berry dipole on Landau levels, anomalous Hall conductivity, and magnetoconductivity. We first consider these effects for a single multifold crossing described by a continuum model $H_N^\xi(\mathbf{q})$, cf. Eq. (\ref{mods}). Below, we will present tight-binding models for MMHSs (with one or more Berry dipoles in the Brillouin zone) and discuss these effects on the lattice.

\subsubsection{Landau levels}

 Consider Eq. (\ref{mods}) for a strong magnetic field
 \begin{equation}
\mathbf{B}=B\hat{\mathbf{B}},
\label{Borient}
 \end{equation}
 where $\hat{\mathbf{B}}=(0,\sin\theta,\cos\theta)$ without loss of generality due to the axial rotation symmetry (\ref{axsym}).
The LLs form a 1D dispersion in terms of the conserved momentum $q_0=\hat{\mathbf{B}}\cdot\mathbf{q}$, and are particle-hole symmetric due to the fact that the magnetic field does not break the chiral symmetry (\ref{chirsym}) of the zero-field Hamiltonian.

\begin{figure}
	\centering
	\includegraphics[width=\columnwidth]{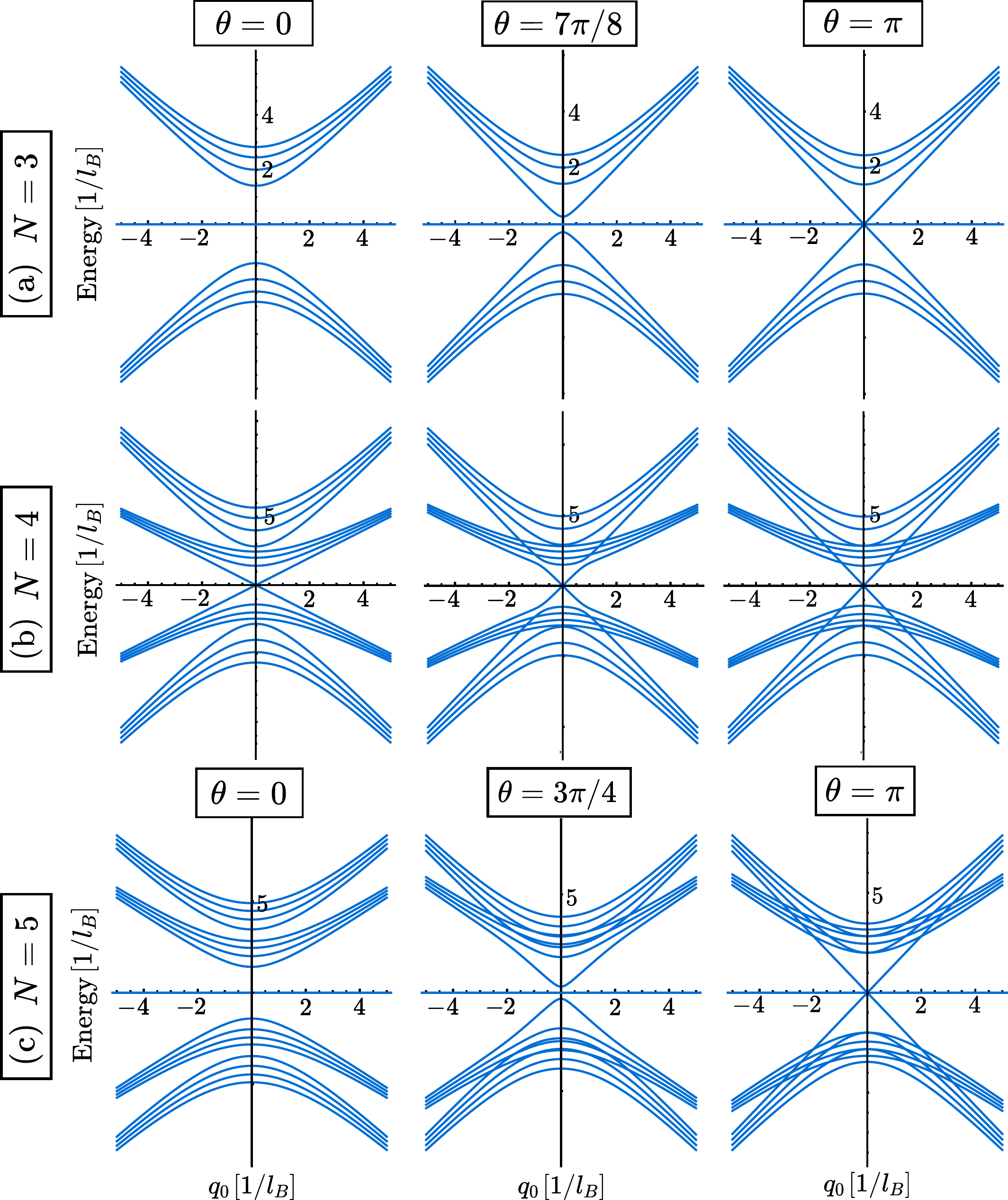}
	\caption{(a) LL spectrum  (\ref{LL3}) for $\xi=+$ for three different $\theta$, in units of the inverse magnetic length $1/l_B\equiv\sqrt{eB}$. It can be tuned from gapped to gapless by rotating $\mathbf{B}$. (b) LLs (\ref{LL4}) for $\xi=+$, $a/2=b=1$. They are gapless for any $\theta$. (c) LLs (\ref{LL5}) for $\xi=+$. They behave similarly to the three-band case.}
	\label{fig:LLs_H3}
\end{figure}

The exact Landau levels (LLs) for a threefold crossing (\ref{Berrydip3}) are given by
\begin{equation}
	\begin{aligned}
		\epsilon_\alpha^{n,\xi}&=c_\alpha\sqrt{2eB\left(n+\frac{1-\kappa_\alpha\xi\cos\theta}{2}\right)+q_0^2},
	\end{aligned}
	\label{LL3}
\end{equation}
with $c_\alpha=0,\pm1$ and $n\in\{0,1,...\}$ the LL index (see Appendix \ref{AppLL} for details of the calculation). As expected, the flat band is maintained under the magnetic field. More importantly, the dispersive bands carry a clear signature of the Berry dipole's charge ($\kappa_\alpha$) and orientation ($\xi\cos\theta\equiv\hat{\mathbf{B}}\cdot\mathbf{d}$). As a consequence, the LL spectrum strongly depends on the magnetic field direction: it is gapped for $\mathbf{B}\upharpoonleft\upharpoonright\mathbf{d}$, and gapless for $\mathbf{B}\upharpoonleft\downharpoonright\mathbf{d}$, see Fig. \ref{fig:LLs_H3}(a). 
We should like to emphasize that this tunability is a \emph{pure quantum geometric effect}. Indeed, the magnetic field couples to the eigenstates of the Hopf semimetal and thus unveils the Berry dipole via the magnetic energy levels; in contrast, the existence of the Berry dipole is invisible in the fully isotropic zero-field energy spectrum.

Similarly, we find the exact LLs for a fourfold crossing (\ref{Berrydip4}):
\begin{equation}
	\begin{aligned}
		\epsilon_\alpha^{n,\xi}&=\alpha_1\sqrt{\eta_++\eta_-+\alpha_2\sqrt{(\eta_+-\eta_-)^2+\nu^2}}, \\
		\eta_\pm&=\frac{c_\pm^2}{2}\left[2eB\left(n+\frac{1-\kappa_\pm\xi\cos\theta}{2}\right)+q_0^2\right],
	\end{aligned}
	\label{LL4}
\end{equation}
where $n\in\{0,1,...\}$.
Here we use a band index tuple $\alpha=(\alpha_1,\alpha_2)$ to capture the four families of Landau bands, with $\alpha_1=\pm$ and $\alpha_2=\pm$; $c_+=a$ and $c_-=b$ are the band velocities of the two cones of the zero-field spectrum, and \smash{$\kappa_\pm=\mp1$} the corresponding dipole charges; moreover, $\nu=c_+c_-e(\mathbf{B}\times\mathbf{d})_x=abeB\xi \sin\theta$. Again, the Berry dipole (\ref{berryungapped}) explicitly appears and the LLs can be tuned by rotating $\mathbf{B}$, see Fig. \ref{fig:LLs_H3}(b). The precise character of this tunability, however, is now quite different, in particular the LL spectrum remains gapless for any orientation of $\mathbf{B}$.

Finally, the LLs of the fivefold crossing (\ref{Berrydip5}) are given by
\begin{equation}
	\begin{aligned}
		\e_\alpha^{n,\xi}&=\alpha_1\sqrt{\eta_++\eta_-+\alpha_2\sqrt{(\eta_+-\eta_-)^2+\tilde{\nu}^2}},\\
		\eta_\pm&=\frac{c_\pm^2}{2}\left[2eB\left(n+\frac{1-\kappa_\pm\xi\cos\theta}{2}\right)+q_0^2\right],
	\end{aligned}
	\label{LL5}
\end{equation}
where $n\in\{1,2,...\}$.
Now we have $\alpha_1=0,\pm$ and $\alpha_2=\pm$. The band velocities of the two cones are $c_+=\sqrt{2}$ and $c_-=1$, the corresponding dipole charges are $\kappa_+=-3$ and $\kappa_-=1$, and 
$\tilde{\nu}=2\sqrt{3}(\mathbf{B}\times\mathbf{d})_x=2\sqrt{3}\xi eB\sin\theta$. This five-band spectrum behaves similarly to the three-band spectrum (\ref{LL3}), notably it can be tuned from gapped to gapless by rotating the magnetic field relative to the Berry dipole direction, see Fig. \ref{fig:LLs_H3}(c).

To contextualize these results, it is useful to compare to the LL spectrum of several known systems with ``Dirac-like" band crossings. First, there is clearly a big difference with the LLs of pseudospin-$s$ systems $H_\text{s}(\mathbf{q})$, which are \emph{independent} of the magnetic field orientation $\hat{\mathbf{B}}$ due to full rotation symmetry. Moreover, for a pseudospin-like crossing, the topological character of the Berry monopole (\ref{berrymon}) is reflected in the LL spectrum via the existence of chiral LLs. These are modes connecting two families of LLs with different band index $\alpha$. The number of chiral LLs is directly determined by the Chern number $C_\alpha$, see for example Ref. \cite{Bradlyn_2016} for the pseudospin-1 case, and Refs. \cite{Ezawa_2017b,Delplace_2022} for the case of general $s$. Also, for a pseudospin-$s$ crossing with integer $s$, the flat band of the zero-field spectrum is destroyed since $\mathbf{B}$ breaks the CP symmetry.

There is also a big difference with the LLs of Dirac fermions \cite{Dirac_1928}. Since the Dirac Hamiltonian has full rotation symmetry, the LLs are independent of $\hat{\mathbf{B}}$. However, in contrast to pseudospin fermions, Dirac fermions feature chiral symmetry and thus the Landau level spectrum remains particle-hole symmetric. Indeed, if we allow the case $a=b$ (which we have so far excluded) in the model (\ref{Berrydip4}), then this model becomes a Dirac semimetal \smash{$H_\text{D}(\mathbf{q})=\mathbf{q}\cdot\boldsymbol{\Gamma}$} with anticommuting matrices $\Gamma_{x,y,z}$. Accordingly, in the limit $a=b$ we recover from Eq. (\ref{LL4}) the famous LL spectrum of Dirac fermions $\e_\pm^n=\pm(2eBn+q_0^2)^{1/2}$, established a long time ago by Rabi \cite{Rabi_1928}. 

Finally, one can also compare the LLs (\ref{LL3})--(\ref{LL5}) of the Hopf semimetals to those of an extended Berry dipole $\mathbf{d}_0$ formed from two Weyl nodes (or more generally two Berry monopole crossings in a chiral multifold semimetal). The latter obviously depend on the direction of $\hat{\mathbf{B}}$ since the dipole axis $\mathbf{d}_0$ induces an anisotropy \cite{Saykin_2018}. However, this dependence is quite distinct from the one of Eqs. (\ref{LL3})--(\ref{LL5}), in particular due to the broken particle-hole symmetry of the spectrum and the presence of connected chiral LLs originating from the two valleys.

As a physical consequence of these differences, one can expect that quantum oscillations (de Haas-van Alphen or Shubnikov-de Haas effects) in a system with band crossings of Berry dipole type should be fundamentally different from those encountered in systems with Berry monopole crossings, in particular regarding the angular dependence of the oscillation frequency.

To close this discussion on the LLs of the Hopf semimetals (\ref{mods}), we emphasize that some useful insight can also be obtained from a semiclassical analysis. Indeed, aside from the quantum approach described above, it is possible to establish Eq. (\ref{LL3}) using Onsager's semiclassical quantization condition \cite{Onsager_1952}. More precisely, one needs to employ an extended Onsager condition that takes into account \emph{intraband} corrections due to Berry curvature and orbital magnetic moment \cite{Roth_1966,Mikitik_1999,Fuchs_2010,GaoNiu_2017,Fuchs_2018}. More importantly, the semiclassical approach helps to understand the origin of the terms $\nu\sim|\mathbf{B}\times\mathbf{d}|$ in Eq. (\ref{LL4}) and $\tilde{\nu}\sim|\mathbf{B}\times\mathbf{d}|$ in Eq. (\ref{LL5}). These terms remain unexplained in the quantum approach but find an intuitive semiclassical interpretation in terms of \emph{interband} coupling between degenerate orbits. Such coupling arises whenever a constant energy curve intersects more than one band, as is unavoidable for a zero-field spectrum consisting of two or more cones. For more details, see Appendix \ref{AppSemi}, where we develop an original approach to semiclassical Landau quantization of degenerate orbits.

\subsubsection{Anomalous Hall effect and magnetotransport} 

Consider now a multifold crossing (\ref{mods}) in the presence of weak electric and magnetic fields $\mathbf{E}$ and $\mathbf{B}$. We adopt a standard approach to describe the linear response of the system, by solving the semiclassical Boltzmann equation in the relaxation time ($\tau$) approximation to first order in $\mathbf{E}$ and $\mathbf{B}$ \cite{Ziman_1960}. This approach is reviewed in detail in Appendix \ref{Appboltzi}.

Taking into account important corrections due to Berry curvature and orbital magnetic moment \cite{Xiao_2010}, and working at zero temperature, we find several nontrivial effects (see Appendix \ref{Apptransp} for a derivation):

(i) At zeroth order in $\mathbf{B}$, a single multifold crossing point described by a Hamiltonian of the form (\ref{mods}) causes a non-dissipative anomalous Hall (AH) current
\begin{equation}
\mathbf{j}_\text{AH}=\sigma_\text{AH}\mathbf{E}\times\mathbf{d},
\label{AHcur}
\end{equation}
which is orthogonal both to the electric field and the Berry dipole direction. This is intuitively understood as a consequence of the Berry dipole acting as a dual magnetic field in momentum space.
Indeed, the form of the current (\ref{AHcur}) is reminiscent of the anomalous Hall current \smash{$\mathbf{j}_\text{AH}^\text{W}=\sigma_\text{AH}^\text{W}\mathbf{E}\times\mathbf{d}_0$} that is known to be created by an extended Berry dipole consisting of a pair of coupled Weyl nodes \cite{Klinkhamer_2005,Burkov_2011,Yang_2011}; here $\mathbf{d}_0$ may represent the distance between Weyl nodes or a tilt. 

\begin{figure}
	\centering
	\includegraphics[width=\columnwidth]{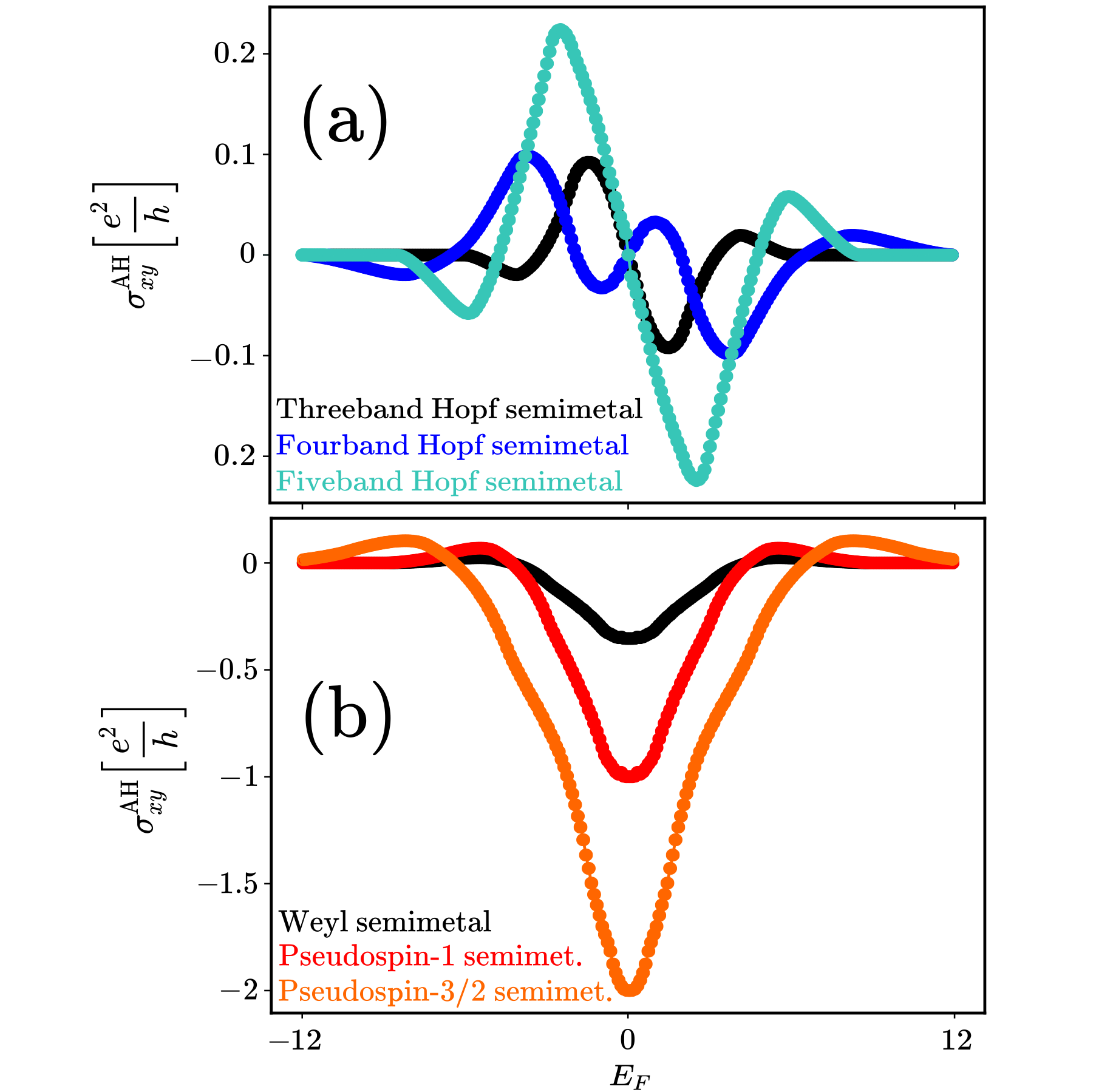}
	\caption{(a) Anomalous Hall conductivity for lattice models $\mathfrak{h}_N(\mathbf{k})$ [cf. Eq. (\ref{truesem})] of { massless} multifold Hopf semimetals. The parameter $\Delta_0=-3$ is chosen such as to ensure the existence of a single Berry dipole at the $\Gamma$ point, described by a continuum Hamiltonian (\ref{mods}). The odd parity of the AH conductivity is due to the property $\kappa_\alpha=\kappa_{-\alpha}$ of the geometric charges of this Berry dipole. (b) Typical anomalous Hall conductivity for lattice models of chiral topological semimetals with a single pair of Berry monopoles [cf. Eq. (\ref{weyleq})]. The even parity is due to the property $C_\alpha=-C_{-\alpha}$ of the Chern number associated to each Berry monopole.}
	\label{fig:AHeffect}
\end{figure}

Importantly, however, the current (\ref{AHcur}) is opposite to \smash{$\mathbf{j}_\text{AH}^\text{W}$} in parity: $\sigma_\text{AH}$ is \emph{odd} in the Fermi level $E_F$, that is, $\sigma_\text{AH}(E_F)=-\sigma_\text{AH}(-E_F)$, while \smash{$\sigma_\text{AH}^\text{W}$} is an \emph{even} function of $E_F$. This striking property predicted by the continuum models will be confirmed by a numerical lattice calculation below and is visualized in Fig. \ref{fig:AHeffect}.
 It can be understood as a direct consequence of the different symmetries of the Berry curvature ($\boldsymbol{\Omega}_\alpha=\boldsymbol{\Omega}_{-\alpha}$ for a MMHS vs. $\boldsymbol{\Omega}_\alpha=-\boldsymbol{\Omega}_{-\alpha}$ for a Weyl semimetal), as discussed in detail in Appendix \ref{Apptransp}.

 (ii) At first order in $\mathbf{B}$, a single multifold crossing point described by a Hamiltonian of the form (\ref{mods}) causes a dissipative quantum geometric current
 \begin{equation}
 	\mathbf{j}_\text{geo}(\mathbf{B})=A_1(\mathbf{E}\cdot\mathbf{B})\mathbf{d}+A_2(\mathbf{E}\cdot\mathbf{d})\mathbf{B}+A_3(\mathbf{B}\cdot\mathbf{d})\mathbf{E}.
 	\label{magcur}
 \end{equation}
 This current is absent if the electric field, magnetic field and Berry dipole direction form an orthogonal tripod, but non-vanishing for any other configuration.
 
 The explicit expressions for the coefficients $A_i$ ($i=1,2,3$) in units of $A_0\equiv -e^3\tau/(12\pi^2)$ are as follows (see Appendix \ref{Apptransp} for a derivation): $A_i=A_0$ for the continuum model (\ref{Berrydip3}) describing a threefold Hopf semimetal; similarly $A_i=A_0(a-b)$ for model (\ref{Berrydip4}); and $A_{1,2}=(21\sqrt{2}-17)A_0/5$, $A_3=(23+\sqrt{2})A_0/5$ for model (\ref{Berrydip5}). As we can see, these coefficients are independent of the Fermi level $E_F$. This is of course not true on the lattice, but it implies that the magnetoconductivity can be expected to be an \emph{even function} of $E_F$. Indeed, as for the anomalous Hall current above, this parity property can be readily understood from general symmetry arguments, more precisely from the combined effect of a particle-hole symmetric spectrum and a Berry curvature $\boldsymbol{\Omega}_\alpha=\boldsymbol{\Omega}_{-\alpha}$. See Appendix \ref{Apptransp} for a short proof.
 
 Again, let us compare the current (\ref{magcur}) to that caused by an extended Berry dipole $\mathbf{d}_0$ formed from a pair of Berry monopoles. As a matter of fact, it is well known that a pair of coupled Weyl nodes gives rise to a current $\mathbf{j}_\text{CA}\sim(\mathbf{E}\cdot\mathbf{B})\mathbf{d}_0$, which is attributed to the chiral anomaly \cite{Nielsen_1983,Zyuzin_2017,Sharma_2017}. Moreover, it gives rise to a current $\mathbf{j}_\text{CME}\sim\delta\e\,\mathbf{B}$, where $\delta\e$ is an energy difference between the valleys. This is known as the chiral magnetic effect \cite{Vilenkin_1980,Fukushima_2008,Zyuzin_2012,Zyuzin_2017}. Finally, a pair of Weyl nodes exhibits a current $\mathbf{j}_\text{MCE}\sim(\mathbf{B}\cdot\mathbf{d}_0)\mathbf{E}$, known as the magnetochiral effect \cite{Cortijo_2016,Kundu_2020}. Similar kinds of currents exist for pairs of pseudospin crossings with $s>1/2$. The three current contributions $\mathbf{j}_\text{CA}$, $\mathbf{j}_\text{CME}$, and $\mathbf{j}_\text{MCE}$ are \emph{odd functions} of $E_F$, essentially due to the Berry curvature property $\boldsymbol{\Omega}_\alpha=-\boldsymbol{\Omega}_{-\alpha}$. Again a short proof is provided in Appendix \ref{Apptransp}.
 
 In summary, each term of the linear magnetocurrent (\ref{magcur}) caused by a point-like Berry dipole crossing (\ref{mods}) has a counterpart in the response of an extended Berry dipole formed from two topological monopoles. However, just like for the anomalous Hall effect, the currents in both systems have opposite parity as a function of the filling. 

 \section{Tight-binding models for massless multifold Hopf semimetals}
 \label{Seclat}
 
 \subsection{Description of the lattice models}
 
 We now demonstrate how the continuum models discussed above can be obtained as the low-energy limit of lattice models. In particular, we introduce two different classes of tight-binding models for { massless} multifold Hopf semimetals, both of which recover Eq. (\ref{mods}) in the vicinity of high-symmetry points of the Brillouin zone.
 
\begin{figure}
 	\centering
 	\includegraphics[width=\columnwidth]{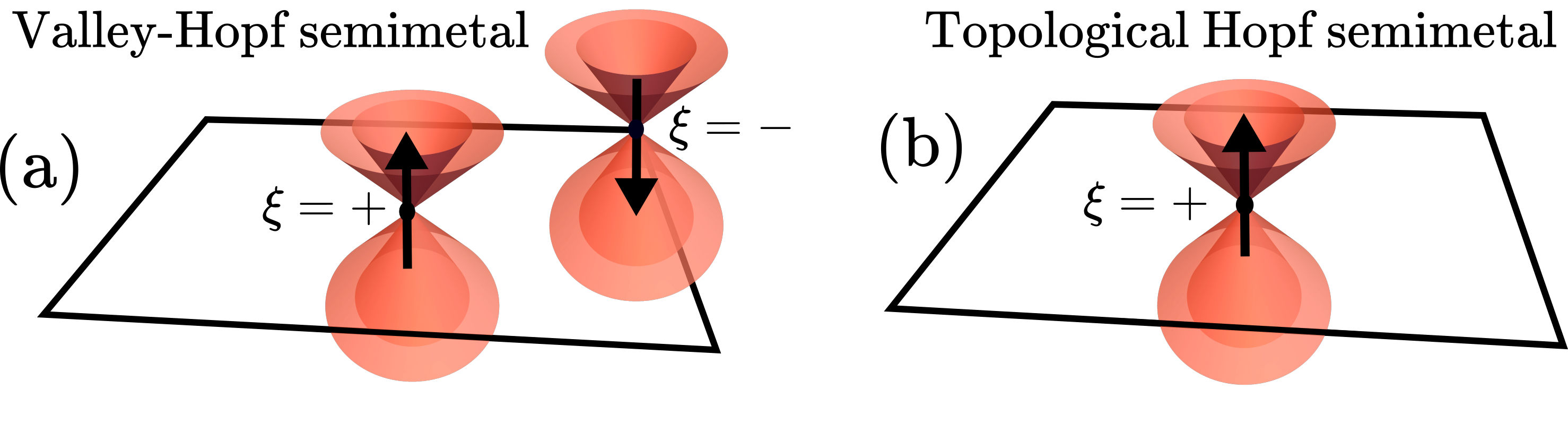}
 	\caption{Valley-Hopf semimetals (topological Hopf semimetals) have an even (odd) number of linear multifold crossings with Berry dipole.}
 	\label{fig:valley_notvalley}
 \end{figure}

 The first class consists of semimetals that have an even number of crossing points with Berry dipole in the Brillouin zone. The crossings can be arranged in pairs $\xi=\pm$, with opposite dipole orientation in each valley [Fig. \ref{fig:valley_notvalley}(a)]. 
 These systems will be called \emph{valley-Hopf semimetals}. The second class comprises semimetals with an odd number of Berry dipole crossings in the Brillouin zone. These systems will be called \emph{topological Hopf semimetals} for reasons that will become clear in Section \ref{insulsec}.
 We first present a selection of examples for the two classes, and then describe their physical properties.

 \subsubsection{Valley-Hopf semimetals}
 
Among lattice models with an even number of Berry dipole crossings, we can further distinguish between models $h_N(\mathbf{k})$ with preserved time-reversal symmetry and models  $\tilde{h}_N(\mathbf{k})$ with broken time-reversal symmetry. 
 
\begin{figure*}
 	\centering
 	\includegraphics[width=\textwidth]{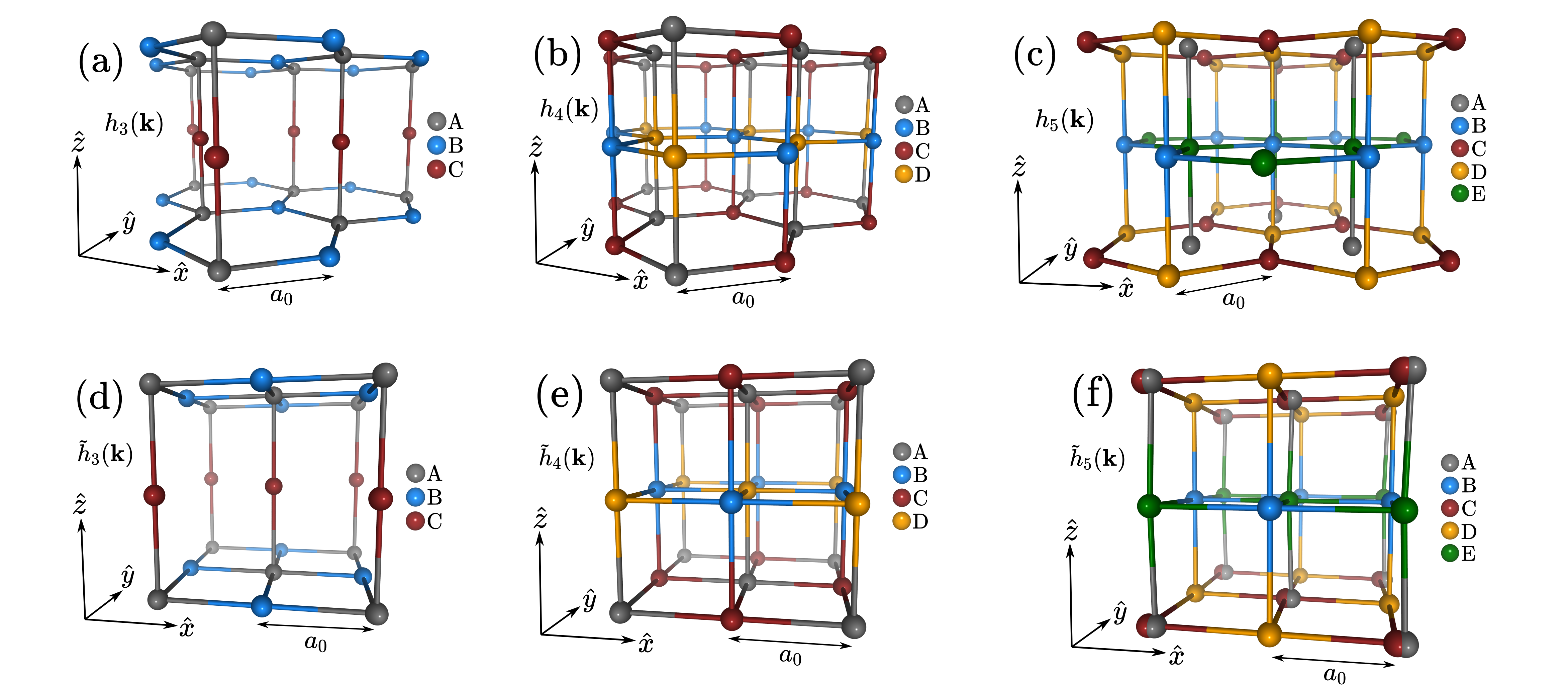}
 	\caption{Tight-binding models for threefold, fourfold and fivefold valley-Hopf semimetals (a)--(c) with time-reversal symmetry on a 3D hexagonal lattice. Links are non-zero hoppings of the tight-binding Hamiltonian (\ref{subst}). (d)--(f) with broken time-reversal symmetry on a 3D tetragonal lattice, as given by the tight-binding Hamiltonian (\ref{subst2}).}
 	\label{fig:4and5hex}
 \end{figure*}

 \emph{Valley-Hopf semimetals with time-reversal symmetry.} As examples for this kind of lattice models we choose Bloch Hamiltonians of the form
 \begin{equation}
 	h_N(\mathbf{k})=\begin{pmatrix}
 		0 & \mathcal{Q}_N\\
 		\mathcal{Q}_N^\dagger & 0
 	\end{pmatrix},
 	\label{subst}
 \end{equation}
 where
 \begin{equation}
 \begin{aligned}
 	\mathcal{Q}_3&=\begin{pmatrix}
 		w_\mathbf{k}& -i \sin k_z
 	\end{pmatrix}, \\
 	\mathcal{Q}_4&=\begin{pmatrix}
 		aw_\mathbf{k} & ia \sin k_z\\
 		ib\sin k_z & bw^*_\mathbf{k}
 	\end{pmatrix}, \\
 	\mathcal{Q}_5&=\begin{pmatrix}
 		0 & i\sqrt{2}\sin k_z\\
 		i\sin k_z&  w^*_\mathbf{k}\\
 		\sqrt{2}w^*_\mathbf{k} & 0
 	\end{pmatrix}.
 \end{aligned}
 \end{equation}
 Here $w_\mathbf{k}\equiv\frac{2}{3}\sum_j\text{exp}(i\mathbf{k}\cdot\boldsymbol{\delta}_j)$, where $\boldsymbol{\delta}_{1,2}=\frac{1}{2}(\pm\sqrt{3},1,0)$ and $\boldsymbol{\delta}_3=(0,-1,0)$.
  The Hamiltonians (\ref{subst}) represent nearest-neighbor tight-binding models on a hexagonal Bravais lattice, with Bravais vectors $\mathbf{a}_1=\sqrt{3}\hat{x}$, $\mathbf{a}_2=\frac{1}{2}(\sqrt{3}\hat{x}+3\hat{y})$, $\mathbf{a}_3=2\hat{z}$. Indeed, the models describe 2D honeycomb layers (as in graphene), stacked in a particular way along the $\hat{z}$ direction, as visualized in Fig. \ref{fig:4and5hex}(a)--(c). 
 The tight-binding models have the following nearest-neighbor hoppings in real space. The three-band model $h_3(\mathbf{k})$  has isotropic hoppings $t_{AB}=2/3$ in the A-B planes and alternating hoppings $t_\text{AC}=\pm 1/2$ along the A-C direction, see Fig. \ref{fig:4and5hex}(a). The four-band model $h_4(\mathbf{k})$ has isotropic hoppings $t_\text{AC}=2a/3$ and $t_\text{BD}=2b/3$ in the A-C and B-D planes, respectively, as well as alternating hoppings $t_\text{AD}=\pm a/2$ and $t_\text{BC}=\pm b/2$ in the perpendicular direction, see Fig. \ref{fig:4and5hex}(b). Finally,  
 the five-band model $h_5(\mathbf{k})$ has isotropic hoppings $t_\text{BE}=2/3$ and $t_\text{CD}=2\sqrt{2}/3$, as well as alternating hoppings $t_\text{BD}=\pm1/2$ and $t_\text{AE}=\pm1/\sqrt{2}$ along the vertical direction, see Fig. \ref{fig:4and5hex}(c). 
 
 The band structure of the models $h_N(\mathbf{k})$ is given by
 \begin{equation}
 	\epsilon_\alpha(\mathbf{k})=\frac{2}{3}c_\alpha\sqrt{f(k_x,k_y)+\frac{9}{4}\sin^2k_z}.
 	\label{valleyspec}
 \end{equation}
 It is particle-hole symmetric due to an obvious chiral symmetry (\ref{chirsym}) of the Bloch Hamiltonian, and the coefficients $c_\alpha$ are the same as for the continuum models (\ref{mods}), see Table \ref{tab0}. The function $f(k_x,k_y)=3+2\cos(\sqrt{3}k_x)+4\cos(\sqrt{3}k_x/2)\cos(3k_y/2)$ describing in-plane hopping is exactly the same as for graphene \cite{Castro_2009}.

 Since, for all models $h_N(\mathbf{k})$, the Bravais period is doubled along the $\hat{z}$ direction, such that the Brilloun zone extends from $k_z=-\pi/2$ to $k_z=\pi/2$, it is clear that nodal points in the spectrum (\ref{valleyspec}) can appear only in the $k_z=0$ plane. 
 Indeed, the hexagonal Brillouin zone contains one $N$-fold nodal point at the K ($\xi=+$) and one at the K' ($\xi=-$) valley, with coordinates $\mathbf{K}_\xi=-\xi\frac{4\pi}{3\sqrt{3}}(1,0)$. These nodal points are described exactly by the continuum models (\ref{mods}) at low energy, as can be easily seen by noting that $w_\mathbf{k}\rightarrow \xi q_x-iq_y$ around these points. Thus, there is a Berry dipole pointing up in the K valley and one pointing down in the K' valley.

 \emph{Valley-Hopf semimetals without time-reversal symmetry.} As examples for this class of lattice models we consider Bloch Hamiltonians of the form
 \begin{equation}
 	\tilde{h}_N(\mathbf{k}) =\begin{pmatrix}
 		0 & \tilde{\mathcal{Q}}_N\\
 		\tilde{\mathcal{Q}}_N^\dagger& 0
 	\end{pmatrix},
 	\label{subst2}
 \end{equation}
 where
 \begin{equation}
 \begin{aligned}
 	\tilde{\mathcal{Q}}_3&=\begin{pmatrix}
 		s_- & -i \sin k_z
 	\end{pmatrix}, \\
 	\tilde{\mathcal{Q}}_4&=\begin{pmatrix}
 		as_- & ia \sin k_z\\
 		ib\sin k_z & bs_+
 	\end{pmatrix},  \\
 	\tilde{\mathcal{Q}}_5&=\begin{pmatrix}
 		0 & i\sqrt{2}\sin k_z\\
 		i\sin k_z&  s_+\\
 		\sqrt{2}s_+ & 0
 	\end{pmatrix}.
 \end{aligned}
 \end{equation}
 Here we use shorthand notations $s_\pm\equiv \sin k_x\pm i \sin k_y$. The Hamiltonians $\tilde{h}_N(\mathbf{k})$ represent nearest-neighbor tight-binding models on a tetragonal Bravais lattice, with Bravais vectors $\mathbf{a}_1=\hat{x}+\hat{y}$, $\mathbf{a}_2=\hat{x}-\hat{y}$, and $\mathbf{a}_3=2\hat{z}$. Indeed, the models describe 2D square layers stacked in a particular way along the $\hat{z}$ direction, as shown in Fig. \ref{fig:4and5hex}(d)--(f).
 Note that the five-band model has two types of orbitals (A and C) located at the same site.
 
 The three-band model $\tilde{h}_3(\mathbf{k})$ has alternating hoppings $t_\text{AB}=\pm i/2$ ($t_\text{AB}=\pm 1/2$) in the $\hat{x}$-direction ($\hat{y}$-direction) within the A-B plane and alternating hoppings $t_\text{AC}=\pm 1/2$ along the A-C direction, see Fig. \ref{fig:4and5hex}(d). Similarly, the model $\tilde{h}_4(\mathbf{k})$ has alternating hoppings $t_\text{AC}=\pm ia/2$ ($t_\text{AC}=\pm a/2$) in the $\hat{x}$ direction ($\hat{y}$ direction) within the A-C plane, alternating hoppings $t_\text{BD}=\pm ib/2$ ($t_\text{BD}=\pm b/2$) in the $\hat{x}$ direction ($\hat{y}$ direction) within the B-D plane, and alternating hoppings $t_\text{AD}=\pm a/2$ and $t_\text{BC}=\pm b/2$ in the perpendicular direction, see Fig. \ref{fig:4and5hex}(e). Finally, the five-band model $\tilde{h}_5(\mathbf{k})$
 has alternating hoppings $t_\text{BE}=\pm i/2$ ($t_\text{BE}=\pm 1/2$) in the $\hat{x}$ direction ($\hat{y}$ direction) within the B-E plane, alternating hoppings $t_\text{CD}=\pm i/\sqrt{2}$ ($t_\text{CD}=\pm 1/\sqrt{2}$) in the $\hat{x}$ direction ($\hat{y}$ direction) within the C-D plane, as well as alternating hoppings $t_\text{AE}=\pm1/\sqrt{2}$ and $t_\text{BD}=\pm1/2$ along the vertical direction, see Fig. \ref{fig:4and5hex}(f).
 
 The corresponding band structure of the models $\tilde{h}_N(\mathbf{k})$ is given by
 \begin{equation}
 	\epsilon_\alpha(\mathbf{k})=c_\alpha\sqrt{\sin^2k_x+\sin^2k_y+\sin^2k_z},
 \end{equation}
 which is again particle-hole symmetric due to a chiral symmetry, and where the band velocities $c_\alpha$ are the same as for the continuum models (\ref{mods}), see Table \ref{tab0}.	
 Again nodal points can appear only in the $k_z=0$ plane. Indeed, the tetragonal Brillouin zone contains two $N$-fold nodal points, namely one located at the $\Gamma$ point ($\xi=+$) and one at the M point ($\xi=-$), where $\mathbf{k}_\Gamma=0$ and $\mathbf{k}_\text{M}=(\pi,0,0)$. It is straightforward to see that the low-energy theory around these points is exactly described by the continuum models (\ref{mods}).

\subsubsection{Topological Hopf semimetals}

\begin{figure}
	\centering
	\includegraphics[width=\columnwidth]{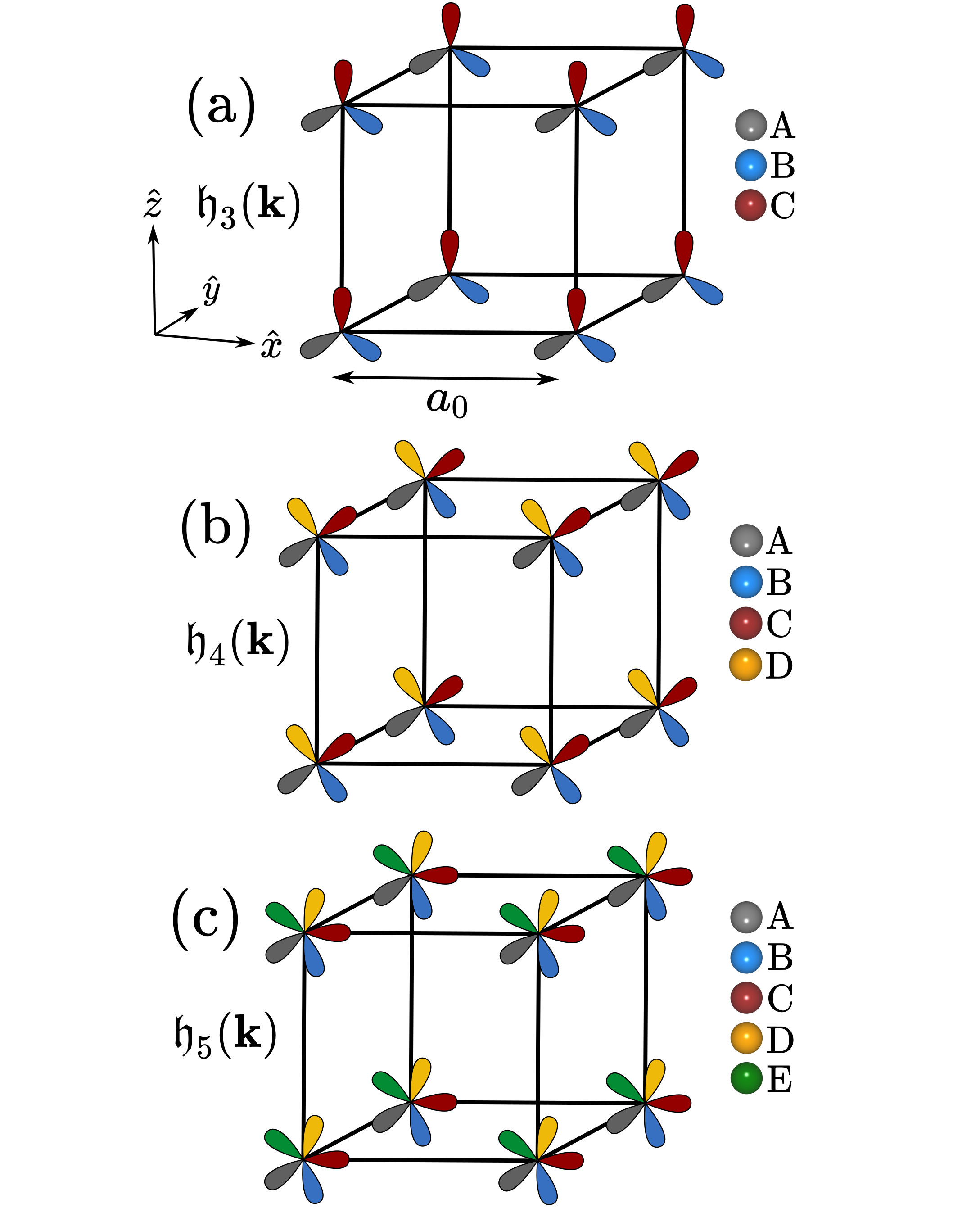}
	\caption{(a)--(c) Tight-binding models for threefold, fourfold and fivefold topological Hopf semimetals can be constructed on a multiorbital cubic lattice, with hoppings as given by the tight-binding Hamiltonian (\ref{truesem}). The same lattice structure allows to realize the tight-binding models (\ref{hopfdef}) for multiband Hopf insulators.}
	\label{fig:multihopf}
\end{figure}

We now come to a second class of MMHSs, characterized by an odd number of Berry dipoles in the Brillouin zone. As examples for such topological Hopf semimetals, we consider Bloch Hamiltonians of the form 
\begin{equation}
	\begin{aligned}
		\mathfrak{h}_N(\mathbf{k})&=\begin{pmatrix}
			0 &\mathfrak{Q}_N\\
			\mathfrak{Q}^\dagger_N & 0
		\end{pmatrix},
	\end{aligned}
	\label{truesem}
\end{equation}
where 
\begin{equation}
\begin{aligned}
	\mathfrak{Q}_3&=\begin{pmatrix}
		s_- & g_{\Delta_0}
	\end{pmatrix}, \\
 \mathfrak{Q}_4&=\begin{pmatrix}
		as_- & -ag_{\Delta_0}\\
		bg^*_{\Delta_0} & bs_+
	\end{pmatrix},\\
 \mathfrak{Q}_5&=\begin{pmatrix}
		0 & \sqrt{2}g^*_{\Delta_0}\\
		g^*_{\Delta_0} & s_+\\
		\sqrt{2}s_+ & 0
	\end{pmatrix}.
\end{aligned}
\label{Q345}
\end{equation}
Here we use again $s_\pm=\sin k_x\pm i \sin k_y$ and 
\begin{equation}
	g_{\Delta_0}\equiv \Delta_0+\cos k_x+\cos k_y+e^{-ik_z},
\end{equation}
where $\Delta_0$ is a real parameter.
These Hamiltonians are difficult to realize as pure hopping models with only one orbital per site, however since they contain only terms $\sim \sin k_i$ or $\sim \cos k_i$ they may be constructed assuming a simple cubic Bravais lattice with $N$ orbitals per site and appropriate couplings, see Fig. \ref{fig:multihopf}. 

\begin{figure}
	\centering
	\includegraphics[width=\columnwidth]{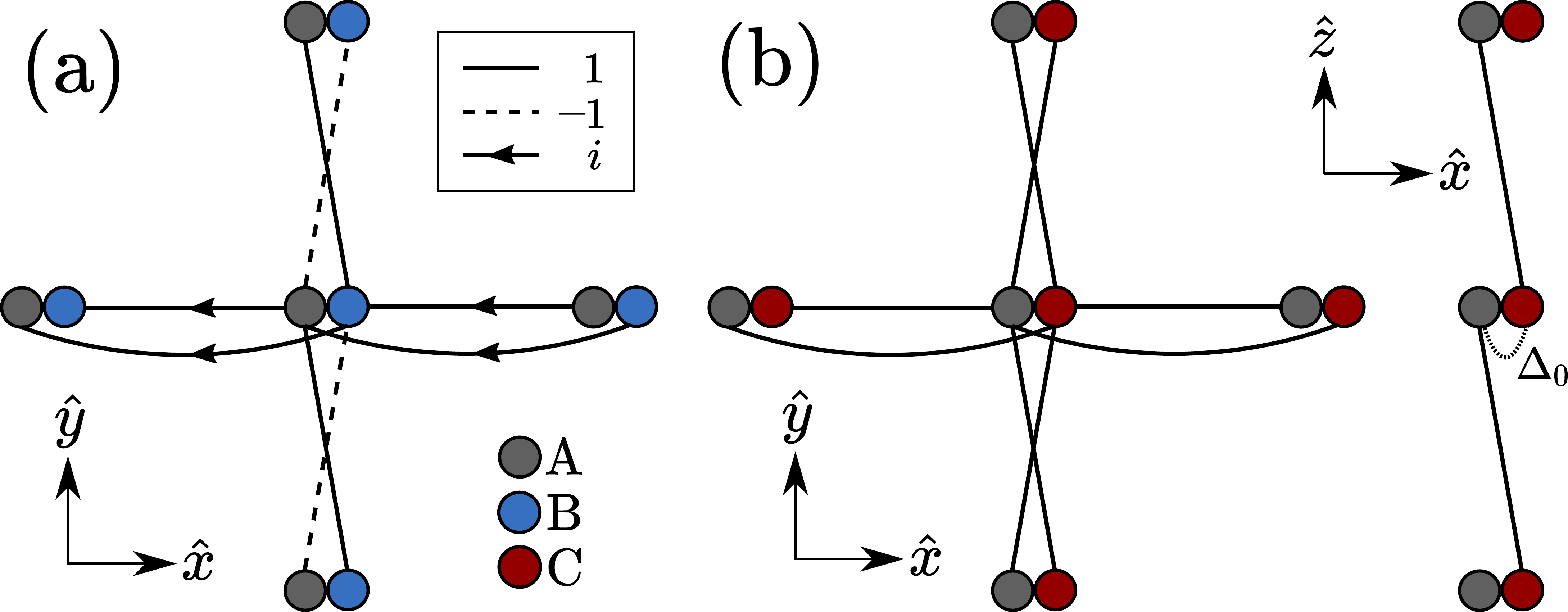}
	\caption{Real-space hopping structure giving rise to the Bloch Hamiltonian $\mathfrak{h}_3(\mathbf{k})$ for a threefold topological Hopf semimetal [and similarly to the Bloch Hamiltonian \smash{$h_3^\text{Hopf}(\mathbf{k})$} for a threeband Hopf insulator]. Solid lines correspond to hoppings $1$, dashed lines to hoppings $-1$ and lines with arrows to hoppings $i$ (in units of $1/2$). (a) Coupling between orbitals A and B. (b) Coupling between orbitals A and C.}
	\label{fig:MHShoppings}
\end{figure}

For example, the three-band semimetal $\mathfrak{h}_3(\mathbf{k})$ is characterized by the following hoppings in real space. Orbitals A are coupled to orbitals B by imaginary nearest-neighbor hoppings $\pm i/2$ along the $\hat{x}$-direction and real nearest-neighbor hoppings $\pm1/2$ along the $\hat{y}$-direction, as visualized in Fig. \ref{fig:MHShoppings}(a). Orbitals A and C are coupled by hoppings $1/2$ along the $\hat{x}$- and $\hat{y}$-directions [left panel of Fig. \ref{fig:MHShoppings}(b)], as well as alternating hoppings $0,1/2$ along the $\hat{z}$-direction and an on-site hopping $\Delta_0$ [right panel of Fig. \ref{fig:MHShoppings}(b)]. Orbitals B and C are uncoupled, which is the reason for the chiral symmetry of the Bloch Hamiltonian.
 In a similar way, one can use Eq. (\ref{Q345}) to read off the precise hopping structure for the four- and five-band semimetals $\mathfrak{h}_4(\mathbf{k})$ and $\mathfrak{h}_5(\mathbf{k})$.

The Hamiltonians $\mathfrak{h}_N(\mathbf{k})$ have an energy spectrum
\begin{equation}
	\e_\alpha(\mathbf{k})=c_\alpha\sqrt{\sum_i\sin^2 k_i+(\Delta_0+\sum_i \cos k_i)^2},
\end{equation}
where again a chiral symmetry is evident, the sum runs over $i=x,y,z$, and the coefficients $c_\alpha$ are as in Table \ref{tab0}. This spectrum
becomes gapless only for $\Delta_0=\pm1,\pm3$. For the moment, since we are interested in the semimetallic phase, we only allow the parameter $\Delta_0$ to take one of these four discrete values. 

We now show that, as desired, the semimetals that are obtained for different values of $\Delta_0$ are all described by continuum Hamiltonians of the form (\ref{mods}) at low energy. Namely, for $\Delta_0=-3$, there is a single $N$-fold crossing at the $\Gamma$ point of the cubic Brillouin zone, described by a low-energy theory $H_N^+(\mathbf{q})$, corresponding to a Berry dipole pointing up. Similarly, for $\Delta_0=3$, there is a single $N$-fold crossing at the R point, $\mathbf{k}_\text{R}=(\pi,\pi,\pi)$, described by a continuum Hamiltonian $-H_N^+(\mathbf{q})$, and thus corresponding to a Berry dipole pointing down. 
For $\Delta_0=-1$, there are three $N$-fold crossings at the inequivalent X points: $\mathbf{k}_{\text{X}1}=(\pi,0,0)$, $\mathbf{k}_{\text{X}2}=(0,\pi,0)$, $\mathbf{k}_{\text{X}3}=(0,0,\pi)$. They are described by $H_{N,\text{X1}}(\mathbf{q})=H_N^+(-q_x,q_y,q_z)$, $H_{N,\text{X2}}(\mathbf{q})=H_N^+(q_x,-q_y,q_z)$, and $H_{N,\text{X3}}(\mathbf{q})=H_N^+(q_x,q_y,-q_z)$, respectively, thus corresponding to Berry dipoles pointing down, down and up. Finally, for $\Delta_0=1$, there are three $N$-fold crossings at the M points: $\mathbf{k}_{\text{M}1}=(0,\pi,\pi)$, $\mathbf{k}_{\text{M}2}=(\pi,0,\pi)$, $\mathbf{k}_{\text{M}3}=(\pi,\pi,0)$. They are described by $H_{N,\text{Mi}}(\mathbf{q})=-H_{N,\text{Xi}}(\mathbf{q})$, thus corresponding to Berry dipoles pointing up, up, and down. The four types of high-symmetry points mentioned here ($\Gamma$, R, X, M) will play an important role below when we discuss topological phase transitions of multiband Hopf insulators.

\subsection{Physical properties of the lattice models}

 	For a Fermi level close to the nodal points, the physical properties of the semimetallic lattice models (\ref{subst}), (\ref{subst2}) and (\ref{truesem}) are simply obtained by summing the continuum results, as obtained in Section \ref{physcon}, over all valleys. 
 	
 	It is then clear that both the anomalous Hall current $\mathbf{j}_\text{AH}$, see Eq. (\ref{AHcur}), and the magnetocurrent $\mathbf{j}_\text{geo}$, see Eq. (\ref{magcur}), cancel for a valley-Hopf semimetal, since there is an even number of crossings with opposite Berry dipoles, cf. Fig. \ref{fig:valley_notvalley}(a). However, anomalous Hall and magnetocurrents are non-trivial for the topological Hopf semimetals (\ref{truesem}), since there is one net Berry dipole in the Brillouin zone. 	
 	To confirm this, we have numerically calculated the anomalous Hall conductivity of $\mathfrak{h}_N(\mathbf{k})$ for parameters such that a single Berry dipole crossing exists at the $\Gamma$ point, as shown in Fig. \ref{fig:AHeffect}(a). The numerical results confirm the previous claim that $\mathbf{j}_\text{AH}$ is odd in $E_F$.
 	A similar lattice calculation (not shown) confirms the existence of weak-field magnetocurrents $\mathbf{j}_\text{geo}$ for a topological Hopf semimetal (\ref{truesem}), which are even in $E_F$. For details of the calculation, see Appendix \ref{Apptransp}.
 
 Despite the fact that the anomalous Hall conductivity and magnetoconductivity cancel for the valley-Hopf semimetals, it is possible to conceive of other ways to unveil the presence of the Berry dipoles in these systems. For example, consider the lattice model $h_3(\mathbf{k})$ [shown in Fig. \ref{fig:4and5hex}(a)], and assume the presence of a strong magnetic field (\ref{Borient}). For $\theta=0$, we know from Fig. \ref{fig:LLs_H3}(a) that the Landau level spectrum at the K valley ($\xi=+$) will be gapped out, while the Landau level spectrum at the K' valley ($\xi=-$) will be gapless. For $\theta=\pi$, the situation is reversed. Flipping the magnetic field thus provides a means to completely switch the valley polarization of dispersive states in an energy window $E_F\in[-eB,eB]$ around the highly degenerate flat band. This effect is purely due to the Berry dipole. A similar effect exists for the lattice model $h_5(\mathbf{k})$. From a more general perspective, it appears promising to study optical and magnetooptical responses of the valley-Hopf semimetals \footnote{Interesting results along this direction were obtained in Ref. \cite{Habe_2022} after submission of this manuscript.}.

\section{Relation to multiband Hopf insulators}
\label{insulsec}
 
 In this section we connect the semimetals introduced above to the theory of Hopf insulators. The concept of a two-band Hopf insulator is by now well known \cite{Moore_2008, Deng_2013,Liu_2017}, and some suggestions for its experimental realization have appeared \cite{Deng_2018,Unal_2019,Schuster_2021,Schuster_2021a}. Moreover, a formal generalization of the Hopf insulator
 to the multiband ($N > 2$) case was achieved very recently by Lapierre
 et al. \cite{Lapierre_2021}. We now propose concrete lattice models for such multiband Hopf insulators (MHIs).
 
 In the Hopf semimetal Hamiltonian (\ref{truesem}), let us replace the discrete values $\Delta_0$ by a continuous parameter $\Delta$. We now claim that the Hamiltonians 
\begin{equation}
h_N^\text{Hopf}(\mathbf{k})\equiv\mathfrak{h}_N(\mathbf{k},\Delta\neq\Delta_0)
\label{hopfdef}
\end{equation}
are nearest-neighbor tight-binding models for MHIs. The orbital hopping structure of these models is the same as previously considered for the semimetals (\ref{truesem}), see Figs. \ref{fig:multihopf} and \ref{fig:MHShoppings}, except for the fact that the allowed values of the parameter $\Delta$, which describes on-site hopping, are now different. The corresponding energy spectrum
\begin{equation}
	\e_\alpha(\mathbf{k})=c_\alpha\sqrt{\sum_i\sin^2 k_i+(\Delta+\sum_i \cos k_i)^2}
	\label{enspec}
\end{equation}
is comprised of $N$ fully gapped bands (if $\Delta\neq\Delta_0$), where again the band velocities $c_\alpha$ are as listed in Table \ref{tab0}. 

 To confirm that Eq. (\ref{hopfdef}) defines MHIs, we have to proceed in two steps according to the rules of the topological classification for MHIs \cite{Lapierre_2021}. First, we need to verify that the three Chern numbers (weak topological invariants)
 	\begin{equation}
 	\mathcal{N}^\text{Chern}_{\alpha,ij}=\frac{1}{2\pi}\int_{\mathbb{T}^2} dk_i dk_j \Omega_{\alpha,ij}(k_i,k_j;k_l=\text{const.})
 	\label{subchern}
 \end{equation}
vanish, where $i,j\in\{x,y,z\}$. This is indeed the case and implies that the homotopy classification of the insulators under consideration is of type $\mathbb{Z}$. Second, we need to compute the Hopf number
 \begin{equation}
 	\mathcal{N}_\text{Hopf}=W_3[U(\mathbf{k})]=\int_\text{BZ}\frac{d^3k}{24\pi^2}\,\chi(\mathbf{k})
 	\label{Hopf}
 \end{equation} 
as the third winding number $W_3$ of the unitary matrix $U(\mathbf{k})$ that diagonalizes \smash{$h_N^\text{Hopf}(\mathbf{k})$}, which can be written as an integral of the \emph{Hopf density}
\begin{equation}
\chi(\mathbf{k})\equiv\e_{ijl}\Tr[u_i(\mathbf{k})u_j(\mathbf{k})u_l(\mathbf{k})]
\label{hopfdens}
\end{equation}
over the Brillouin zone, with $u_i(\mathbf{k})\equiv U^\dagger(\mathbf{k})\partial_i U(\mathbf{k})$ and the partial derivative $\partial_i\equiv\partial/\partial k_i$. 

From Eq. (\ref{Hopf}), we find a quantized Hopf number for $\Delta\neq\Delta_0$, as visualized in Fig. \ref{fig:Hopfinvariant} and as derived in more detail in Appendix \ref{Apphopf}. 
\begin{figure}
	\centering
	\includegraphics[width=\columnwidth]{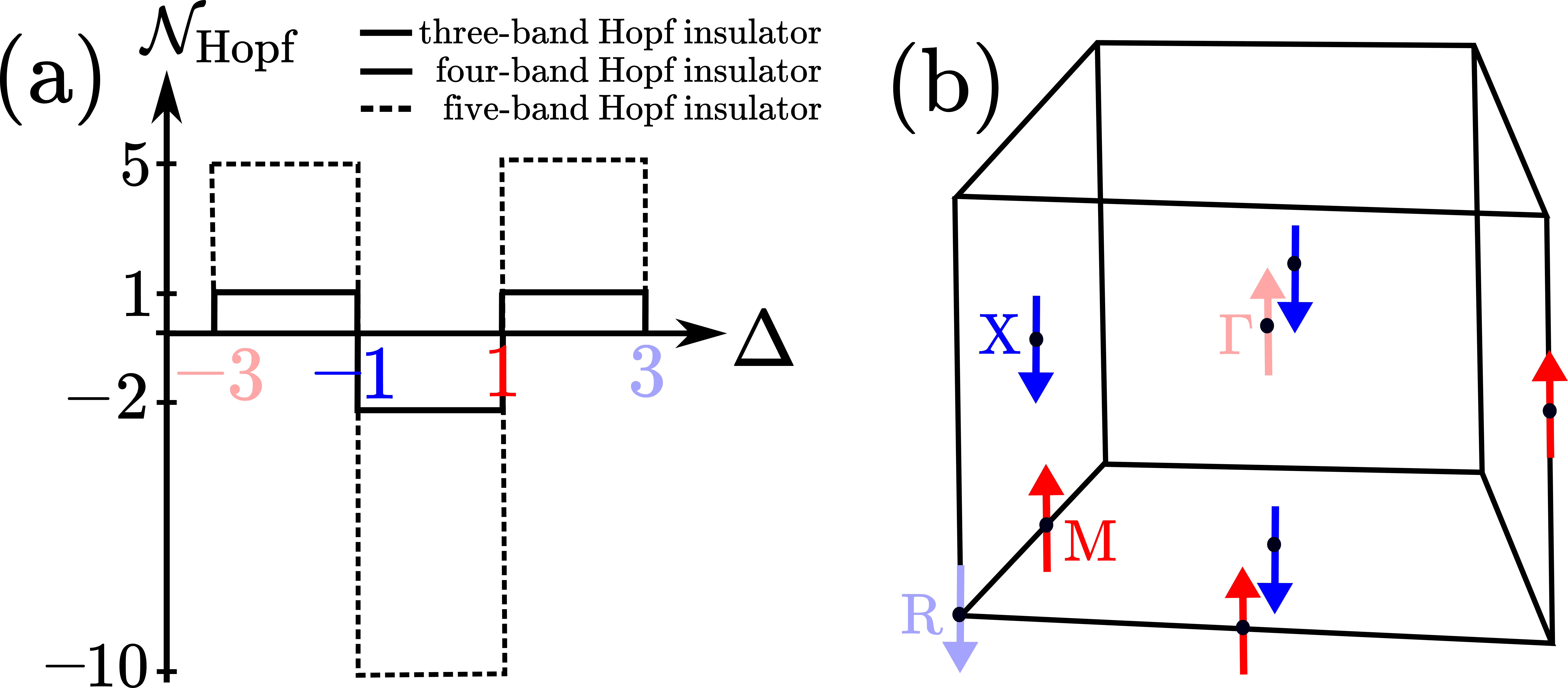}
	\caption{(a) Hopf invariant (\ref{Hopf}) for the tight-binding models (\ref{hopfdef}). Topological phase transitions occur for $\Delta=\Delta_0$, where gaps close at the $\Gamma$, X, M or R points of the Brillouin zone, see (b). At these transitions, the semimetals (\ref{truesem}) are recovered. Red (blue) colors denote a positive (negative) sign of the jump $\delta\mathcal{N}_\text{Hopf}$ of the Hopf number.}
	\label{fig:Hopfinvariant}
\end{figure} 
At the topological phase transitions $\Delta=\Delta_0$, \emph{all $N-1$ band gaps} that are in general present in the spectrum (\ref{enspec})
close simultaneously at one or three points of the Brillouin zone.
  Clearly, the topological MMHSs (\ref{truesem}) are critical points of the MHIs (\ref{hopfdef}). This is somewhat analogous to how quadratic band touchings with Berry dipole mediate topological phase transitions between two-band Hopf insulators \cite{Alexandradinata_2021,Nelson_2022}.

Two remarks are in order. First, the model $\smash{h_3^\text{Hopf}(\mathbf{k})}$ is actually well known as a \emph{chiral topological insulator} in the literature \cite{Wang_2014,Lian_2019}. It was introduced for constructing fractional topological phases \cite{Neupert_2012}, and we here identify it as a three-band Hopf insulator. Indeed, it appears that the topological invariant defined in Eq. (3.7) of Ref. \cite{Neupert_2012} is nothing else than the Hopf number, however calculated from a simplified formula that makes explicit use of the fact that the system has $N=3$ bands and chiral symmetry. However, when chiral symmetry is broken without a gap-closing, the topological invariant in Eq. (3.7) of Ref. \cite{Neupert_2012} should become ill-defined. In contrast, the Hopf number (\ref{Hopf}) remains well-defined and its values shown in Fig. \ref{fig:Hopfinvariant} should remain unchanged.
 
Second, one may be tempted to analyze the jump $\delta\mathcal{N}_\text{Hopf}$ of the Hopf number at a topological phase transition by using a ``continuum Hopf number" and summing over all band crossing points. This appears however as a delicate task, see Appendix \ref{Apphopf} for a more detailed discussion. 

\section{Conclusions \& Perspectives}
\label{conclusion}

In this paper we have emphasized that linear band crossings in 3D exhibit rich physical properties if more than two bands cross simultaneously. 
In particular, beyond the well-known Berry monopoles (\ref{berrymon}) which occur in Weyl semimetals \cite{Armitage_2018} and chiral multifold semimetals \cite{Lv_2021}, 
other types of linear band crossings with a more exotic quantum geometric structure are possible. 
{ We have focused on the case where each crossing point acts as a Berry dipole (\ref{berryungapped}), which we call \emph{massless multifold Hopf semimetals (MMHSs)}, 
however we emphasize that Berry quadrupoles and Berry octupoles can also exist. }
Indeed, preliminary results of Appendix \ref{Appmulti} indicate the possibility to establish a full hierarchy of Berry multipole crossings.

{ To study MMHSs, we have introduced several lattice models which are characterized by a low-energy theory of the form (\ref{mods}) that feature linear $N$-fold crossings with Berry dipole.}
Such MMHSs can be distinguished according to whether they exhibit an even (valley-Hopf semimetal) 
or odd (topological Hopf semimetal) number of Berry dipoles in the Brillouin zone.

From an experimental point of view, it appears possible that the valley-Hopf semimetals, 
and in particular the models (\ref{subst}) which are quite close to the graphene tight-binding model, 
may exist in a crystalline setup. To make progress in this regard, one should conduct a precise symmetry analysis 
of the models and check in which materials they might occur. Although the anomalous Hall and magnetoconductivities 
caused by each Berry dipole cancel for a valley-Hopf semimetal, the effect of the Berry dipole is still 
clearly visible in the peculiar depence of the Landau level spectrum on the magnetic field orientation (Fig. \ref{fig:LLs_H3}). 
More generally, it would be interesting to study optical or magneto-optical responses for which the contributions 
from the different valleys do not cancel, and a first step in this direction was made very recently in Ref. \cite{Habe_2022}. 

For the topological { massless} Hopf semimetals (\ref{truesem}) there are clear signatures of the Berry dipole in the anomalous Hall current and magnetoconductivity. 
The fact that these currents exhibit a parity opposite to those caused by a pair of Weyl nodes (Fig. \ref{fig:AHeffect}) might be probed 
by varying the electron density close to half filling.  Nevertheless, it has to be mentioned that our models (\ref{truesem}) 
for topological Hopf semimetals are probably hard to realize in a real crystal. A different possible route involves artificial systems 
such as ultracold atoms, photonic crystals, or superconducting circuits. Indeed, those have been suggested and used many times 
to realize semimetallic phases with two- and multifold crossings  \cite{Lu_2015,Chen_2016,Riwar_2016,Wang_2017,Zhu_2017,Zhang_2018a,Tan_2018,Fulga_2018,Hu_2018b}. 
Moreover, a so-called tensor monopole crossing, which is very similar to the Hamiltonian (\ref{Berrydip3}), 
was recently observed experimentally using a transmon in a cavity \cite{Tan_2021}. In such artificial systems, 
the magnetic responses that we focused on in this paper are likely irrelevant, but similar signatures of
the Berry dipole should be present also in a host of physical responses routinely studied in artificial systems.

Going beyond the semimetallic case, we introduced the first concrete lattice models (\ref{hopfdef}) for multiband Hopf insulators (MHIs),
which become exactly equivalent to the Hopf semimetals (\ref{truesem}) at topological phase transitions. One considerable advantage 
of these models is that they require only nearest-neighbor hoppings, thus avoiding the complicated second-neighbor hoppings 
that are necessarily present in any model for a two-band Hopf insulator \cite{Moore_2008,Deng_2013,Nelson_2022}. 
Due to their relative simplicity, the MHI models might provide a fertile platform to test theoretical predictions 
for the bulk-boundary correspondence of delicate topological insulators \cite{Alexandradinata_2021,Lapierre_2021}. 

Also for the MHIs an experimental realization would be desirable, and in this context it is noteworthy that
there is considerable activity regarding the observation of Hopf numbers in two-band insulators \cite{Deng_2018,Unal_2019,Schuster_2021a,Schuster_2021}. 
Such proposals could potentially be extended to the MHIs (\ref{hopfdef}). Most notably, for the three-band Hopf insulator \smash{$h_3^\text{Hopf}(\mathbf{k})$}
(known as chiral topological insulator in the literature) there already exists not only a proposal based on ultracold atoms \cite{Wang_2014}, 
but also a claimed experimental realization based on machine learning analysis of a nitrogen-vacancy center in diamond \cite{Lian_2019}. 
In such experiments, the topological Hopf semimetal (\ref{truesem}) may be reached at critical parameter values corresponding to topological phase transitions.

To close,  we observe that there appear to be interesting connections between the systems introduced in this paper 
and systems of different spatial dimensions. For example, all multiband Hopf semimetals and Hopf insulators that we described are, 
in a sense, 3D analogs of 2D Dirac semimetals and Chern insulators. Indeed, they all have a 2D counterpart in Haldane's model \cite{Haldane_1988}, 
in the sense that the relevant topological densities (Berry curvature in 2D vs. Hopf density in 3D) and topological numbers (Chern number in 2D vs.
Hopf number in 3D) behave very similarly. This analogy is discussed in more detail in Appendix \ref{AppHaldane}.
To make such an analogy more precise and complete, one should systematically analyze all possible ways to perturb MMHSs 
and establish the corresponding phase diagrams, similar to recent work on the conversion between Weyl points, nodal lines, 
quadratic Berry dipole touchings and two-band Hopf insulators \cite{Liu_2017,Sun_2018,Bouhon_2020,Nelson_2022}. 
Such an analysis is also important to determine in more detail the stability of the MMHS band crossings.

Further, the Hamiltonians (\ref{Berrydip3})-(\ref{Berrydip4}) seem to be related to 4D semimetals with tensor monopoles \cite{Palumbo_2018,Zhu_2020} by dimensional reduction.
Given such connections to the 2D Haldane model as well as to 4D semimetals, it appears very intriguing to fully develop the corresponding dimensional hierarchies.

\section*{Acknowledgements} We thank M. O. Goerbig for useful comments on the manuscript and are indebted to A. Mesaros for very enlightening discussions on Hopf insulators.

\appendix

\section{Models with quadrupolar and octupolar Berry curvature}
\label{Appmulti}

\begin{figure}
	\centering
	\includegraphics[width=\columnwidth]{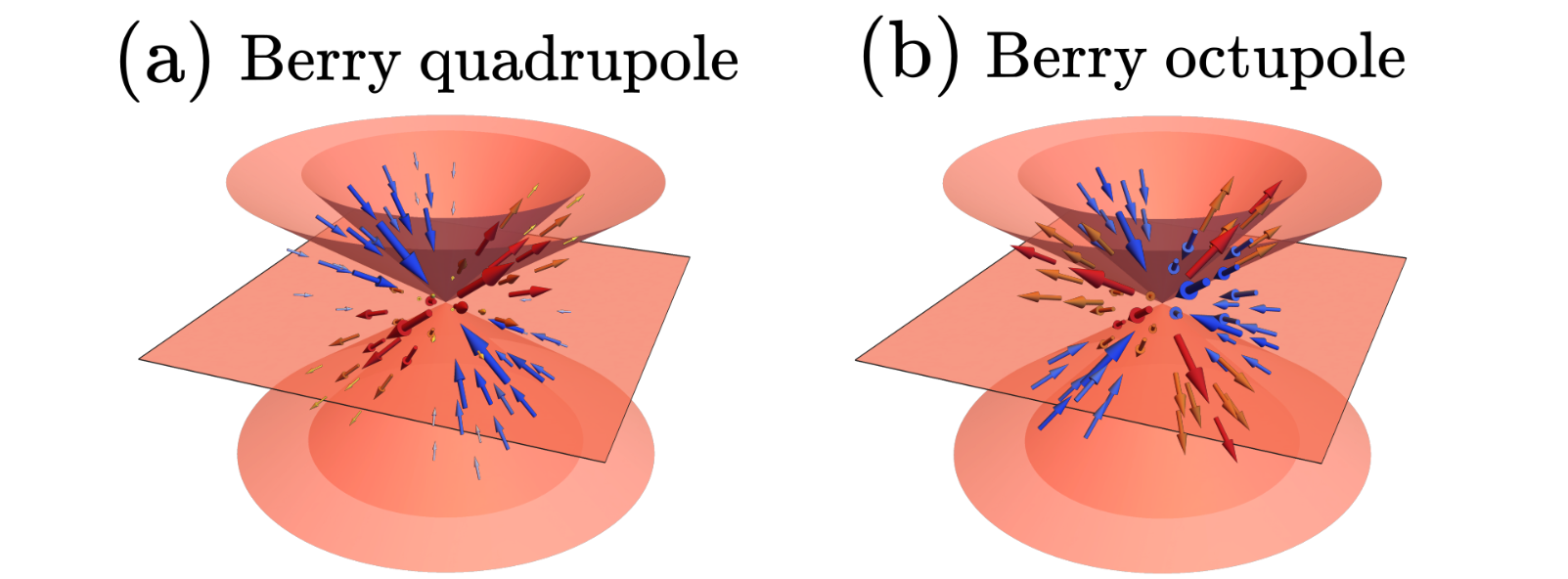}
	\caption{A fivefold linear crossing acting as (a) a Berry quadrupole. (b) a Berry octupole.}
	\label{fig:multifold_new}
\end{figure}

As it turns out, linear multiband crossings even offer the possibility for quantum geometric structures that go beyond the Berry dipole. To see this, consider the fivefold crossing
\begin{equation}
	H_{5a}^\xi(\mathbf{q})=\begin{pmatrix}
		0 & q_+^\xi & 0 & iq_z & 0\\
		q_-^\xi & 0 & -iq_z & 0 & iq_z\\
		0 & iq_z & 0 & q_-^\xi& 0 \\
		-i q_z & 0 & q_+^\xi & 0 & q_+^\xi\\
		0 & -i q_z &0 & q_-^\xi & 0
	\end{pmatrix},
	\label{H8}
\end{equation}
where $q_\pm^\xi=\xi q_x\pm i q_y$.
The spectrum is exactly the same as for the five-band model (\ref{Berrydip5}), namely $E_\alpha(\mathbf{q})=c_\alpha|\mathbf{q}|$, with $c_\alpha=0,\pm1,\pm\sqrt{2}$. However, the Berry curvature takes the form
\begin{equation}
	\boldsymbol{\Omega}_\alpha(\mathbf{q})=\kappa_\alpha(\mathbf{q}\cdot\mathbf{d})\frac{\mathbf{q}}{|\mathbf{q}|^4}+\dbtilde{\kappa}_\alpha(\mathbf{q}\cdot\mathbf{d})^3\frac{\mathbf{q}}{|\mathbf{q}|^6},
\end{equation}
where $\mathbf{d}=(0,0,\xi)$, which corresponds to a dipolar together with an octupolar term. The octupolar Berry curvature is visualized in Fig. \ref{fig:multifold_new}(b).
The dipole and octupole charges are $\kappa_\alpha=1,-3,4,-3,1$ and $\dbtilde{\kappa}_\alpha=-4,8,-8,8,-4$ from the lowest to the highest band.

As a second example, consider the fivefold crossing
\begin{equation}
	H_{5b}^\xi(\mathbf{q})=\begin{pmatrix}
		0 & q_-^\xi & iq_z & iq_z & 0\\
		q_+^\xi & 0 & q_+^\xi & 0 & -i q_z\\
		-i q_z & q_-^\xi & 0 & -q_+^\xi & -iq_z\\
		-iq_z & 0 & -q_-^\xi & 0 & q_-^\xi\\
		0 & iq_z & iq_z & q_+^\xi & 0
	\end{pmatrix}.
	\label{H9}
\end{equation}
The spectrum is again of the form (\ref{spec}) where now $c_\alpha=0,\pm1,\pm\sqrt{3}$. The Berry curvature takes the exotic form
\begin{equation}
	\boldsymbol{\Omega}_\alpha(\mathbf{q})=\tilde{\kappa}_\alpha q_xq_z\frac{\mathbf{q}}{|\mathbf{q}|^5}+\xi\dbtilde{\kappa}_\alpha q_yq_z^2\frac{\mathbf{q}}{|\mathbf{q}|^6},
\end{equation}
with $\tilde{\kappa}_\alpha=-\sqrt{3},3,0,-3,\sqrt{3}$ from lowest to highest band, and $\dbtilde{\kappa}_\alpha=-4/3,4,-16/3,4,-4/3$. This corresponds to a quadrupolar and an octupolar term. The quadrupolar Berry curvature is visualized in Fig. \ref{fig:multifold_new}(a).


	Considering these examples, it appears interesting to speculate that the Berry curvature of any multifold linear crossing in 3D takes the form of a multipole expansion
\begin{equation}
	\boldsymbol{\Omega}_\alpha(\mathbf{q})=\sum_{n=0}^\infty \kappa_\alpha^{(n)}\prod_{i=1}^n(\mathbf{q}\cdot\mathbf{d}_i)\frac{\mathbf{q}}{|\mathbf{q}|^{n+3}},
	\label{multiex}
\end{equation}
where $\kappa_\alpha^{(n)}$ are geometric charges and $\mathbf{d}_i$ are 3D unit vectors. The allowed terms in Eq. (\ref{multiex}) should be selected depending on the symmetries and the number $N$ of bands involved in the crossing. Indeed, we know that for $N=2$ only the $n=0$ term in Eq. (\ref{multiex}) is allowed. For $N=3$ and $N=4$ we know that the $n=0$ term is allowed in the presence of CP symmetry (e.g. for a pseudospin Hamiltonian with $s=1$ or $s=3/2$), whereas the $n=1$ term is allowed in the presence of chiral symmetry (e.g. for the MMHSs \ref{mods}). The examples (\ref{H8}) and (\ref{H9}) show that for $N=5$ all terms $n=0,1,2,3$ are in principle possible, depending on the symmetries. For example, the model (\ref{H8}) has a chiral symmetry $\mathcal{S}=\text{diag}(1,-1,1,-1,1)$, which appears to select the terms with odd $n$. 

There are many open questions, such as whether Berry quadrupoles and octupoles are possible for $N<5$, whether hexadecapoles are possible for $N=5$, and so on. Moreover, very rich physical properties can be expected for such exotic crossings, for example in the Landau level spectrum, anomalous Hall conductivity, et cetera. The possibility to establish a full hierarchy of Berry multipole crossings thus appears quite intriguing.

\section{Model with tunable Berry dipole vector}
\label{Apptune}

Consider the continuum model
$$
	H_3(\mathbf{q},\varphi)=\begin{pmatrix}0 & q_x+is_\varphi\, q_z& q_y-ic_\varphi \,q_z\\
		q_x-is_\varphi	\,q_z&0 & 0\\
		q_y+ic_\varphi \,q_z	&0&0
	\end{pmatrix},
$$
where $c_\varphi=\cos\varphi$, $s_\varphi=\sin\varphi$, and $\varphi$ is a free parameter.
It has an energy spectrum (\ref{spec}) with $c_\alpha=0,\pm1$ and a dipolar Berry curvature (\ref{berryungapped}) with $\kappa_\alpha=2-3\alpha^2$ and $\mathbf{d}=(\cos\varphi,\sin\varphi,0)$. Clearly, a lattice realization of this model will involve hopping amplitudes $\sim\sin\varphi$ and $\sim\cos\varphi$, and tuning the value of $\varphi$ will rotate the Berry dipole vector in the $\hat{x}$-$\hat{y}$-plane, despite a fixed lattice geometry. Note that the presence of a tunable Berry curvature in terms of a parameter that leaves the spectrum invariant is somewhat reminiscent of the two-dimensional $\alpha-\mathcal{T}_3$ model that allows to interpolate between honeycomb and dice lattices \cite{Raoux_2014}.

\section{Landau levels of MMHS continuum models}
\label{AppLL}

In order to describe the behavior of the electrons modeled by the MMHS continuum models (\ref{mods}) in the presence of a strong magnetic field (\ref{Borient}), we replace the canonical momentum $\mathbf{q}$ by the gauge-invariant kinetic momentum \cite{Jackson_2012}, $\mathbf{q}\rightarrow \Pi=\mathbf{q}+e\mathbf{A}$, with the gauge choice $\mathbf{A}=Bx(0,\cos\theta,-\sin\theta)$ for the electromagnetic vector potential. Note that the momentum along the magnetic field, $q_0=\hat{\mathbf{B}}\cdot\mathbf{q}=\sin\theta q_y+\cos\theta q_z$, is conserved. Using the canonical commutation relations $[x_j,q_k]=i\delta_{jk}$, one finds $[\Pi_x,\Pi_y]=-i\cos\theta/l_B^2$, $[\Pi_y,\Pi_z]=0$, $[\Pi_z,\Pi_x]=-i\sin\theta/l_B^2$, where $l_B\equiv1/\sqrt{eB}$ is the magnetic length. It is further convenient to introduce ladder operators as \cite{Li_2016}
\begin{equation}
	\begin{aligned}
		\hat{d}&=\frac{l_B}{\sqrt{2}}(\Pi_x-i\cos\theta\, \Pi_y+i\sin \theta\, \Pi_z),\\
		\hat{d}^\dagger&=\frac{l_B}{\sqrt{2}}(\Pi_x+i\cos\theta \,\Pi_y-i\sin\theta \,\Pi_z),
	\end{aligned}
\end{equation}
such that  $[\hat{d},\hat{d}^\dagger]=1$, which act on number states $\ket{n}$ as $\hat{d}^\dagger\ket{n}=\sqrt{n+1}\ket{n+1}$ and $\hat{d}\ket{n}=\sqrt{n}\ket{n-1}$. Reversing the above relations, we have
$$
	\begin{aligned}
		\Pi_x&=\frac{1}{\sqrt{2}l_B}(\hat{d}+\hat{d}^\dagger),\\
		\Pi_y&=q_0\sin\theta+\frac{i}{\sqrt{2}l_B}\cos\theta(\hat{d}-\hat{d}^\dagger),\\
		\Pi_z&=q_0\cos\theta-\frac{i}{\sqrt{2}l_B}\sin\theta(\hat{d}-\hat{d}^\dagger),\\
		\Pi_\pm^\xi&=\xi\Pi_x\pm i\Pi_y\\
		&=\frac{1}{\sqrt{2}l_B}\left[(\xi\pm\cos\theta)\hat{d}^\dagger+(\xi\mp\cos\theta)\hat{d}\right]\pm i \sin\theta\,q_0.
	\end{aligned}
$$

\subsection{Threefold Hopf semimetal}

The LL spectrum of the threefold Hopf semimetal (\ref{Berrydip3}) is easily computed analytically. Replacing $q_i\rightarrow\Pi_i$, we have
\begin{equation}
	\hat{H}_3^\xi=\begin{pmatrix}
		0 & Q\\
		Q^\dagger & 0_2
	\end{pmatrix},\hspace{.5cm}Q=\begin{pmatrix}
		\Pi_-^\xi & -i \Pi_z
	\end{pmatrix}.
\end{equation}
Making an ansatz $
\hat{H}_3^\xi\Psi_\alpha=\e_\alpha\Psi_\alpha$,
where $\e_\alpha=\alpha\e$, $\alpha=0,\pm$ and $\Psi_\alpha=(\psi_1^\alpha,\Psi_2^\alpha)$ with $\Psi_2^\alpha$ a two-component spinor, it follows
\begin{equation}
	\begin{aligned}
		\e_\alpha\psi_1^\alpha&=Q\Psi_2^\alpha,\\
		\e_\alpha\Psi_2^\alpha&=Q^\dagger\psi_1^\alpha,\\
		\e_\alpha^2\psi_1^\alpha&=QQ^\dagger \psi_1^\alpha.
	\end{aligned}
	\label{conds}
\end{equation}
Using $\Pi_-^\xi\Pi_+^\xi=\Pi_x^2+\Pi_y^2+\xi\cos\theta/l_B^2$ and $\Pi^2=(2\hat{d}^\dagger\hat{d}+1)/l_B^2+q_0^2$, one easily finds 
\begin{equation}
	QQ^\dagger=[eB(2\hat{d}^\dagger\hat{d}+1+\xi\cos\theta)+q_0^2],
\end{equation}
implying that $\psi_1^\alpha\sim\ket{n}$. From the second line of Eq. (\ref{conds}), one immediately obtains
$$
	\e_\alpha\Psi_2^\alpha\sim\begin{pmatrix}
		\xi-\cos\theta & i\sin\theta\,q_0 & \xi+\cos\theta\\
		\sin\theta & i\cos\theta\,q_0& -\sin \theta
	\end{pmatrix}\begin{pmatrix}
		\beta_n\ket{n-1}\\
		\ket{n}\\
		\beta_{n+1}\ket{n+1}
	\end{pmatrix},
$$
where $\beta_n=\sqrt{eBn/2}$,
and finally the full solution for the LLs is given by
$$
		\begin{aligned}
			\epsilon_\alpha^{n,\xi}&=\alpha\sqrt{2eB\left(n+\frac{1+\xi\cos\theta}{2}\right)+q_0^2},\hspace{.5cm}n=0,1,2,...\\
				\end{aligned}
$$
with corresponding eigenstates
$$
			\Psi_\alpha^{n,\xi}\sim\begin{pmatrix}
				0 & \epsilon_\alpha^{n,\xi}& 0 \\
				\xi-\cos\theta & i\sin\theta\, q_0 & \xi+\cos\theta\\
				\sin\theta & i\cos\theta \,q_0 & -\sin\theta
			\end{pmatrix}\begin{pmatrix}
				\beta_n\ket{n-1}\\
				\ket{n}\\
				\beta_{n+1}\ket{n+1}
			\end{pmatrix}.
$$

\subsection{Fourfold Hopf semimetal}

We proceed to compute the LL spectrum of the fourfold Hopf semimetal (\ref{Berrydip4}). Replacing $q_i\rightarrow \Pi_i$, we have
\begin{equation}
	\hat{H}_4^\xi=\begin{pmatrix}
		0_2 & Q\\
		Q^\dagger & 0_2
	\end{pmatrix},\hspace{.5cm}Q=\begin{pmatrix}
		a\Pi_-^\xi & ia \Pi_z\\
		ib \Pi_z & b\Pi_+^\xi
	\end{pmatrix}.
\end{equation}
We make an ansatz $\hat{H}_4^\xi\Psi_\alpha=\e_\alpha\Psi_\alpha$,
where $\e_\alpha=\alpha_1\e_{\alpha_2}$ with $\alpha_1=\pm$, $\alpha_2=\pm$, and where $\Psi_\alpha=(\Psi_1^\alpha,\Psi_2^\alpha)$ with $\Psi_1^\alpha$ and $\Psi_2^\alpha$ being two-component spinors. It follows
\begin{equation}
	\begin{aligned}
		\e_\alpha\Psi_1^\alpha&=Q\Psi_2^\alpha,\\
		\e_\alpha\Psi_2^\alpha&=Q^\dagger\Psi_1^\alpha,\\
		\e_\alpha^2\Psi_1^\alpha&=QQ^\dagger \Psi_1^\alpha.
	\end{aligned}
	\label{conds2}
\end{equation}
We first focus on the last line. Straightforward computation yields
\begin{equation}
	QQ^\dagger=\begin{pmatrix}
		a^2\hat{D}_+ & \xi abeB\sin\theta\\
		\xi abeB\sin\theta& b^2\hat{D}_-
	\end{pmatrix},
\end{equation}
where $\hat{D}_\pm\equiv eB(2\hat{d}^\dagger\hat{d}+1\pm\xi\cos\theta)+q_0^2$,
and it is clear that the spinor $\Psi_1^\alpha$ has to be of the form $\Psi_1^\alpha\sim(u_{\alpha_2},v_{\alpha_2})\ket{n}$, with some functions $u_{\alpha_2},v_{\alpha_2}$ to be determined. Solving the eigenvalue problem
\begin{equation}
	\begin{pmatrix}
		2\eta_a & \nu\\
		\nu & 2\eta_b
	\end{pmatrix}\begin{pmatrix}
		u_{\alpha_2}\\
		v_{\alpha_2}
	\end{pmatrix}=\e_\alpha^2\begin{pmatrix}
		u_{\alpha_2}\\
		v_{\alpha_2}
	\end{pmatrix},
\end{equation}
where
\begin{equation}
	\begin{aligned}
		\eta_a&=\frac{a^2}{2}\left[2eB\left(n+\frac{1+\xi\cos\theta}{2}\right)+q_0^2\right],\\
		\eta_b&=\frac{b^2}{2}\left[2eB\left(n+\frac{1-\xi\cos\theta}{2}\right)+q_0^2\right],\\
		\nu&= \xi abeB\sin\theta,
	\end{aligned}
\end{equation}
leads to 
\begin{equation}
	\begin{aligned}
		\e_\alpha^2&=\eta_a+\eta_b+\alpha_2\sqrt{(\eta_a-\eta_b)^2+\nu^2},\\
		u_{\alpha_2}&=\eta_a-\eta_b+\alpha_2\sqrt{(\eta_a-\eta_b)^2+\nu^2},\\
		v_{\alpha_2}&=\nu.
	\end{aligned}
\end{equation}
 Thus, the complete solution for the LL energies is given by
$$
\epsilon_\alpha^{n,\xi}=\alpha_1\sqrt{	\eta_a+\eta_b+\alpha_2\sqrt{(\eta_a-\eta_b)^2+\nu^2}},\hspace{.2cm}n=0,1,2,...
$$
To determine the corresponding eigenstates one can now turn to the second line of Eq. (\ref{conds2}), yielding
$$
			\epsilon_\alpha\Psi_2^\alpha\sim\begin{pmatrix}
	a\Pi_+^\xi & -ib\Pi_z\\
	-ia\Pi_z & b\Pi_-^\xi
\end{pmatrix}\begin{pmatrix}
	u_{\alpha_2}\ket{n}\\
	v_{\alpha_2}\ket{n}
\end{pmatrix}.
$$
Thus, the spinor $\Psi_2^\alpha$ has contributions from number states $|n-1\rangle$, $|n\rangle$ as well as $|n+1\rangle$, while the spinor $\Psi_1^\alpha$ has contributions only from $|n\rangle$.
	
	\subsection{Fivefold Hopf semimetal}
	
	The LL spectrum of the fivefold Hopf semimetal (\ref{Berrydip5}) can be derived from the Hamiltonian
	\begin{equation}
		\hat{H}_5^\xi=\begin{pmatrix}
			0_3 & Q\\
			Q^\dagger & 0_2
		\end{pmatrix},\hspace{.5cm}Q=\begin{pmatrix}
			0& i\sqrt{2}\Pi_z\\
			i\Pi_z & \Pi_+^\xi\\
			\sqrt{2}\Pi_+^\xi & 0
		\end{pmatrix}.
	\end{equation}
	We make an ansatz $\hat{H}_5^\xi\Psi_\alpha=\e_\alpha\Psi_\alpha$,
	where $\e_\alpha=\alpha_1\e_{\alpha_2}$, $\alpha_1=0,\pm$ and $\alpha_2=\pm$. The wave function is of the form $\Psi_\alpha=(\Psi_1^\alpha,\Psi_2^\alpha)$ with $\Psi_1^\alpha$ a three-component and $\Psi_2^\alpha$ a two-component spinor.
	
	 We again have to solve 
	\begin{equation}
		\begin{aligned}
			\e_\alpha\Psi_1^\alpha&=Q\Psi_2^\alpha,\\
			\e_\alpha\Psi_2^\alpha&=Q^\dagger\Psi_1^\alpha,\\
			\e_\alpha^2\Psi_2^\alpha&=Q^\dagger Q \Psi_2^\alpha,
		\end{aligned}
		\label{conds3}
	\end{equation}
	and thus consider first the matrix
	\begin{equation}
		Q^\dagger Q=\begin{pmatrix}
			2\Pi_-^\xi\Pi_+^\xi+\Pi_z^2 & -i\Pi_z\Pi_+^\xi\\
			(-i\Pi_z\Pi_+^\xi)^\dagger & \Pi_-^\xi\Pi_+^\xi+2\Pi_z^2 
		\end{pmatrix},
	\label{qdagq}
	\end{equation}
	whose components are given as follows in terms of ladder operators:
	$$
		\begin{aligned}
			(Q^\dagger Q)_{11}&=\frac{eB}{2}s_\theta^2 (\hat{d})^2+\frac{eB}{2}s_\theta^2(\hat{d}^\dagger)^2+eB(3+c_\theta^2)\hat{d}^\dagger\hat{d}\\
			&+\sqrt{\frac{eB}{2}}iq_0s_{2\theta}\hat{d}-\sqrt{\frac{eB}{2}}iq_0s_{2\theta}\hat{d}^\dagger\\
			&+\frac{eB}{2}(3+c_\theta^2+4\xi c_\theta)+(1+s_\theta^2)q_0^2,
			\end{aligned}
$$
		$$
	\begin{aligned}
			(Q^\dagger Q)_{12}&=\frac{eB}{2}s_\theta(c_\theta-\xi)(\hat{d})^2+\frac{eB}{2}s_\theta(\xi+c_\theta)(\hat{d}^\dagger)^2\\
			&-\frac{eB}{2}s_{2\theta}\hat{d}^\dagger\hat{d}+\sqrt{\frac{eB}{2}}iq_0(c_{2\theta}-\xi c_\theta)\hat{d}\\
			&-\sqrt{\frac{eB}{2}}iq_0(c_{2\theta}+\xi c_\theta)\hat{d}^\dagger-\frac{eB}{2}s_\theta(\xi+c_\theta)+s_\theta c_\theta q_0^2,
		\end{aligned}
	$$
$$
\begin{aligned}
			(Q^\dagger Q)_{22}&=-\frac{eB}{2}s_\theta^2(\hat{d})^2-\frac{eB}{2}s_\theta^2(\hat{d}^\dagger)^2+eB(2+s_\theta^2)\hat{d}^\dagger\hat{d}\\
			&-\sqrt{\frac{eB}{2}}iq_0s_{2\theta}\hat{d}+\sqrt{\frac{eB}{2}}iq_0s_{2\theta}\hat{d}^\dagger\\
			&+\frac{eB}{2}(2+s_\theta^2+2\xi c_\theta)+(1+c_\theta^2)q_0^2,
		\end{aligned}
	$$
with shorthand notations $c_\theta=\cos\theta$ and $s_\theta=\sin\theta$.

To proceed, we make an ansatz for the spinor $\Psi_2^\alpha$:
	\begin{equation*}
		\Psi_2^\alpha=\begin{pmatrix}U \\ V
		\end{pmatrix}=\begin{pmatrix}
			u_1\ket{n-1}+u_2\ket{n}+u_3\ket{n+1}\\
			v_1\ket{n-1}+v_2\ket{n}+v_3\ket{n+1}
		\end{pmatrix}.
	\end{equation*}
This state gets projected into the space of number states $N=(\ket{n-3},...,\ket{n+3})$ by the matrix (\ref{qdagq}); however, we will ensure that the coefficients of $|n\pm3\rangle$ and $|n\pm2\rangle$ in the product $Q^\dagger Q \Psi_2^\alpha$ vanish, such that one obtains a solution for the last line of Eq. (\ref{conds3}).
A lengthy calculation yields 
\[
\begin{aligned}
(Q^\dagger Q)_{11}U=M_U^T N^T,\\
(Q^\dagger Q)_{12} V=M_V^T N^T,
\end{aligned}
\]
where
\newpage
\begin{widetext}
	$$
M_U=\begin{pmatrix}
			s_\theta^2\beta_{n-1}\beta_{n-2}u_1\\
			\beta_{n-1}\left[s_{2\theta}iq_0u_1+s_\theta^2\beta_nu_2\right]\\
			\beta_n\left[s_{2\theta}iq_0u_2+s_\theta^2\beta_{n+1}u_3\right]+eB\left[(3+c_\theta^2)(n-\frac{1}{2})+2\xi c_\theta\right]u_1+(1+s_\theta^2)q_0^2u_1\\
			s_{2\theta}iq_0\left[\beta_{n+1}u_3-\beta_n u_1\right]+eB\left[(3+c_\theta^2)(n+\frac{1}{2})+2\xi c_\theta\right]u_2+(1+s_\theta^2)q_0^2u_2\\
			\beta_{n+1}\left[-s_{2\theta}iq_0u_2+s_\theta^2\beta_nu_1\right]+eB\left[(3+c_\theta^2)(n+\frac{3}{2})+2\xi c_\theta\right]u_3+(1+s_\theta^2)q_0^2u_3\\
			\beta_{n+2}\left[-s_{2\theta}iq_0u_3+s_\theta^2\beta_{n+1}u_2\right]\\
			s_\theta^2\beta_{n+2}\beta_{n+3}u_3
		\end{pmatrix},
	$$
	$$
		M_V=\begin{pmatrix}
			s_\theta(c_\theta-\xi)\beta_{n-1}\beta_{n-2}v_1\\
			\beta_{n-1}\left[(c_{2\theta}-\xi c_\theta)iq_0v_1+s_\theta(c_\theta-\xi)\beta_nv_2\right]\\
			\beta_n\left[(c_{2\theta}-\xi c_\theta)iq_0v_2+s_\theta(c_\theta-\xi)\beta_{n+1}v_3\right]-eBs_\theta\left[(n-\frac{1}{2})c_\theta+\frac{\xi}{2}\right]v_1+s_\theta c_\theta q_0^2v_1\\
			iq_0\left[(c_{2\theta}-\xi c_\theta)\beta_{n+1}v_3-(c_{2\theta}+\xi c_\theta)\beta_n v_1\right]-eBs_\theta\left[(n+\frac{1}{2})c_\theta+\frac{\xi}{2}\right]v_2+s_\theta c_\theta q_0^2v_2\\
			\beta_{n+1}\left[-(c_{2\theta}+\xi c_\theta)iq_0v_2+s_\theta(c_\theta+\xi)\beta_nv_1\right]-eBs_\theta\left[(n+\frac{3}{2})c_\theta+\frac{\xi}{2}\right]v_3+s_\theta c_\theta q_0^2v_3\\
			\beta_{n+2}\left[-(c_{2\theta}+\xi c_\theta)iq_0v_3+s_\theta(c_\theta+\xi)\beta_{n+1}v_2\right]\\	s_\theta(c_\theta+\xi)\beta_{n+2}\beta_{n+3}v_3\\
		\end{pmatrix}.
	$$
\end{widetext}
Solving the linear system $(M_U^T+M_V^T)N^T=\e_\alpha^2U$, one finds the LL spectrum
$$
			\begin{aligned}
				\e_\alpha^{n,\xi}&=\alpha_1\sqrt{\eta_++\eta_-+\alpha_2\sqrt{(\eta_+-\eta_-)^2+\tilde{\nu}^2}},\\
				\eta_\pm&=\frac{c_\pm^2}{2}\left[2eB\left(n+\frac{1-\kappa_\pm\xi\cos\theta}{2}\right)+q_0^2\right],\\
				\tilde{\nu}&=2\sqrt{3}\xi eB\sin\theta,
		\end{aligned}
$$
	where $n=1,2,3,...$ is the LL index, $c_\pm=\sqrt{2},1$ are the band velocities of the two cones and $\kappa_\pm=-3,1$ are the corresponding Berry dipole charges. The corresponding eigenfunctions are very complicated and are not written here explicitly.
	
\section{Semiclassical approach to Landau levels: extended Onsager quantization with intraband and interband coupling}
\label{AppSemi}

Here we derive the Landau level spectrum of the MMHS continuum models (\ref{mods}) in the presence of a field (\ref{Borient}) using an alternative method: semiclassical quantization based on a generalized Onsager condition. We shall find excellent agreement with the exact quantum mechanical results.
 More precisely, we try to recover the exact Landau levels by increasing the complexity of the semiclassical quantization condition in three steps: first, using Onsager's method \cite{Onsager_1952} for a single closed orbit; second, using Onsager quantization for a single closed orbit, extended by intraband quantum geometric corrections which are important in a multiband system \cite{Roth_1966,Mikitik_1999,Fuchs_2010,GaoNiu_2017,Fuchs_2018}; finally, we develop an approach to Landau quantization of degenerate orbits, taking into account also interband matrix elements of Berry curvature and orbital magnetic moment, similar in spirit to Ref. \cite{Wang_2019}.

\subsection{Onsager quantization of a single closed orbit}

Consider a band dispersion relation $E_\alpha(\mathbf{q})$. Let's denote $q_0$ the component of $\mathbf{q}$ parallel to the magnetic field
and $\mathbf{q}_\perp$ the momentum perpendicular to the magnetic field such that $\mathbf{q}\equiv (q_x,q_y,q_z)=q_0 \hat{\mathbf{B}} +\mathbf{q}_\perp$,
with $\mathbf{q}_\perp=q_\perp [\cos\phi_\perp \hat{\mathbf{x}}+\sin \phi_\perp (\hat{\mathbf{x}}\times \hat{\mathbf{B}})]$. We can then rewrite $q_x=q_\perp \cos \phi_\perp$, $q_y=q_0 \sin \theta - q_\perp \sin \phi_\perp \cos \theta$, $q_z=q_0\cos \theta +q_\perp \sin \phi_\perp \sin \theta$.
Let's assume that for a fixed $q_0$, the constant energy curve
$E_\alpha(\mathbf{q}_\perp,q_0)=E$ defines a closed orbit $\mathcal{O}_\alpha$ in the $\mathbf{q}_\perp$ plane.
Onsager quantization \cite{Onsager_1952} then corresponds to postulate that the Landau level energies $\epsilon_n$ (with LL index $n$) are obtained by quantizing the $k$-space area $S_\alpha(\epsilon_n,q_0)$ of the orbit $\mathcal{O}_\alpha$ 
according to
\begin{equation}
	S_\alpha(\epsilon_n,q_0)l_B^2 =2\pi (n+\gamma)
\end{equation}
with $l_B=1/\sqrt{eB}$ the magnetic length and
where $\gamma$ is the Maslov index of the orbit, in particular for an orbit deformable to a circle $\gamma=1/2$.
Since $S_\alpha(E,q_0)=4\pi^2 N_\alpha(E,q_0)$, where $N_\alpha(E,q_0)=\int \frac{d\mathbf{q}_\perp}{4\pi^2} \Theta(E-E_\alpha(\mathbf{q}_\perp,q_0))$ 
is equivalent to the effective 2D zero-field integrated density of states of the band $E_\alpha(\mathbf{q})$,
the previous relation can be rewritten as
\begin{equation}
	N_\alpha(\epsilon_n,q_0)=\left(n+\frac{1}{2}\right)\frac{eB}{2\pi},
\end{equation}
where now $eB/(2\pi)$ is the degeneracy (per unit area) of each Landau level (note that $\hbar=1$).

For a zero-field spectrum of the form $E_\alpha(\mathbf{q})=c_\alpha |\mathbf{q}|=c_\alpha \sqrt{|\mathbf{q}_\perp|^2+q_0^2}$
one immediately obtains 
\begin{equation}
	N_\alpha(E,q_0)=\frac{1}{4\pi} \left(\frac{E^2}{c_\alpha^2}-q_0^2\right),
\end{equation}
from which we deduce 
\begin{equation}
	\epsilon_n=\pm|c_\alpha| \sqrt{2eB\left(n+\frac{1}{2}\right) +q_0^2}.
\end{equation}
For Weyl or chiral multifold topological semimetals $H_\text{s}(\mathbf{q})$, and also for Hopf semimetals (\ref{mods}), this
last expression does not recover the correct results because it misses quantum geometric effects.

\subsection{Quantization of a single closed orbit in a multiband system}

To linear order in the magnetic field, the  modified Onsager quantization rule (which takes care of intraband effects but still ignores coupling between degenerate orbits) reads \cite{GaoNiu_2017,Fuchs_2018}
\begin{equation}
	N_\alpha(\epsilon_n,q_0)+\mathbf{M}_\alpha'(\epsilon_n,q_0)\cdot\mathbf{B}=\left(n+\frac{1}{2}\right)\frac{eB}{2\pi},
\end{equation}
where $\mathbf{M}_\alpha'(E,q_0)=\partial/\partial_E \mathbf{M}_\alpha(E,q_0)$ with $\mathbf{M}_\alpha(E,q_0)$ the orbital magnetization (for spinless particles) of the band $\alpha$ at fixed $(E,q_0)$.
This orbital magnetization may be written as (at $T=0$)
$$
	\begin{aligned}
	\mathbf{M}_\alpha(E,q_0)=\int & \frac{d\mathbf{q}_\perp}{4\pi^2} [\mathbf{m}_\alpha(\mathbf{q})+e(E-E_\alpha(\mathbf{q})){\bm \Omega}_\alpha(\mathbf{q})]\\
	&\times\Theta(E-E_\alpha(\mathbf{q})),
	\end{aligned}
$$
with $\mathbf{m}_\alpha(\mathbf{q})$ and ${\bm \Omega}_\alpha(\mathbf{q})$ the \emph{intraband} contributions of orbital magnetic moment (OMM) and Berry curvature, respectively. More precisely, the Berry curvature pseudovector is given by $\boldsymbol{\Omega}_\alpha(\mathbf{q})=(\Omega_{\alpha,yz},\Omega_{\alpha,zx},\Omega_{\alpha,xy})$ with  \cite{Xiao_2010}
\begin{equation}
	\begin{aligned}
		\Omega_{\alpha,ij}&=i\sum_{\beta\neq\alpha}\langle\psi_\alpha|\frac{H^iP_\beta H^j-H^jP_\beta H^i}{(E_\alpha-E_\beta)^2}|\psi_\alpha\rangle,
	\end{aligned}
\label{Berry}
\end{equation}
where $H(\mathbf{q})$ is the Hamiltonian with eigenfunctions $|\psi_\alpha(\mathbf{q})\rangle$ and eigenenergies $E_\alpha(\mathbf{q})$, $P_\alpha=|\psi_\alpha\rangle\langle\psi_\alpha|$ is an eigenprojector, and $H^i\equiv \partial H/\partial q_i$. Similarly, the OMM pseudovector is defined as 
$\mathbf{m}_\alpha(\mathbf{q})=(m_{\alpha,yz},m_{\alpha,zx},m_{\alpha,xy})$, with
\begin{equation}
	\begin{aligned}
		m_{\alpha,ij}&=\frac{ie}{2}\sum_{\beta\neq\alpha}\langle\psi_\alpha|\frac{H^i P_\beta H^j-H^j P_\beta H^i}{E_\alpha-E_\beta}|\psi_\alpha\rangle.
	\end{aligned}
	\label{OMM}
\end{equation}
For our purposes, it is instructive to introduce an alternative notation that will ensure a seamless transition to the discussion of interband coupling below:
\begin{equation}
	\begin{aligned}
		\mathbf{m}_\alpha(\mathbf{q})&=-\frac{e}{2}\sum_{\gamma \ne \alpha} \boldsymbol{\mathcal{A}}_{\alpha \gamma} \times \boldsymbol{\mathcal{V}}_{\gamma \alpha},\\
		{\bm \Omega}_\alpha(\mathbf{q})&=\boldsymbol{\nabla}_\mathbf{q} \times \boldsymbol{\mathcal{A}}_{\alpha \alpha}=i \sum_{\gamma} \boldsymbol{\mathcal{A}}_{\alpha \gamma} \times \boldsymbol{\mathcal{A}}_{\gamma \alpha}.
	\end{aligned}
	\label{nonAbOmBer}
\end{equation}
Here, $ \boldsymbol{\mathcal{A}}_{\alpha \gamma}(\mathbf{q})=i \langle \psi_\alpha|\boldsymbol{\nabla}_\mathbf{q} |\psi_\gamma \rangle$ is the Berry connection and
$\boldsymbol{\mathcal{V}}_{\alpha \gamma}=\langle \psi_\alpha| \boldsymbol{\nabla}_\mathbf{q}  H(\mathbf{q}) |\psi_\gamma\rangle$ is the velocity operator
such that $\boldsymbol{\mathcal{V}}_{\alpha \alpha}(\mathbf{q})=\boldsymbol{\nabla}_\mathbf{q}E_\alpha$. 
Note that, using the identity  $\boldsymbol{\mathcal{V}}_{\alpha \gamma}=i (E_\alpha -E_\gamma)\boldsymbol{\mathcal{A}}_{\alpha \gamma}$, valid for $\alpha \ne \gamma$, one immediately recovers from Eq. (\ref{nonAbOmBer}) the textbook formulas (\ref{Berry}) and (\ref{OMM}) for intraband Berry curvature and OMM.

In particular, for our MMHS continuum models with energy spectrum (\ref{spec}), these intraband contributions read
\begin{equation}
	\begin{aligned}
	\mathbf{m}_\alpha(\mathbf{q})&=\frac{e}{2}\omega_\alpha \frac{(\mathbf{q}\cdot\mathbf{d})\mathbf{q}}{|\mathbf{q}|^3},\\
	{\bm \Omega}_\alpha(\mathbf{q})&=\kappa_\alpha \frac{(\mathbf{q}\cdot\mathbf{d})\mathbf{q}}{|\mathbf{q}|^4},
	\end{aligned}
	\label{omeq}
\end{equation}
with coefficients as listed in Table \ref{tab0}.
From these expressions, using $\mathbf{q}=q_0 \hat{\mathbf{B}}+\mathbf{q}_\perp$, one first obtains
$$
\begin{aligned}
\mathbf{m}_\alpha(\mathbf{q})\cdot \hat{\mathbf{B}}&=\frac{e}{2}\omega_\alpha \frac{q_0^2 {(\mathbf{d}\cdot \hat{\mathbf{B}})}+q_0{(\mathbf{d}\cdot \mathbf{q}_\perp)} }{(|\mathbf{q}_\perp|^2+q_0^2)^{3/2}},\\
{\bm \Omega}_\alpha(\mathbf{q})\cdot \hat{\mathbf{B}}&=\kappa_\alpha \frac{q_0^2 {(\mathbf{d}\cdot\hat{\mathbf{B}})}+q_0{(\mathbf{d}\cdot \mathbf{q}_\perp)} }{(|\mathbf{q}_\perp|^2+q_0^2)^{2}}.
\end{aligned}
$$
Performing the $\mathbf{q}_\perp$ integration at fixed $(E,q_0)$, the contributions proportional to $(\mathbf{d}\cdot \mathbf{q}_\perp)$ average to zero and one finds
$$
\begin{aligned}
\mathbf{M}_\alpha(E,q_0)\cdot\mathbf{B}&=\frac{eB}{2\pi}(\mathbf{d}\cdot \hat{\mathbf{B}})
\left[\frac{\kappa_\alpha }{2} E+\left(\frac{\omega_\alpha}{2c_\alpha} - \kappa_\alpha\right)c_\alpha q_0\right.\\
&\left.-\left(\frac{\omega_\alpha}{c_\alpha} - \kappa_\alpha\right)\frac{(c_\alpha q_0)^2}{2E}\right].
\end{aligned}
$$
By differentiating with respect to $E$ one finally obtains
$$\mathbf{M}'_\alpha(E,q_0)\cdot\mathbf{B}=\frac{e}{2\pi}{(\mathbf{d}\cdot \mathbf{B})}\left[\frac{\kappa_\alpha }{2}
+\left(\frac{\omega_\alpha}{c_\alpha} - \kappa_\alpha\right)\frac{(c_\alpha q_0)^2}{2E^2}\right].$$
If we ignore the contribution $\sim1/E^2$ then 
the semiclassical LL spectrum reads
\begin{equation}
	\epsilon_n
	= \pm |c_\alpha| \sqrt{2eB\left(n+\frac{1}{2}-\frac{\kappa_\alpha}{2}\xi\cos\theta\right)+q_0^2},
\end{equation}
where $\xi\cos \theta = (\mathbf{d} \cdot \hat{\mathbf{B}})$.
For $N=3$ since $\omega_\alpha/c_\alpha - \kappa_\alpha= 0$ this expression is identical to the exact LL spectrum (\ref{LL3}).
For $N = 4$, despite the fact that $\omega_\alpha/c_\alpha - \kappa_\alpha= 0$ this expression still fails to recover the exact LL spectrum (\ref{LL4})
as it misses the coupling between degenerate orbits.
For $N=5$ since $\omega_\alpha/c_\alpha - \kappa_\alpha\neq 0$ it is strictly speaking no longer valid to ignore the contribution $\sim1/E^2$.
A more physical argument in favor of neglecting this contribution anyway consists in acknowledging that the semiclassical calculation 
should only be trustworthy in the large $n$ or large $E$ limit, where $1/E^2\ll1$. We adopt this approximation hereafter.

\subsection{Quantization of two degenerate closed orbits in a multiband system}

The previous calulation of the quantity $\mathbf{M}_\alpha(E,q_0)$ takes into account only the {\it intraband} diagonal element of the orbital magnetization.
As recently put forward in Ref. \cite{Wang_2019}, for systems with two (or more) bands exhibiting quasi-degenerate orbits (i.e. orbits degenerate simultaneously in $k$-space 
and energy space) it is necessary to also consider off-diagonal (interband) elements of the orbital magnetization. We here adopt a similar approach and define these interband contributions as
$$
\begin{aligned}
	\mathbf{m}_{\alpha \beta}(\mathbf{q})&=-\frac{e}{4}\left(\sum_{\gamma \ne \alpha} \boldsymbol{\mathcal{A}}_{\alpha \gamma} \times \boldsymbol{\mathcal{V}}_{\gamma \beta}-
	\sum_{\gamma \ne \beta} \boldsymbol{\mathcal{V}}_{\alpha \gamma} \times \boldsymbol{\mathcal{A}}_{\gamma \beta}\right),\\
	{\bm \Omega}_{\alpha \beta}(\mathbf{q})&=\boldsymbol{\nabla}_\mathbf{q}\times \boldsymbol{\mathcal{A}}_{\alpha \beta}= i \sum_{\gamma} \boldsymbol{\mathcal{A}}_{\alpha \gamma} \times \boldsymbol{\mathcal{A}}_{\gamma \beta},
\end{aligned}
$$
which is a generalization of Eq. (\ref{nonAbOmBer}) that satisfies $\mathbf{m}_{\alpha \beta}=\mathbf{m}_{\beta\alpha }^*$ and ${\bm \Omega}_{\alpha \beta}={\bm \Omega}_{\beta\alpha}^*$.

Considering the MMHS models (\ref{mods}) with $N=4,5$ and $\xi=+$,
denoting $E_\pm(\mathbf{q})>0$ the two bands that are associated to  degenerate electron orbits either in energy or $k$-space,
explicit calculation yields
$$
\begin{aligned}
\mathbf{m}_{+-}(\mathbf{q})\cdot \hat{\mathbf{B}}&=\\
\frac{e}{2}\omega_{+-}& \frac{\sin\theta q_0^2+q_0q_\perp (i\cos \phi_\perp- \sin \phi_\perp \cos \theta) }{(|\mathbf{q}_\perp|^2+q_0^2)^{3/2}},\\
{\bm \Omega}_{+-}(\mathbf{q})\cdot \hat{\mathbf{B}}&=\\
\kappa_{+-}  & \frac{\sin\theta q_0^2+q_0q_\perp (i\cos \phi_\perp-\sin \phi_\perp \cos \theta) }{(|\mathbf{q}_\perp|^2+q_0^2)^{2}},
\end{aligned}
$$
where $\omega_{+-}=\omega_{-+}$ and $\kappa_{+-}=\kappa_{-+}$
are effective parameters that play a similar role as their diagonal conterparts $\omega_\alpha,\kappa_\alpha$.
More quantitatively for $N=4$ we obtain $\omega_{+-}=-(a+b)/4$ and $\kappa_{+-}=-1$ whereas for $N=5$ we find $\omega_{+-}=-\sqrt{2}$ and $\kappa_{+-}=(1+\sqrt{2})$. 
We now integrate these expressions over $\mathbf{q}_\perp$ on a constant energy contour $E=(E_++E_-)/2=c_{+-}|\mathbf{q}|$ with $c_{+-}= (c_++c_-)/2$ at fixed $q_0$. 
Very similarly to the previous intraband calculation we obtain
$$
	\begin{aligned}
		\mathbf{M}_{+-}(E,q_0)\cdot \mathbf{B}&=\\
		\frac{eB}{2\pi}\sin \theta& \left[\frac{\kappa_{+-} }{2} E+\left(\frac{\omega_{+-} }{2c_{+-}} - \kappa_{+-}\right)c_{+-} q_0\right.\\
		&\left.-\left(\frac{\omega_{+-} }{c_{+-}} - \kappa_{+-}\right)\frac{(c_{+-} q_0)^2}{2E}\right],\\
		\mathbf{M}'_{+-}(E,q_0)\cdot \mathbf{B}&=\\
		\frac{eB}{2\pi}\sin \theta & \left[\frac{\kappa_{+-} }{2} 
		+\left(\frac{\omega_{+-} }{c_{+-}} - \kappa_{+-}\right)\frac{(c_{+-} q_0)^2}{2E^2}\right].
	\end{aligned}
$$
The most striking qualitative feature is that this off-diagonal contribution is proportional to $\sin \theta=|(\mathbf{d} \times \hat{\mathbf{B}})|$, whereas the diagonal contributions are proportional to
$\cos \theta =\mathbf{d} \cdot \hat{\mathbf{B}}$.
Taking this quantitative form of the off-diagonal term, the modified semiclassical LL quantization rule now takes a $2\times 2$ matrix form for the two coupled orbits. 
More precisely, the LLs are found as solutions of
$$
	\det
	\begin{pmatrix}
		X_+&\mathbf{M}_{+-} '(\epsilon_n,q_0)\cdot\mathbf{B}\\
		\mathbf{M}_{-+} '(\epsilon_n,q_0)\cdot\mathbf{B}&X_-\\
	\end{pmatrix}=0,
$$
where $X_\pm\equiv N_\pm(\epsilon_n,q_0)+\mathbf{M}_\pm '(\epsilon_n,q_0)\cdot\mathbf{B} -\frac{eB}{2\pi}(n+\frac{1}{2})$.
Within the approximation $1/E^2 \ll 1$ we only consider the simplified forms 
$\mathbf{M}_\pm '(\epsilon_n,q_0)\cdot\mathbf{B}=\frac{eB}{2\pi}\frac{\kappa_{\pm}}{2} \cos \theta $
and $\mathbf{M}_{+-} '(\epsilon_n,q_0)\cdot\mathbf{B}=\frac{eB}{2\pi}\frac{\kappa_{+-}}{2} \sin \theta $.
Within that simplified scheme the semiclassical LLs are solutions of 
$$
\det
\begin{pmatrix}
	x_+ & eB\kappa_{+-}\sin \theta\\
	eB\kappa_{+-} \sin \theta & x_-\\
\end{pmatrix}=0,
$$
where $x_\pm\equiv\epsilon_n^2/c_\pm^2-[2eB(n+\frac{1}{2}-\frac{\kappa_\pm}{2}\cos\theta)+q_0^2]$. From this we finally obtain
\begin{equation}
	\begin{aligned}
		\epsilon_n&=\pm \sqrt{\eta_++\eta_- \pm \sqrt{(\eta_+-\eta_-)^2+\eta_{+-}^2}},\\
		\eta_\pm&=\frac{c_\pm^2}{2}\left[2eB\left(n+\frac{1}{2}-\frac{\kappa_\pm}{2}\cos\theta\right)+q_0^2\right],\\
		\eta_{+-}&=c_+c_-eB \kappa_{+-} \sin \theta
	\end{aligned}
\end{equation}
Repeating the same calculation for $\xi=-$, we exactly recover the LLs (\ref{LL4}) and (\ref{LL5}) if we choose the effective interband parameters as $\kappa_{+-}=1$ for $N=4$ and 
$\kappa_{+-}=\sqrt{6}$ for $N=5$.

\section{Boltzmann formalism for multiband systems to linear order in the magnetic field}
\label{Appboltzi}

Here we review semiclassical magnetotransport to first order in the magnetic field. Below we will apply the formalism to MMHS continuum and lattice models.

Consider a multiband system with Bloch Hamiltonian $H(\mathbf{k})$, eigenfunctions $|\psi_\alpha(\mathbf{k})\rangle$ and eigenenergies $\e_\alpha(\mathbf{k})$. A typical strategy to address DC magnetotransport properties of such systems consists in considering the Boltzmann equation in the (constant) relaxation time approximation,
\begin{equation}
	\begin{aligned}
	\left(\frac{\partial}{\partial t}+\dot{\mathbf{r}}\cdot\boldsymbol{\nabla_\mathbf{r}}+\dot{\mathbf{k}}\cdot\boldsymbol{\nabla_\mathbf{k}}\right)&f_\alpha(\mathbf{k},\mathbf{r},t)=\\
	-\frac{1}{\tau}&[f_\alpha(\mathbf{k},\mathbf{r},t)-f_\alpha^\text{eq}(\mathbf{k},\mathbf{r},t)],
	\end{aligned}
	\label{boltzi}
\end{equation}
where $\mathbf{r}$ and $\mathbf{k}$ are the position and momentum of the semiclassical wave packet in the band $\alpha$, $f_\alpha(\mathbf{k},\mathbf{r},t)$ is the distribution function and $f_\alpha^\text{eq}(\mathbf{k},\mathbf{r},t)$ its equilibrium part in the absence of an electric field. The second and third terms on the left-hand side of Eq. (\ref{boltzi}) describe diffusion and drift of the wave packet, respectively, while the right-hand side takes account of scattering with a phenomenological scattering rate $1/\tau$. 

The goal is now to combine the Boltzmann equation with the semiclassical equations of motion in order to solve for the distribution function $f_\alpha(\mathbf{k})$, and thus to obtain the full electrical current.

\subsection{Derivation of the linear response electrical current}

It is known that the semiclassical equations of motion entering the Boltzmann equation (\ref{boltzi}) should be extended by terms due to the Berry curvature $\boldsymbol{\Omega}_\alpha(\mathbf{k})$ and the orbital magnetic moment (OMM) $\mathbf{m}_\alpha(\mathbf{k})$ of the band $\alpha$ \cite{Xiao_2010}, see for example Ref. \cite{Cortijo_2016} for a recent discussion. The Berry curvature pseudovector is given by Eq. (\ref{Berry}) and the OMM by Eq. (\ref{OMM}) [with $\mathbf{q}\rightarrow\mathbf{k}$].
In the presence of Berry curvature and OMM, the semiclassical equations of motion read as
\begin{equation}
	\dot{\mathbf{r}}=\mathbf{w}_\alpha-\dot{\mathbf{k}}\times\boldsymbol{\Omega}_\alpha,\hspace{.5cm}\dot{\mathbf{k}}=-e(\mathbf{E}+\dot{\mathbf{r}}\times\mathbf{B}),
\end{equation}
where $e$ is the electron charge, and $\mathbf{w}_\alpha=\mathbf{v}_\alpha-\boldsymbol{\nabla_\mathbf{k}}$($\mathbf{m}_\alpha\cdot\mathbf{B}$) is the band velocity in the presence of a Zeeman-like energy shift $\e_\alpha(\mathbf{k})\rightarrow\e_\alpha(\mathbf{k})-\mathbf{m}_\alpha\cdot\mathbf{B}$, with $\mathbf{v}_\alpha=\boldsymbol{\nabla_\mathbf{k}}\e_\alpha$ the band velocity of the zero-field spectrum. The equations of motion can be fully decoupled as
\begin{equation*}
	\begin{aligned}
		(1+e\boldsymbol{\Omega}_\alpha\cdot\mathbf{B})\dot{\mathbf{r}}&=\mathbf{w}_\alpha-e\boldsymbol{\Omega}_\alpha\times\mathbf{E}+e(\mathbf{w}_\alpha\cdot\boldsymbol{\Omega}_\alpha)\mathbf{B},\\
		(1+e\boldsymbol{\Omega}_\alpha\cdot\mathbf{B})\dot{\mathbf{k}}&=-e\mathbf{E}-e\mathbf{w}_\alpha\times\mathbf{B}-e^2(\mathbf{E}\cdot\mathbf{B})\boldsymbol{\Omega}_\alpha.
	\end{aligned}
\end{equation*}
Now, for a homogeneous system in the steady state, and to first order in the electric field (linear response regime), the Boltzmann equation (\ref{boltzi}) becomes
$$
	\begin{aligned}
	e[\mathbf{E}+\mathbf{w}_\alpha\times\mathbf{B}+e(\mathbf{E}\cdot\mathbf{B})\boldsymbol{\Omega}_\alpha]\cdot\boldsymbol{\nabla_\mathbf{k}}f_\alpha(\mathbf{k})&=\\
	(1+e\boldsymbol{\Omega}_\alpha\cdot\mathbf{B})&\frac{f_\alpha(\mathbf{k})-f^\text{eq}_\alpha(\mathbf{k})}{\tau}.
	\end{aligned}
$$
One can now solve for the distribution function $f_\alpha(\mathbf{k})$ to obtain the electrical current as a power series in the magnetic field.

To proceed, it is convenient to rewrite the Boltzmann equation as
$$
\begin{aligned}
	[1+e\boldsymbol{\Omega}_\alpha\cdot\mathbf{B}+e\tau\mathbf{B}\cdot(\mathbf{v}_\alpha\times\boldsymbol{\nabla_\mathbf{k}})]f_\alpha^\text{neq}(\mathbf{k})&=\\
	[e\tau\mathbf{E}+e^2\tau(\mathbf{E}\cdot\mathbf{B})\boldsymbol{\Omega}_\alpha]\cdot\boldsymbol{\nabla_\mathbf{k}}f_\alpha^\text{eq}(\mathbf{k})&,
\end{aligned}
$$
where $f_\alpha^\text{neq}(\mathbf{k})=f_\alpha(\mathbf{k})-f_\alpha^\text{eq}(\mathbf{k})$ is the nonequilibrium part of the distribution function, and $f_\alpha^\text{eq}(\mathbf{k})\equiv f(\tilde{\e}_\alpha)$ is the equilibrium part with
\begin{equation}
f(x)\equiv1/(1+\text{exp}[\beta (x-\mu)])
\end{equation}
 the Fermi-Dirac distribution function with inverse temperature $\beta=1/(k_BT)$ and chemical potential $\mu$. To first order in $B$, the nonequilibrium part can be obtained as
\begin{widetext}
$$
	\begin{aligned}
		f_\alpha^\text{neq}(\mathbf{k})&=e\tau[1-e\boldsymbol{\Omega}_\alpha\cdot\mathbf{B}-e\tau\mathbf{B}\cdot(\mathbf{v}_\alpha\times\boldsymbol{\nabla_\mathbf{k}})][\mathbf{E}+e(\mathbf{E}\cdot\mathbf{B})\boldsymbol{\Omega}_\alpha]\cdot\mathbf{w}_\alpha f'(\tilde{\e}_\alpha)\\
		&=e\tau(\mathbf{E}\cdot\mathbf{v}_\alpha)f'(\e_\alpha)+e^2\tau[(\mathbf{E}\cdot\mathbf{B})(\boldsymbol{\Omega}_\alpha\cdot\mathbf{v}_\alpha)-(\boldsymbol{\Omega}_\alpha\cdot\mathbf{B})(\mathbf{E}\cdot\mathbf{v}_\alpha)]f'(\e_\alpha)\\
		&-e\tau\left\{\mathbf{E}\cdot[\boldsymbol{\nabla}_\mathbf{k}(\mathbf{m}_\alpha\cdot\mathbf{B})]f'(\e_\alpha)+(\mathbf{E}\cdot\mathbf{v}_\alpha)(\mathbf{m}_\alpha\cdot\mathbf{B})f''(\e_\alpha)\right\}-e^2\tau^2[\mathbf{B}\cdot(\mathbf{v}_\alpha\times\boldsymbol{\nabla}_\mathbf{k})](\mathbf{E}\cdot\mathbf{v}_\alpha)f'(\e_\alpha).
	\end{aligned}
$$
\end{widetext}
The electrical current is obtained by integrating over the full phase space using the full distribution function:
$$
	\begin{aligned}
		\mathbf{j}&=
		-e\sum_\alpha\int\frac{d^3k}{(2\pi)^3}(1+e\boldsymbol{\Omega}_\alpha\cdot\mathbf{B})\dot{\mathbf{r}}[f(\tilde{\e}_\alpha)+f_\alpha^\text{neq}(\mathbf{k})]\\
		&=-e\sum_\alpha\int\frac{d^3k}{(2\pi)^3}[\mathbf{w}_\alpha-e\boldsymbol{\Omega}_\alpha\times\mathbf{E}+e(\mathbf{v}_\alpha\cdot\boldsymbol{\Omega}_\alpha)\mathbf{B}]\\
		&\hspace{2.5cm}\times[f(\tilde{\e}_\alpha)+f_\alpha^\text{neq}(\mathbf{k})].
	\end{aligned}
$$

\subsection{Decomposition of the electrical current}

The current $\mathbf{j}$ consists of three parts,
\begin{equation}
\mathbf{j}=\mathbf{j}_0+\mathbf{j}_\text{AH}+\mathbf{j}_\text{out}.
\end{equation}
The first term $\mathbf{j}_0$ is an equilibrium current (typically vanishing) independent of $\mathbf{E}$, which we will not consider here. The second term $\mathbf{j}_\text{AH}$ describes the (non-dissipative) anomalous Hall (AH) current, which requires the presence of an electrical field but is determined by the equilibrium distribution function; the third term $\mathbf{j}_\text{out}$ is the true out-of-equilibrium current with both dissipative and non-dissipative contributions, determined by the nonequilibrium part of the distribution function. Quantitatively, these two current contributions are given by
$$
	\begin{aligned}
		\mathbf{j}_\text{AH}&=e^2\sum_\alpha\int\frac{d^3k}{(2\pi)^3}(\boldsymbol{\Omega}_\alpha\times\mathbf{E})f(\tilde{\e}_\alpha),\\
		\mathbf{j}_\text{out}&=-e\sum_\alpha\int\frac{d^3k}{(2\pi)^3}[\mathbf{w}_\alpha+e(\mathbf{v}_\alpha\cdot\boldsymbol{\Omega}_\alpha)\mathbf{B}]f_\alpha^\text{neq}(\mathbf{k}).
	\end{aligned}
$$
Let us first focus on $\mathbf{j}_\text{AH}$.
Expanding to linear order in $B$ and introducing an anomalous Hall conductivity tensor as
\begin{equation}
j^\text{AH}_i=\sum_j[\sigma_{ij}^\text{AH}+\sigma_{ij}^\text{AH1}(\mathbf{B})]E_j,
\end{equation}
  we have
\begin{equation}
		\begin{aligned}
			\sigma_{ij}^\text{AH}&=-e^2\sum_\alpha\int\frac{d^3k}{(2\pi)^3}f(\e_\alpha)\epsilon_{ijl}(\Omega_\alpha)_l,\\
			\hspace{-.4cm}\sigma_{ij}^\text{AH1}(\mathbf{B})&=e^2\sum_\alpha\int\frac{d^3k}{(2\pi)^3}f'(\e_\alpha)(\mathbf{m}_\alpha\cdot\mathbf{B})\epsilon_{ijl}(\Omega_\alpha)_l,
	\end{aligned}
	\label{AH}
\end{equation}
where $\sigma_{ij}^\text{AH}$ is the true anomalous Hall effect (a consequence of the Berry curvature) and $\sigma_{ij}^\text{AH1}(\mathbf{B})$ is a magnetic field dependent quantum geometric correction that also involves the OMM. 

Similarly, the out-of-equilibrium conductivity tensor can be defined as
\begin{equation}
j^\text{out}_i=\sum_j\sigma_{ij}(\mathbf{B})E_j,
\end{equation}
with
\begin{equation}
\begin{aligned}
	\sigma_{ij}(\mathbf{B})&=\sigma_{ij}^{\text{Drude}}+\sigma_{ij}^\text{Lorentz}(\mathbf{B})\\
	&+\sigma_{ij}^\text{Berry}(\mathbf{B})+\sigma_{ij}^\text{OMM}(\mathbf{B}).
\end{aligned}
\end{equation}
Here $\sigma_{ij}^{\text{Drude}}$ is the Drude conductivity, $\sigma_{ij}^\text{Lorentz}(\mathbf{B})$ the classical Hall conductivity induced by the Lorentz force, and $\sigma_{ij}^\text{Berry}(\mathbf{B})$ and $\sigma_{ij}^\text{OMM}(\mathbf{B})$ are interband contributions induced by Berry curvature and orbital magnetic moment. These tensors take the following explicit form in terms of band velocity, Berry curvature and OMM:
\begin{widetext}
\begin{equation}
		\begin{aligned}		
			\sigma_{ij}^{\text{Drude}}&=	-e^2\tau\sum_\alpha\int\frac{d^3k}{(2\pi)^3}f'(\e_\alpha)(v_\alpha)_i(v_\alpha)_j,\\
			\sigma_{ij}^\text{Lorentz}(\mathbf{B})&=e^3\tau^2\sum_\alpha\int\frac{d^3k}{(2\pi)^3}f'(\e_\alpha)(v_\alpha)_i[\mathbf{B}\cdot(\mathbf{v}_\alpha\times\boldsymbol{\nabla_\mathbf{k}})](v_\alpha)_j,\\
			\sigma_{ij}^\text{Berry}(\mathbf{B})&=-e^3\tau\sum_\alpha\int\frac{d^3k}{(2\pi)^3}f'(\e_\alpha)\left\{(\boldsymbol{\Omega}_\alpha\cdot\mathbf{v}_\alpha)\left[(v_\alpha)_iB_j+(v_\alpha)_jB_i\right]-(\boldsymbol{\Omega}_\alpha\cdot\mathbf{B})(v_\alpha)_i(v_\alpha)_j\right\},\\
			\sigma_{ij}^\text{OMM}(\mathbf{B})&=e^2\tau\sum_\alpha\int\frac{d^3k}{(2\pi)^3}\left\{f'(\e_\alpha)\left[(v_\alpha)_i\partial_j(\mathbf{m}_\alpha\cdot\mathbf{B})+(v_\alpha)_j\partial_i(\mathbf{m}_\alpha\cdot\mathbf{B})\right]+f''(\e_\alpha)(\mathbf{m}_\alpha\cdot\mathbf{B})(v_\alpha)_i(v_\alpha)_j\right\}\\
			&=e^2\tau\sum_\alpha\int\frac{d^3k}{(2\pi)^3}f'(\e_\alpha)\left\{\frac{1}{2}\left[(v_\alpha)_i\partial_j(\mathbf{m}_\alpha\cdot\mathbf{B})+(v_\alpha)_j\partial_i(\mathbf{m}_\alpha\cdot\mathbf{B})\right]-(\mathbf{m}_\alpha\cdot\mathbf{B})\partial_j(v_\alpha)_i\right\}.
	\end{aligned}
	\label{siggy}
\end{equation}
\end{widetext}
Note that these results for Berry curvature and OMM contributions appear to agree exactly with a microscopic (fully quantum mechanical) approach, see Eqs. (8) and (12) of Ref. \cite{Konye_2021}. 

\section{Semiclassical magnetotransport theory for MMHS continuum and lattice models}
\label{Apptransp}

We now apply the formalism of Appendix \ref{Appboltzi} to MMHS continuum and lattice models.

\subsection{Application to MMHS continuum models}

 We now wish to evaluate the conductivity tensors (\ref{AH}) and (\ref{siggy}) for the MMHS continuum models (\ref{mods}). More concretely , we ignore the Drude and Lorentz conductivities, as they depend only on the energy spectrum and are the same as for a pseudospin-$s$ Hamiltonian $H_\text{s}(\mathbf{q})$. We also leave aside the ``magnetic field dependent anomalous Hall contribution", as it is a higher-order quantum geometric effect that couples Berry curvature and OMM. 
 The three remaining interesting contributions are then the true anomalous Hall contribution $\sigma_{ij}^\text{AH}$ as well as the quantum geometric contributions \smash{$\sigma_{ij}^\text{Berry}$} and \smash{$\sigma_{ij}^\text{OMM}$}.

\subsubsection{Symmetries of anomalous Hall and quantum geometric currents}

Before evaluating these three contributions explicitly, we can already infer how they behave as a function of the Fermi level $E_F$, simply by applying general symmetry considerations to Eqs. (\ref{AH}) and (\ref{siggy}).

In particular, for a Weyl semimetal, or in fact for any two-band system, the Berry curvature has opposite signs in the two bands, such that the anomalous Hall conductivity of Eq. (\ref{AH}) is $\sigma_{ij}^\text{AH}\sim\e_{ijl}\int d^3k[f(\epsilon_+)-f(\epsilon_-)](\Omega_+)_l$. If the spectrum is particle-hole symmetric, $\epsilon_+=-\epsilon_-$, then the AH conductivity is necessarily even in $E_F$. 

We now show that the AH conductivity is odd in $E_F$ for a MMHS. First, for $N$ even, such that there is no flat band, each pair of bands $\pm\epsilon_n$ contributes an AH conductivity $\sigma_{ij,n}^\text{AH}\sim\e_{ijl}\int d^3k[f(\epsilon_n)+f(-\epsilon_n)](\Omega_n)_l$, where we have used that the Berry curvature is symmetric with respect to zero energy. Rewriting $f(\epsilon_n)+f(-\epsilon_n)=1+g_n$, where $g_n$ is an odd function of $E_F$, we have
$$
\sigma_{ij}^\text{AH}\sim\e_{ijl} \int d^3k\sum_{n=1}^{N/2}g_n(\Omega_n)_l,
$$
where we have used $\sum_{n=1}^{N/2}(\Omega_n)_l=0$. Second, for $N$ odd, such that there is a flat band, we can conduct a similar procedure to find
$$
\begin{aligned}
	\sigma_{ij}^\text{AH}&\sim\e_{ijl} \int d^3k\left[\sum_{n=1}^{(N-1)/2}(1+g_n)(\Omega_n)_l+f(0)(\Omega_0)_l\right]\\
	&=\e_{ijl} \int d^3k\sum_{n=1}^{(N-1)/2}[g_n+1-2f(0)](\Omega_n)_l,
\end{aligned}
$$
where we have used $\sum_{n=1}^{(N-1)/2}(\Omega_n)_l=-\frac{1}{2}(\Omega_0)_l$. The two above equations are odd in $E_F$, such that $\sigma_{ij}^\text{AH}$ is odd as claimed in the main text.

Similar arguments can be applied to $\sigma_{ij}^\text{Berry}$ and $\sigma_{ij}^\text{OMM}$ in Eq. (\ref{siggy}). For a particle-hole symmetric two-band system (e.g. Weyl semimetal), since the Berry curvature (orbital magnetic moment) is antisymmetric (symmetric) with respect to zero energy, both these contributions are $\sim(f'(\epsilon_+)-f'(-\epsilon_+))$, i.e. odd functions of $E_F$. In contrast, for the MMHS models, the Berry curvature (orbital magnetic moment) is symmetric (antisymmetric) with respect to zero energy, such that $\sigma_{ij}^\text{Berry}$ and $\sigma_{ij}^\text{OMM}$ are $\sim\sum_{n=1}^{\lfloor N/2\rfloor}(f'(\epsilon_n)+f'(-\epsilon_n))\mathcal{F}_n$, that is, even in $E_F$, as claimed in the main text. Here, $\mathcal{F}_n$ is some function independent of $E_F$, and the flat band for $N$ odd plays no role.

\subsubsection{Anomalous Hall current}

Let us now come to the evaluation of the conductivities (at zero temperature).  For the AH conductivity (\ref{AH}) we obtain
\begin{equation}
		\sigma_{ij}^{\text{AH}}=-\xi\frac{e^2}{6\pi^2}\sum_\alpha \kappa_\alpha\int_0^{q_c} dq\Theta(E_F-c_\alpha q)\e_{ijz},
		\label{coldAH}
\end{equation}
where we have introduced a momentum cutoff $q_c$ to the integral in order to avoid divergence. This issue is an artefact of the continuum model and is absent for a lattice model, as will be shown below. To explicitly compute Eq. (\ref{coldAH}), recall the coefficients $c_\alpha$, $\kappa_\alpha$ and $\omega_\alpha$ listed in Table \ref{tab0}. 

\begin{figure}
	\centering
	\includegraphics[width=\columnwidth]{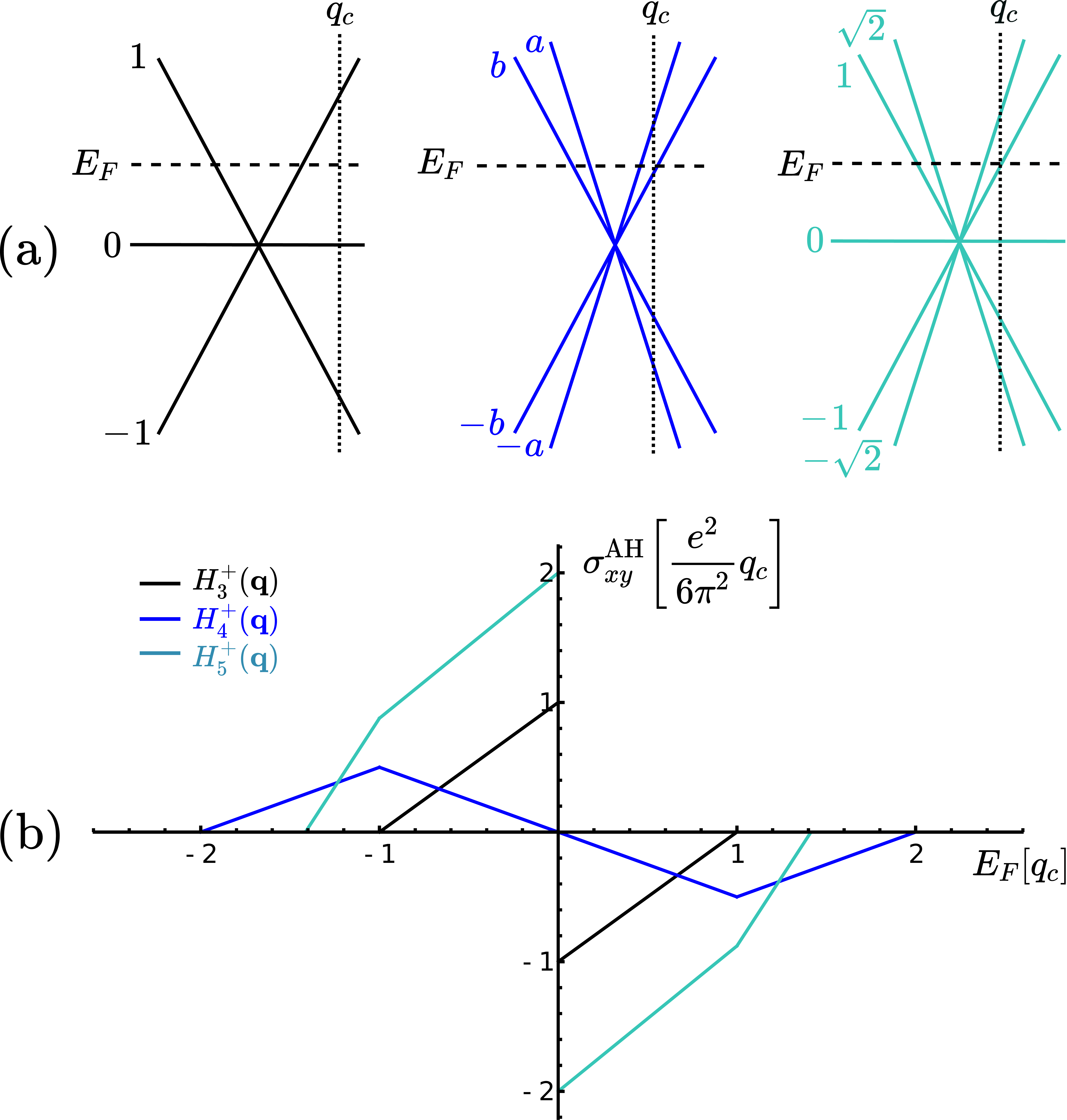}
	\caption{(a) Schematic energy spectrum of the Hopf semimetal continuum models $H_{N=3,4,5}^{\xi=+}(\mathbf{q})$, with Fermi energy $E_F$, momentum cutoff $q_c$ and band velocities $c_\alpha$ indicated. (b) Corresponding anomalous Hall conductivity. For $N=4$, we have used band velocities $a=2$, $b=1$. For $N$ odd, there is an abrupt sign change at half filling, with a jump equal to the Berry dipole charge $\kappa_0$ of the flat band.}
	\label{fig:AH}
\end{figure}

Consider first the three-band Hopf semimetal (\ref{Berrydip3}), whose spectrum is shown in the left column of Fig. \ref{fig:AH}(a).
 When tuning $E_F$, three regions can be distinguished and the AH conductivity becomes
$$
	\begin{aligned}
		\sigma_{xy}^\text{AH}&=-\xi\frac{e^2}{6\pi^2}\begin{cases}
			0\\
			\kappa_{-1}(E_F+q_c)\\
			\kappa_{1}(E_F-q_c)
		\end{cases}\\
	&=\xi\frac{e^2}{6\pi^2}\begin{cases}
			0, &\hspace{.2cm} |E_F|>q_c\\
			E_F+q_c, &\hspace{.2cm}  -q_c<E_F<0\\
			E_F-q_c, &\hspace{.2cm}  0<E_F<q_c
		\end{cases}
	\end{aligned}
$$
where the Berry dipole charges are $\kappa_1=\kappa_{-1}=-\kappa_0/2=-1$, cf. Table \ref{tab0}. This conductivity is plotted in Fig. \ref{fig:AH}(b) and changes sign abruptly at the flat band, with a jump equal to the flat band Berry dipole charge $\kappa_0=2$.

 Similarly, for the four-band Hopf semimetal (\ref{Berrydip4}), see Fig. \ref{fig:AH}(a), four regions can be distinguished as a function of $E_F$. We find
$$
	\begin{aligned}
		\sigma_{xy}^\text{AH}&=-\xi\frac{e^2}{6\pi^2}\begin{cases}
			0\\
			\kappa_{-a}(E_F/a+q_c)\\
			E_F(\kappa_{-a}/a+\kappa_{-b}/b)\\
			\kappa_a(E_F/a-q_c)
		\end{cases}\\
	&=\xi\frac{e^2}{6\pi^2}\begin{cases}
			0, &\hspace{1cm}  |E_F|>aq_c\\
			E_F/a+q_c, &\hspace{.2cm}  -aq_c<E_F<-bq_c\\
			E_F(1/a-1/b), & \hspace{.2cm} -bq_c<E_F<bq_c\\
			E_F/a-q_c, &\hspace{.2cm}  bq_c<E_F<aq_c
		\end{cases}
	\end{aligned}
$$
with Berry dipole charges $\kappa_{\pm a}=-1$, $\kappa_{\pm b}=1$. This conductivity is plotted in Fig. \ref{fig:AH}(b).

 Finally, for the five-band Hopf semimetal (\ref{Berrydip5}), five regions can be distinguished, with conductivities
$$
	\begin{aligned}
		\sigma_{xy}^\text{AH}&=-\xi\frac{e^2}{6\pi^2}\begin{cases}
			0\\
			\kappa_{-\sqrt{2}}(E_F/\sqrt{2}+q_c)\\
			E_F(\kappa_{-\sqrt{2}}/\sqrt{2}+\kappa_{-1})+(\kappa_0/2)q_c\\
			E_F(\kappa_{\sqrt{2}}/\sqrt{2}+\kappa_1)-(\kappa_0/2)q_c\\
			\kappa_{\sqrt{2}}(E_F/\sqrt{2}-q_c)\\
		\end{cases}\\
	&=\xi\frac{e^2}{6\pi^2}\begin{cases}
			0, & \hspace{.2cm}|E_F|>\sqrt{2}q_c\\
			3(E_F/\sqrt{2}+q_c), &\hspace{.2cm}-\sqrt{2}q_c<E_F<-q_c\\
			E_F(3/\sqrt{2}-1)+2q_c, &  \hspace{.2cm}-q_c<E_F<0\\
			E_F(3/\sqrt{2}-1)-2q_c, & \hspace{.2cm} 0<E_F<q_c\\
			3(E_F/\sqrt{2}-q_c), & \hspace{.2cm} q_c<E_F<\sqrt{2}q_c\\
		\end{cases}
	\end{aligned}
$$
where $\kappa_{\pm\sqrt{2}}=-3$, $\kappa_{\pm1}=1$ and $\kappa_0=4$, cf. Table \ref{tab0}. This is also plotted in Fig. \ref{fig:AH}(b). Again, the jump of the conductivity at half filling is equal to the flat-band Berry dipole charge.

The anomalous Hall conductivity clearly shows three main features: it is of opposite sign for opposite Berry dipole orientation $\xi$, its amplitude explicitly depends on the Berry dipole charges $\kappa_\alpha$, and it is odd in $E_F$ as anticipated from the above symmetry considerations. For all continuum models, we may therefore write the general anomalous Hall current $\mathbf{j}_\text{AH}$ with components $j_i^\text{AH}=\sum_j\sigma_{ij}^\text{AH}E_j$ in the form (\ref{AHcur}),
where $\sigma_\text{AH}(E_F)=-\sigma_\text{AH}(-E_F)$. Note that we used $\mathbf{d}=(0,0,\xi)$.

\subsubsection{Dissipative quantum geometric current}

Beyond the AH effect, we also consider the dissipative quantum geometric current $\mathbf{j}_\text{geo}$ with components 
\begin{equation}
j_i^\text{geo}=\sum_j\left[\sigma_{ij}^\text{Berry}(\mathbf{B})+\sigma_{ij}^\text{OMM}(\mathbf{B})\right]E_j,
\end{equation}
cf. Eq. (\ref{siggy}).
At zero temperature, the tensors take the form
$$
		\begin{aligned}
			\sigma_{ij}^{\text{Berry}}(\mathbf{B})&=\xi\frac{e^3\tau}{30\pi^2}\left(\sum_{c_\alpha>0}\kappa_\alpha c_\alpha\right)\begin{pmatrix}
				-B_z & 0 & 4B_x\\
				0 & -B_z & 4B_y\\
				4B_x & 4B_y & 7B_z
			\end{pmatrix},\\
			\sigma_{ij}^{\text{OMM}}(\mathbf{B})&=\xi\frac{e^3\tau}{60\pi^2}\left(\sum_{c_\alpha>0}\omega_\alpha\right)\begin{pmatrix}
				7B_z & 0 & -3B_x\\
				0 & 7B_z & -3B_y\\
				-3B_x & -3B_y & B_z
			\end{pmatrix},
		\end{aligned}
$$
where the coefficients are again listed in Table \ref{tab0}.
 For the threefold HS (\ref{Berrydip3}), we get
$$
	\sigma_{ij}^\text{Berry}(\mathbf{B})+\sigma_{ij}^\text{OMM}(\mathbf{B})=-\xi\frac{e^3\tau}{12\pi^2}\begin{pmatrix}
		B_z & 0 & B_x\\
		0 & B_z & B_y\\
		B_x & B_y & 3B_z
	\end{pmatrix},
$$
which corresponds to
\begin{equation}
	\mathbf{j}_\text{geo}=-\frac{e^3\tau}{12\pi^2}\left[(\mathbf{E}\cdot\mathbf{B})\mathbf{d}+(\mathbf{E}\cdot\mathbf{d})\mathbf{B}+(\mathbf{B}\cdot\mathbf{d})\mathbf{E}\right],
\end{equation}
where we used $\mathbf{d}=(0,0,\xi)$. A similar calculation for the four- and fivefold HSs yields the general result (\ref{magcur}).


\subsection{Comparison to lattice calculations}

In the case of the AH conductivity, some doubts may arise as to the validity of the continuum treatment which requires using a cutoff. To avoid this issue it is instructive to consider the AH conductivity on the lattice.

In particular, let us consider the topological MMHS lattice models $\mathfrak{h}_N(\mathbf{k})$ defined in the main text [see Eq. (\ref{truesem})] for $\Delta_0=-3$, where a single nodal point in the Brillouin zone exists at the $\Gamma$ point. Computing the Berry curvature of these models, one can numerically calculate the AH conductivity at zero temperature from Eq. (\ref{AH}). The resulting conductivity $\sigma_{xy}^\text{AH}(E_F)$ is plotted in Fig. \ref{fig:AHeffect}(a). Its qualitative behavior is exactly like  predicted from the continuum theory and from general symmetry arguments above: it is odd in $E_F$ and its global sign is determined by the direction of the Berry dipole at the $\Gamma$ point.

For comparison and for completeness, we also numerically compute the AH conductivity for a lattice model featuring a pair of Weyl nodes. In particular, we take the model \cite{Yang_2011}
\begin{equation}
	\begin{aligned}
	h_\text{W}(\mathbf{k})&=[m(2-\cos k_x-\cos k_y)+2t_z(\cos k_z-\cos k_0)]\sigma_1\\
	&+2t_x \sin k_x \sigma_2+2 t_y \sin k_y \sigma_3,
	\end{aligned}
	\label{weyleq}
\end{equation} 
which describes a Weyl semimetal with a pair of nodes at $\mathbf{k}_\text{W}=\pm(0,0,k_0)$. Taking (arbitrary) parameters $m=2$, $t_i=1$, and $k_0=\pi/2$, one obtains the anomalous Hall conductivity shown in Fig. \ref{fig:AHeffect}(b). As expected from symmetry arguments, it is even in $E_F$. The same calculation can be repeated for multifold chiral topological semimetals simply upon replacing $\boldsymbol{\sigma}\rightarrow\mathbf{S}$ in the model (\ref{weyleq}), and the results (for $s=1$ and $s=3/2$) are also shown in Fig. \ref{fig:AHeffect}(b).

Finally, we have also conducted numerical calculations of the geometric current $\mathbf{j}_\text{geo}$ on the lattice (not shown). They confirm that $\mathbf{j}_\text{geo}$ is even in $E_F$ for the MMHS models (\ref{truesem}), while it is odd in $E_F$ for the Weyl semimetal (\ref{weyleq}), as claimed in the main text and as expected by symmetry.

\section{Hopf density and Hopf number}
\label{Apphopf}

\subsection{Calculation of the lattice Hopf number}

Here we compute the Hopf number (\ref{Hopf}) for the insulators (\ref{hopfdef}). When computing the Hopf density (\ref{hopfdens}), it is important to note that this density is gauge-dependent, however the integral (\ref{Hopf}) is gauge-invariant under the condition that the weak invariants (\ref{subchern})  vanish.

For the $N=3$ and $N=4$ case, using a convenient gauge choice we find a Hopf density
\begin{equation}
	\chi(\mathbf{k})=\frac{12}{\e^4}\left(c_x c_y+c_y c_z+c_z c_x+\Delta c_x c_y c_z\right),
	\label{hodens}
\end{equation}
where $c_i=\cos k_i$ and $s_i=\sin k_i$. From numerical integration one obtains 
\begin{equation}
	\mathcal{N}_\text{Hopf}=\begin{cases}
		0 & |\Delta|>3\\
		1 & 1<|\Delta|<3\\
		-2 & 0<|\Delta|<1
	\end{cases},
	\label{honum}
\end{equation}
as shown in Fig. \ref{fig:Hopfinvariant}(a). 
For the $N=5$ model the Hopf density is more involved; for our gauge choice we have
$$
	\begin{aligned}
		\chi(\mathbf{k})&=\frac{12}{\e^4}\left\{-c_xc_y+2c_xc_z+2c_yc_z-2(c_x^2+c_y^2)\right.\\
		&\hspace{1cm}-(c_xc_{2y}+c_yc_{2x})c_z+2(c_x-c_y)s_xs_ys_z\\
		&\hspace{1cm}+\Delta\left[c_xc_yc_z-(c_xc_{2y}+c_yc_{2x})-3(c_x+c_y)\right]\\
		&\hspace{1cm}\left.-2\Delta^2c_xc_y\right\}.
	\end{aligned}
$$
This leads to higher Hopf numbers,
\begin{equation}
	\mathcal{N}_\text{Hopf}=\begin{cases}
		0 & |\Delta|>3\\
		5 & 1<|\Delta|<3\\
		-10 & 0<|\Delta|<1
	\end{cases},
\end{equation}
as also shown in Fig. \ref{fig:Hopfinvariant}(a). 

\subsection{Hopf number from a continuum approach}

  One may be tempted to study the jump $\delta\mathcal{N}_\text{Hopf}$ at the topological phase transitions shown in Fig. \ref{fig:Hopfinvariant} in terms of the continuum limit $H_\nu(\mathbf{q})$ of the models (\ref{hopfdef}) around gap-closing momenta $\mathbf{k}_\nu$, where $\nu\in\{\Gamma,\text{X},\text{M},\text{R}\}$.  In particular, one may try to define a continuum Hopf number as
$$
	\begin{aligned}
		\mathcal{N}_\nu^\text{Hopf}(\Delta_\nu)&=\frac{1}{24\pi^2}\int_{\mathbb{R}^3} d^3q\,\chi(\mathbf{q}),\\
		\chi(\mathbf{q})&\equiv\epsilon_{ijk}\Tr[u_i(\mathbf{q})u_j(\mathbf{q})u_k(\mathbf{q})],\\
		&=3 \Tr\left[\boldsymbol{\mathcal{A}}\cdot\left(\boldsymbol{\nabla}\times\boldsymbol{\mathcal{A}}\right)-\frac{2i}{3}\boldsymbol{\mathcal{A}}\cdot\left(\boldsymbol{\mathcal{A}}\times\boldsymbol{\mathcal{A}}\right)\right],
	\end{aligned}
$$
which can be viewed as a multiband generalization of a formula for a ``continuum Hopf number" proposed for two-band systems in Ref. \cite{Nelson_2022}. Here, we have $\Delta_\Gamma=\Delta+3$, $\Delta_\text{R}=\Delta-3$, $\Delta_\text{X}=\Delta+1$ and  $\Delta_\text{M}=\Delta-1$. Moreover, $\boldsymbol{\mathcal{A}}=\boldsymbol{\mathcal{A}}(\mathbf{q})=(\mathcal{A}_x(\mathbf{q}),\mathcal{A}_y(\mathbf{q}),\mathcal{A}_z(\mathbf{q}))$ is a vector formed from $N\times N$ non-Abelian Berry connection matrices with matrix elements $\mathcal{A}_{i,\alpha\beta}(\mathbf{q})=i\langle \psi_\alpha(\mathbf{q})|\partial_i \psi_\beta(\mathbf{q})\rangle$.

We may then aim to compute the jump $\delta\mathcal{N}_\text{Hopf}$ of the Hopf number at a topological phase transition as the difference between the continuum Hopf numbers on both sides of the transition, summed over all points where the gap closes:
$$
	\delta\mathcal{N}_\text{Hopf}=\sum_\nu\left[\mathcal{N}_\nu^\text{Hopf}(\Delta_\nu>0)-\mathcal{N}_\nu^\text{Hopf}(\Delta_\nu<0)\right].
$$
This sum consists of one term for the transitions at $\Gamma$ and R, and of three terms for the transitions at X and M. For a convenient gauge choice,
 we indeed find that the jumps shown in Fig. \ref{fig:Hopfinvariant}(a) can be predicted from the continuum approach described here. However, the utility of this approach remains inconclusive, given that the continuum Hopf numbers  $\mathcal{N}_\nu^\text{Hopf}$ are  gauge-dependent and should be treated with care.

\section{Hopf-Haldane analogy}
\label{AppHaldane}

We here observe that the MMHSs and MHIs presented in this paper can be viewed as a family of 3D systems that is quite analogous to the family of 2D Dirac semimetals and Chern insulators.

\emph{Valley-Dirac semimetal versus valley-Hopf semimetal.} Consider a 2D Dirac semimetal such as graphene \cite{Castro_2009}. In fact, to be precise, graphene is an example for a \emph{valley-Dirac semimetal}, because it has an even number of linear nodal points in the Brillouin zone, with ``opposite" quantum geometric properties in the two valleys. More precisely, the Berry curvature vanishes everywhere in the Brillouin zone except for singularities at the two nodal points \cite{Cayssol_2021},
\begin{equation}
	\Omega_{\alpha,xy}(\mathbf{k})\sim\xi\delta(\mathbf{k}-\mathbf{K}_\xi).
\end{equation}
Accordingly, the Berry phase computed as a line integral around these points is given by $\phi_\alpha\sim\xi\pi$.

\begin{figure}
	\centering
	\includegraphics[width=\columnwidth]{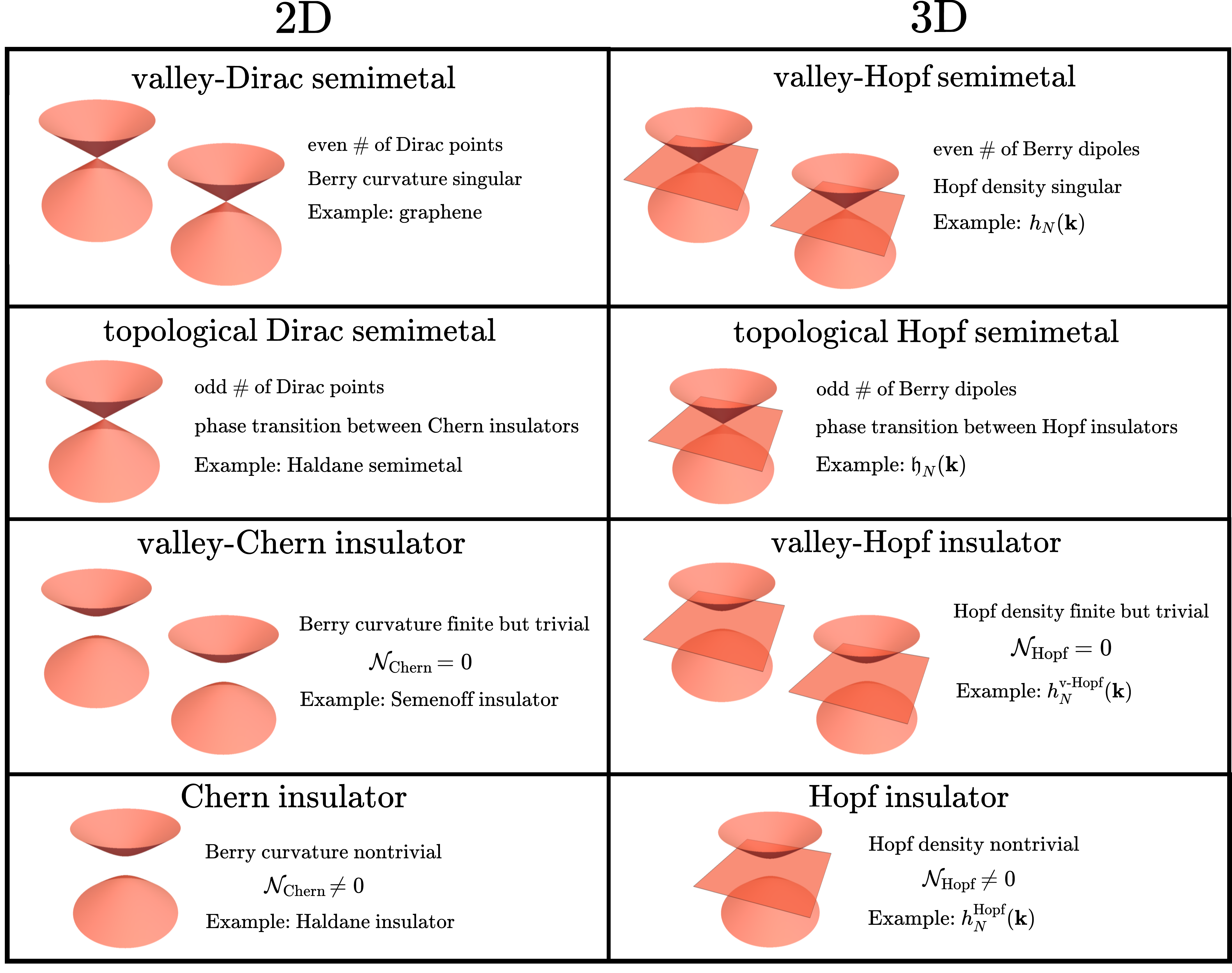}
	\caption{Analogy between 2D Dirac semimetals (2D Chern insulators) and 3D Hopf semimetals (3D Hopf insulators), based on the number of linear nodal points and the properties of the relevant topological densities (Berry curvature versus Hopf density).}
	\label{fig:hopfhald}
\end{figure}

 In a very similar way, the 3D valley-Hopf semimetals (\ref{subst}) and (\ref{subst2}) have two linear nodal points with opposite quantum geometry. To be more precise, the Hopf density vanishes everywhere in the Brillouin zone, except for singularities at these nodal points:
\begin{equation}
	\chi(\mathbf{k})\sim\xi\delta(\mathbf{k}-\mathbf{k}_\nu),
\end{equation}
where $\mathbf{k}_\nu\in\{\mathbf{K}_+,\mathbf{K}_-\}$ for the model (\ref{subst}) based on hexagonal layers and $\mathbf{k}_\nu\in\{\mathbf{k}_\Gamma,\mathbf{k}_\text{M}\}$ for the model (\ref{subst2}) based on square layers. Additionally, the Berry dipoles have opposite orientation in the two valleys, $\mathbf{d}=(0,0,\xi)$. The analogy between these two kinds of semimetals is visualized in the first row of Fig. \ref{fig:hopfhald}.

\emph{Topological Dirac semimetal versus topological Hopf semimetal.} Second, consider the topological phase transition lines in Haldane's model \cite{Haldane_1988}. We may call the semimetallic phase along these lines a \emph{Haldane semimetal}. The Haldane semimetal is an example  for a  2D \emph{topological Dirac semimetal}, as it exists at the transition between two insulators with different Chern number. The Haldane semimetal has only a single nodal point in the Brillouin zone \cite{Thonhauser_2006}. The Berry curvature is peaked at this nodal point, but also non-zero in other regions of the Brillouin zone. 

In a very similar way, the 3D topological Hopf semimetals (\ref{truesem}) exist at the transition between two insulators with different Hopf number. They have an odd number of nodal points in the Brillouin zone. The corresponding Hopf density (for $N=3,4$) is given by 
\begin{equation}
	\chi(\mathbf{k})=\frac{12}{\e^4}\left(c_x c_y+c_y c_z+c_z c_x+\Delta_0 c_x c_y c_z\right),
	\label{semhod}
\end{equation}
as obtained from Eq. (\ref{hodens}), where $\Delta_0\in\{\pm1,\pm3\}$. It is peaked at the nodal points but also non-zero away from them. The analogy is again visualized in the second row of Fig. \ref{fig:hopfhald}.

\emph{Valley-Chern insulator versus valley-Hopf insulator.} Third, consider Semenoff's model \cite{Semenoff_1984}, which we may view as an example for a 2D \emph{valley-Chern insulator}. It has a Berry curvature which is non-zero in each valley \cite{Cayssol_2021}, but distributed such that the Chern number vanishes. By analogy, we may write down a model for a 3D \emph{valley-Hopf insulator} as
\begin{equation}
	h_N^\text{v-Hopf}(\mathbf{k})=\begin{pmatrix}
		0 & Q_N^\text{v-Hopf}\\
		(Q_N^\text{v-Hopf})^\dagger & 0
	\end{pmatrix},
\end{equation}
where the matrices \smash{$Q_N^{\text{v-Hopf}}$} are obtained from Eq. (\ref{hopfdef}) upon deleting the cosine terms, for example \smash{$Q_3^\text{v-Hopf}=(s_- \,\,\,\,\, \Delta-i\sin k_z)$}. We then find a Hopf density
$$
\begin{aligned}
	\chi_{N=3,4}(\mathbf{k})&=\frac{12}{\e^4}\Delta c_x c_y c_z, & \chi_{N=5}(\mathbf{k})&=\frac{60}{\e^4}\Delta c_x c_y c_z,
\end{aligned}
$$
where $\epsilon^2=s_x^2+s_y^2+s_z^2+\Delta^2$. It is non-zero throughout the Brillouin zone but topologically trivial in the sense that the Hopf number vanishes. This analogy is visualized in the third row of Fig. \ref{fig:hopfhald}.

\emph{Chern insulator versus Hopf insulator.}	Finally, a 2D Chern insulator such as Haldane's insulator \cite{Haldane_1988} has a non-trivial Berry curvature that produces a non-zero Chern number. In the same way, the 3D Hopf insulators (\ref{hopfdef}) have a non-trivial Hopf density, cf. Eq. (\ref{hodens}), which produces a non-zero Hopf number. This is visualized in the last row of Fig. \ref{fig:hopfhald}.

\bibliography{ref}

\begin{thebibliography}{75}%
\makeatletter
\providecommand \@ifxundefined [1]{%
 \@ifx{#1\undefined}
}%
\providecommand \@ifnum [1]{%
 \ifnum #1\expandafter \@firstoftwo
 \else \expandafter \@secondoftwo
 \fi
}%
\providecommand \@ifx [1]{%
 \ifx #1\expandafter \@firstoftwo
 \else \expandafter \@secondoftwo
 \fi
}%
\providecommand \natexlab [1]{#1}%
\providecommand \enquote  [1]{``#1''}%
\providecommand \bibnamefont  [1]{#1}%
\providecommand \bibfnamefont [1]{#1}%
\providecommand \citenamefont [1]{#1}%
\providecommand \href@noop [0]{\@secondoftwo}%
\providecommand \href [0]{\begingroup \@sanitize@url \@href}%
\providecommand \@href[1]{\@@startlink{#1}\@@href}%
\providecommand \@@href[1]{\endgroup#1\@@endlink}%
\providecommand \@sanitize@url [0]{\catcode `\\12\catcode `\$12\catcode
  `\&12\catcode `\#12\catcode `\^12\catcode `\_12\catcode `\%12\relax}%
\providecommand \@@startlink[1]{}%
\providecommand \@@endlink[0]{}%
\providecommand \url  [0]{\begingroup\@sanitize@url \@url }%
\providecommand \@url [1]{\endgroup\@href {#1}{\urlprefix }}%
\providecommand \urlprefix  [0]{URL }%
\providecommand \Eprint [0]{\href }%
\providecommand \doibase [0]{https://doi.org/}%
\providecommand \selectlanguage [0]{\@gobble}%
\providecommand \bibinfo  [0]{\@secondoftwo}%
\providecommand \bibfield  [0]{\@secondoftwo}%
\providecommand \translation [1]{[#1]}%
\providecommand \BibitemOpen [0]{}%
\providecommand \bibitemStop [0]{}%
\providecommand \bibitemNoStop [0]{.\EOS\space}%
\providecommand \EOS [0]{\spacefactor3000\relax}%
\providecommand \BibitemShut  [1]{\csname bibitem#1\endcsname}%
\let\auto@bib@innerbib\@empty
\bibitem [{\citenamefont {Armitage}\ \emph {et~al.}(2018)\citenamefont
  {Armitage}, \citenamefont {Mele},\ and\ \citenamefont
  {Vishwanath}}]{Armitage_2018}%
  \BibitemOpen
  \bibfield  {author} {\bibinfo {author} {\bibfnamefont {N.~P.}\ \bibnamefont
  {Armitage}}, \bibinfo {author} {\bibfnamefont {E.~J.}\ \bibnamefont {Mele}},\
  and\ \bibinfo {author} {\bibfnamefont {A.}~\bibnamefont {Vishwanath}},\
  }\bibfield  {title} {\bibinfo {title} {{Weyl and Dirac semimetals in
  three-dimensional solids}},\ }\href
  {https://doi.org/10.1103/RevModPhys.90.015001} {\bibfield  {journal}
  {\bibinfo  {journal} {Rev. Mod. Phys.}\ }\textbf {\bibinfo {volume} {90}},\
  \bibinfo {pages} {015001} (\bibinfo {year} {2018})}\BibitemShut {NoStop}%
\bibitem [{\citenamefont {Lv}\ \emph {et~al.}(2021)\citenamefont {Lv},
  \citenamefont {Qian},\ and\ \citenamefont {Ding}}]{Lv_2021}%
  \BibitemOpen
  \bibfield  {author} {\bibinfo {author} {\bibfnamefont {B.~Q.}\ \bibnamefont
  {Lv}}, \bibinfo {author} {\bibfnamefont {T.}~\bibnamefont {Qian}},\ and\
  \bibinfo {author} {\bibfnamefont {H.}~\bibnamefont {Ding}},\ }\bibfield
  {title} {\bibinfo {title} {Experimental perspective on three-dimensional
  topological semimetals},\ }\href
  {https://doi.org/10.1103/RevModPhys.93.025002} {\bibfield  {journal}
  {\bibinfo  {journal} {Rev. Mod. Phys.}\ }\textbf {\bibinfo {volume} {93}},\
  \bibinfo {pages} {025002} (\bibinfo {year} {2021})}\BibitemShut {NoStop}%
\bibitem [{\citenamefont {Berry}(1984)}]{Berry_1984}%
  \BibitemOpen
  \bibfield  {author} {\bibinfo {author} {\bibfnamefont {M.~V.}\ \bibnamefont
  {Berry}},\ }\bibfield  {title} {\bibinfo {title} {{Quantal phase factors
  accompanying adiabatic changes}},\ }\href
  {https://doi.org/10.1098/rspa.1984.0023} {\bibfield  {journal} {\bibinfo
  {journal} {Proceedings of the Royal Society of London. Series A}\ }\textbf
  {\bibinfo {volume} {392}},\ \bibinfo {pages} {45} (\bibinfo {year}
  {1984})}\BibitemShut {NoStop}%
\bibitem [{\citenamefont {Berry}(1985)}]{Berry_1985a}%
  \BibitemOpen
  \bibfield  {author} {\bibinfo {author} {\bibfnamefont {M.~V.}\ \bibnamefont
  {Berry}},\ }\bibinfo {title} {{Aspects of Degeneracy}},\ in\ \href
  {https://doi.org/10.1007/978-1-4613-2443-0_8} {\emph {\bibinfo {booktitle}
  {Chaotic Behavior in Quantum Systems: Theory and Applications}}},\ \bibinfo
  {editor} {edited by\ \bibinfo {editor} {\bibfnamefont {G.}~\bibnamefont
  {Casati}}}\ (\bibinfo  {publisher} {Springer US},\ \bibinfo {address}
  {Boston, MA},\ \bibinfo {year} {1985})\ pp.\ \bibinfo {pages}
  {123--140}\BibitemShut {NoStop}%
\bibitem [{\citenamefont {Volovik}(1987)}]{Volovik_1987}%
  \BibitemOpen
  \bibfield  {author} {\bibinfo {author} {\bibfnamefont {G.}~\bibnamefont
  {Volovik}},\ }\bibfield  {title} {\bibinfo {title} {Zeros in the fermion
  spectrum in superfluid systems as diabolical points},\ }\href
  {http://jetpletters.ru/ps/1225/article_18512.shtml} {\bibfield  {journal}
  {\bibinfo  {journal} {JETP Lett.}\ }\textbf {\bibinfo {volume} {46}},\
  \bibinfo {pages} {98} (\bibinfo {year} {1987})}\BibitemShut {NoStop}%
\bibitem [{\citenamefont {Fang}\ \emph {et~al.}(2003)\citenamefont {Fang},
  \citenamefont {Nagaosa}, \citenamefont {Takahashi}, \citenamefont {Asamitsu},
  \citenamefont {Mathieu}, \citenamefont {Ogasawara}, \citenamefont {Yamada},
  \citenamefont {Kawasaki}, \citenamefont {Tokura},\ and\ \citenamefont
  {Terakura}}]{Fang_2003}%
  \BibitemOpen
  \bibfield  {author} {\bibinfo {author} {\bibfnamefont {Z.}~\bibnamefont
  {Fang}}, \bibinfo {author} {\bibfnamefont {N.}~\bibnamefont {Nagaosa}},
  \bibinfo {author} {\bibfnamefont {K.~S.}\ \bibnamefont {Takahashi}}, \bibinfo
  {author} {\bibfnamefont {A.}~\bibnamefont {Asamitsu}}, \bibinfo {author}
  {\bibfnamefont {R.}~\bibnamefont {Mathieu}}, \bibinfo {author} {\bibfnamefont
  {T.}~\bibnamefont {Ogasawara}}, \bibinfo {author} {\bibfnamefont
  {H.}~\bibnamefont {Yamada}}, \bibinfo {author} {\bibfnamefont
  {M.}~\bibnamefont {Kawasaki}}, \bibinfo {author} {\bibfnamefont
  {Y.}~\bibnamefont {Tokura}},\ and\ \bibinfo {author} {\bibfnamefont
  {K.}~\bibnamefont {Terakura}},\ }\bibfield  {title} {\bibinfo {title} {{The
  Anomalous Hall Effect and Magnetic Monopoles in Momentum Space}},\ }\href
  {https://doi.org/10.1126/science.1089408} {\bibfield  {journal} {\bibinfo
  {journal} {Science}\ }\textbf {\bibinfo {volume} {302}},\ \bibinfo {pages}
  {92} (\bibinfo {year} {2003})}\BibitemShut {NoStop}%
\bibitem [{\citenamefont {Nielsen}\ and\ \citenamefont
  {Ninomiya}(1981)}]{Nielsen_1981}%
  \BibitemOpen
  \bibfield  {author} {\bibinfo {author} {\bibfnamefont {H.}~\bibnamefont
  {Nielsen}}\ and\ \bibinfo {author} {\bibfnamefont {M.}~\bibnamefont
  {Ninomiya}},\ }\bibfield  {title} {\bibinfo {title} {A no-go theorem for
  regularizing chiral fermions},\ }\href
  {https://doi.org/https://doi.org/10.1016/0370-2693(81)91026-1} {\bibfield
  {journal} {\bibinfo  {journal} {Physics Letters B}\ }\textbf {\bibinfo
  {volume} {105}},\ \bibinfo {pages} {219} (\bibinfo {year}
  {1981})}\BibitemShut {NoStop}%
\bibitem [{\citenamefont {Bradlyn}\ \emph {et~al.}(2016)\citenamefont
  {Bradlyn}, \citenamefont {Cano}, \citenamefont {Wang}, \citenamefont
  {Vergniory}, \citenamefont {Felser}, \citenamefont {Cava},\ and\
  \citenamefont {Bernevig}}]{Bradlyn_2016}%
  \BibitemOpen
  \bibfield  {author} {\bibinfo {author} {\bibfnamefont {B.}~\bibnamefont
  {Bradlyn}}, \bibinfo {author} {\bibfnamefont {J.}~\bibnamefont {Cano}},
  \bibinfo {author} {\bibfnamefont {Z.}~\bibnamefont {Wang}}, \bibinfo {author}
  {\bibfnamefont {M.~G.}\ \bibnamefont {Vergniory}}, \bibinfo {author}
  {\bibfnamefont {C.}~\bibnamefont {Felser}}, \bibinfo {author} {\bibfnamefont
  {R.~J.}\ \bibnamefont {Cava}},\ and\ \bibinfo {author} {\bibfnamefont
  {B.~A.}\ \bibnamefont {Bernevig}},\ }\bibfield  {title} {\bibinfo {title}
  {{Beyond Dirac and Weyl fermions: Unconventional quasiparticles in
  conventional crystals}},\ }\href {https://doi.org/10.1126/science.aaf5037}
  {\bibfield  {journal} {\bibinfo  {journal} {Science}\ }\textbf {\bibinfo
  {volume} {353}},\ \bibinfo {pages} {aaf5037} (\bibinfo {year}
  {2016})}\BibitemShut {NoStop}%
\bibitem [{\citenamefont {Ezawa}(2017)}]{Ezawa_2017b}%
  \BibitemOpen
  \bibfield  {author} {\bibinfo {author} {\bibfnamefont {M.}~\bibnamefont
  {Ezawa}},\ }\bibfield  {title} {\bibinfo {title} {Chiral anomaly enhancement
  and photoirradiation effects in multiband touching fermion systems},\ }\href
  {https://doi.org/10.1103/PhysRevB.95.205201} {\bibfield  {journal} {\bibinfo
  {journal} {Phys. Rev. B}\ }\textbf {\bibinfo {volume} {95}},\ \bibinfo
  {pages} {205201} (\bibinfo {year} {2017})}\BibitemShut {NoStop}%
\bibitem [{\citenamefont {Alexandradinata}\ \emph {et~al.}(2021)\citenamefont
  {Alexandradinata}, \citenamefont {Nelson},\ and\ \citenamefont
  {Soluyanov}}]{Alexandradinata_2021}%
  \BibitemOpen
  \bibfield  {author} {\bibinfo {author} {\bibfnamefont {A.}~\bibnamefont
  {Alexandradinata}}, \bibinfo {author} {\bibfnamefont {A.}~\bibnamefont
  {Nelson}},\ and\ \bibinfo {author} {\bibfnamefont {A.~A.}\ \bibnamefont
  {Soluyanov}},\ }\bibfield  {title} {\bibinfo {title} {{Teleportation of Berry
  curvature on the surface of a Hopf insulator}},\ }\href
  {https://doi.org/10.1103/PhysRevB.103.045107} {\bibfield  {journal} {\bibinfo
   {journal} {Phys. Rev. B}\ }\textbf {\bibinfo {volume} {103}},\ \bibinfo
  {pages} {045107} (\bibinfo {year} {2021})}\BibitemShut {NoStop}%
\bibitem [{\citenamefont {Nelson}\ \emph {et~al.}(2022)\citenamefont {Nelson},
  \citenamefont {Neupert}, \citenamefont {Alexandradinata},\ and\ \citenamefont
  {Bzdu\ifmmode~\check{s}\else \v{s}\fi{}ek}}]{Nelson_2022}%
  \BibitemOpen
  \bibfield  {author} {\bibinfo {author} {\bibfnamefont {A.}~\bibnamefont
  {Nelson}}, \bibinfo {author} {\bibfnamefont {T.}~\bibnamefont {Neupert}},
  \bibinfo {author} {\bibfnamefont {A.}~\bibnamefont {Alexandradinata}},\ and\
  \bibinfo {author} {\bibfnamefont {T.~c.~v.}\ \bibnamefont
  {Bzdu\ifmmode~\check{s}\else \v{s}\fi{}ek}},\ }\bibfield  {title} {\bibinfo
  {title} {{Delicate topology protected by rotation symmetry: Crystalline Hopf
  insulators and beyond}},\ }\href
  {https://doi.org/10.1103/PhysRevB.106.075124} {\bibfield  {journal} {\bibinfo
   {journal} {Phys. Rev. B}\ }\textbf {\bibinfo {volume} {106}},\ \bibinfo
  {pages} {075124} (\bibinfo {year} {2022})}\BibitemShut {NoStop}%
\bibitem [{Note1()}]{Note1}%
  \BibitemOpen
  \bibinfo {note} {See Refs. \cite {Moore_2008,Deng_2013} for an introduction
  to the well-known two-band Hopf insulator; see Ref. \cite {Nelson_2021a} for
  a comparison of stable, fragile and delicate topology}\BibitemShut {NoStop}%
\bibitem [{\citenamefont {Lapierre}\ \emph {et~al.}(2021)\citenamefont
  {Lapierre}, \citenamefont {Neupert},\ and\ \citenamefont
  {Trifunovic}}]{Lapierre_2021}%
  \BibitemOpen
  \bibfield  {author} {\bibinfo {author} {\bibfnamefont {B.}~\bibnamefont
  {Lapierre}}, \bibinfo {author} {\bibfnamefont {T.}~\bibnamefont {Neupert}},\
  and\ \bibinfo {author} {\bibfnamefont {L.}~\bibnamefont {Trifunovic}},\
  }\bibfield  {title} {\bibinfo {title} {{$N$-band Hopf insulator}},\ }\href
  {https://doi.org/10.1103/PhysRevResearch.3.033045} {\bibfield  {journal}
  {\bibinfo  {journal} {Phys. Rev. Research}\ }\textbf {\bibinfo {volume}
  {3}},\ \bibinfo {pages} {033045} (\bibinfo {year} {2021})}\BibitemShut
  {NoStop}%
\bibitem [{\citenamefont {Graf}\ and\ \citenamefont
  {Pi\'echon}(2021)}]{Graf_2021}%
  \BibitemOpen
  \bibfield  {author} {\bibinfo {author} {\bibfnamefont {A.}~\bibnamefont
  {Graf}}\ and\ \bibinfo {author} {\bibfnamefont {F.}~\bibnamefont
  {Pi\'echon}},\ }\bibfield  {title} {\bibinfo {title} {{Berry curvature and
  quantum metric in $N$-band systems: An eigenprojector approach}},\ }\href
  {https://doi.org/10.1103/PhysRevB.104.085114} {\bibfield  {journal} {\bibinfo
   {journal} {Phys. Rev. B}\ }\textbf {\bibinfo {volume} {104}},\ \bibinfo
  {pages} {085114} (\bibinfo {year} {2021})}\BibitemShut {NoStop}%
\bibitem [{Note2()}]{Note2}%
  \BibitemOpen
  \bibinfo {note} {More precisely, we have $\Sigma _d=\protect \frac
  {1}{3}\protect \text {diag}(1,-2,1)$, $\Sigma _d=\protect \frac
  {1}{2}\protect \text {diag}(1,-1,-1,1)$, and $\Sigma _d=\protect \frac
  {1}{5}\protect \text {diag}(4,-1,-6,-1,4)$ for Eqs. (\ref {Berrydip3})--(\ref
  {Berrydip5}), respectively.}\BibitemShut {Stop}%
\bibitem [{\citenamefont {Delplace}(2022)}]{Delplace_2022}%
  \BibitemOpen
  \bibfield  {author} {\bibinfo {author} {\bibfnamefont {P.}~\bibnamefont
  {Delplace}},\ }\bibfield  {title} {\bibinfo {title} {{Berry-Chern monopoles
  and spectral flows}},\ }\href
  {https://doi.org/10.21468/SciPostPhysLectNotes.39} {\bibfield  {journal}
  {\bibinfo  {journal} {SciPost Phys. Lect. Notes}\ ,\ \bibinfo {pages} {39}}
  (\bibinfo {year} {2022})}\BibitemShut {NoStop}%
\bibitem [{\citenamefont {Dirac}(1928)}]{Dirac_1928}%
  \BibitemOpen
  \bibfield  {author} {\bibinfo {author} {\bibfnamefont {P.~A.~M.}\
  \bibnamefont {Dirac}},\ }\bibfield  {title} {\bibinfo {title} {The quantum
  theory of the electron},\ }\href {https://doi.org/10.1098/rspa.1928.0023}
  {\bibfield  {journal} {\bibinfo  {journal} {Proceedings of the Royal Society
  of London. Series A}\ }\textbf {\bibinfo {volume} {117}},\ \bibinfo {pages}
  {610} (\bibinfo {year} {1928})}\BibitemShut {NoStop}%
\bibitem [{\citenamefont {Rabi}(1928)}]{Rabi_1928}%
  \BibitemOpen
  \bibfield  {author} {\bibinfo {author} {\bibfnamefont {I.~I.}\ \bibnamefont
  {Rabi}},\ }\bibfield  {title} {\bibinfo {title} {{Das freie Elektron im
  homogenen Magnetfeld nach der Diracschen Theorie}},\ }\href
  {https://doi.org/10.1007/BF01333634} {\bibfield  {journal} {\bibinfo
  {journal} {Zeitschrift f\"ur Physik}\ }\textbf {\bibinfo {volume} {49}},\
  \bibinfo {pages} {507} (\bibinfo {year} {1928})}\BibitemShut {NoStop}%
\bibitem [{\citenamefont {Saykin}\ \emph {et~al.}(2018)\citenamefont {Saykin},
  \citenamefont {Tikhonov},\ and\ \citenamefont {Rodionov}}]{Saykin_2018}%
  \BibitemOpen
  \bibfield  {author} {\bibinfo {author} {\bibfnamefont {D.~R.}\ \bibnamefont
  {Saykin}}, \bibinfo {author} {\bibfnamefont {K.~S.}\ \bibnamefont
  {Tikhonov}},\ and\ \bibinfo {author} {\bibfnamefont {Y.~I.}\ \bibnamefont
  {Rodionov}},\ }\bibfield  {title} {\bibinfo {title} {{Landau levels with
  magnetic tunneling in a Weyl semimetal and magnetoconductance of a ballistic
  $p\text{\ensuremath{-}}n$ junction}},\ }\href
  {https://doi.org/10.1103/PhysRevB.97.041202} {\bibfield  {journal} {\bibinfo
  {journal} {Phys. Rev. B}\ }\textbf {\bibinfo {volume} {97}},\ \bibinfo
  {pages} {041202} (\bibinfo {year} {2018})}\BibitemShut {NoStop}%
\bibitem [{\citenamefont {Onsager}(1952)}]{Onsager_1952}%
  \BibitemOpen
  \bibfield  {author} {\bibinfo {author} {\bibfnamefont {L.}~\bibnamefont
  {Onsager}},\ }\bibfield  {title} {\bibinfo {title} {{Interpretation of the de
  Haas-van Alphen effect}},\ }\href {https://doi.org/10.1080/14786440908521019}
  {\bibfield  {journal} {\bibinfo  {journal} {The London, Edinburgh, and Dublin
  Philosophical Magazine and Journal of Science}\ }\textbf {\bibinfo {volume}
  {43}},\ \bibinfo {pages} {1006} (\bibinfo {year} {1952})}\BibitemShut
  {NoStop}%
\bibitem [{\citenamefont {Roth}(1966)}]{Roth_1966}%
  \BibitemOpen
  \bibfield  {author} {\bibinfo {author} {\bibfnamefont {L.~M.}\ \bibnamefont
  {Roth}},\ }\bibfield  {title} {\bibinfo {title} {{Semiclassical Theory of
  Magnetic Energy Levels and Magnetic Susceptibility of Bloch Electrons}},\
  }\href {https://doi.org/10.1103/PhysRev.145.434} {\bibfield  {journal}
  {\bibinfo  {journal} {Phys. Rev.}\ }\textbf {\bibinfo {volume} {145}},\
  \bibinfo {pages} {434} (\bibinfo {year} {1966})}\BibitemShut {NoStop}%
\bibitem [{\citenamefont {Mikitik}\ and\ \citenamefont
  {Sharlai}(1999)}]{Mikitik_1999}%
  \BibitemOpen
  \bibfield  {author} {\bibinfo {author} {\bibfnamefont {G.~P.}\ \bibnamefont
  {Mikitik}}\ and\ \bibinfo {author} {\bibfnamefont {Y.~V.}\ \bibnamefont
  {Sharlai}},\ }\bibfield  {title} {\bibinfo {title} {{Manifestation of Berry's
  Phase in Metal Physics}},\ }\href
  {https://doi.org/10.1103/PhysRevLett.82.2147} {\bibfield  {journal} {\bibinfo
   {journal} {Phys. Rev. Lett.}\ }\textbf {\bibinfo {volume} {82}},\ \bibinfo
  {pages} {2147} (\bibinfo {year} {1999})}\BibitemShut {NoStop}%
\bibitem [{\citenamefont {Fuchs}\ \emph {et~al.}(2010)\citenamefont {Fuchs},
  \citenamefont {Pi\'echon}, \citenamefont {Goerbig},\ and\ \citenamefont
  {Montambaux}}]{Fuchs_2010}%
  \BibitemOpen
  \bibfield  {author} {\bibinfo {author} {\bibfnamefont {J.~N.}\ \bibnamefont
  {Fuchs}}, \bibinfo {author} {\bibfnamefont {F.}~\bibnamefont {Pi\'echon}},
  \bibinfo {author} {\bibfnamefont {M.~O.}\ \bibnamefont {Goerbig}},\ and\
  \bibinfo {author} {\bibfnamefont {G.}~\bibnamefont {Montambaux}},\ }\bibfield
   {title} {\bibinfo {title} {{Topological Berry phase and semiclassical
  quantization of cyclotron orbits for two-dimensional electrons in coupledband
  models}},\ }\href {https://doi.org/10.1140/epjb/e2010-00259-2} {\bibfield
  {journal} {\bibinfo  {journal} {The European Physical Journal B}\ }\textbf
  {\bibinfo {volume} {77}},\ \bibinfo {pages} {351} (\bibinfo {year}
  {2010})}\BibitemShut {NoStop}%
\bibitem [{\citenamefont {Gao}\ and\ \citenamefont {Niu}(2017)}]{GaoNiu_2017}%
  \BibitemOpen
  \bibfield  {author} {\bibinfo {author} {\bibfnamefont {Y.}~\bibnamefont
  {Gao}}\ and\ \bibinfo {author} {\bibfnamefont {Q.}~\bibnamefont {Niu}},\
  }\bibfield  {title} {\bibinfo {title} {{Zero-field magnetic response
  functions in Landau levels}},\ }\href
  {https://doi.org/10.1073/pnas.1702595114} {\bibfield  {journal} {\bibinfo
  {journal} {Proceedings of the National Academy of Sciences}\ }\textbf
  {\bibinfo {volume} {114}},\ \bibinfo {pages} {7295} (\bibinfo {year}
  {2017})}\BibitemShut {NoStop}%
\bibitem [{\citenamefont {Fuchs}\ \emph {et~al.}(2018)\citenamefont {Fuchs},
  \citenamefont {Piéchon},\ and\ \citenamefont {Montambaux}}]{Fuchs_2018}%
  \BibitemOpen
  \bibfield  {author} {\bibinfo {author} {\bibfnamefont {J.~N.}\ \bibnamefont
  {Fuchs}}, \bibinfo {author} {\bibfnamefont {F.}~\bibnamefont {Piéchon}},\
  and\ \bibinfo {author} {\bibfnamefont {G.}~\bibnamefont {Montambaux}},\
  }\bibfield  {title} {\bibinfo {title} {{Landau levels, response functions and
  magnetic oscillations from a generalized Onsager relation}},\ }\href
  {https://doi.org/10.21468/SciPostPhys.4.5.024} {\bibfield  {journal}
  {\bibinfo  {journal} {SciPost Phys.}\ }\textbf {\bibinfo {volume} {4}},\
  \bibinfo {pages} {24} (\bibinfo {year} {2018})}\BibitemShut {NoStop}%
\bibitem [{\citenamefont {Ziman}(1960)}]{Ziman_1960}%
  \BibitemOpen
  \bibfield  {author} {\bibinfo {author} {\bibfnamefont {J.}~\bibnamefont
  {Ziman}},\ }\href {https://books.google.fr/books?id=\_HEsAAAAYAAJ} {\emph
  {\bibinfo {title} {Electrons and Phonons: The Theory of Transport Phenomena
  in Solids}}}\ (\bibinfo  {publisher} {Clarendon Press},\ \bibinfo {year}
  {1960})\BibitemShut {NoStop}%
\bibitem [{\citenamefont {Xiao}\ \emph {et~al.}(2010)\citenamefont {Xiao},
  \citenamefont {Chang},\ and\ \citenamefont {Niu}}]{Xiao_2010}%
  \BibitemOpen
  \bibfield  {author} {\bibinfo {author} {\bibfnamefont {D.}~\bibnamefont
  {Xiao}}, \bibinfo {author} {\bibfnamefont {M.-C.}\ \bibnamefont {Chang}},\
  and\ \bibinfo {author} {\bibfnamefont {Q.}~\bibnamefont {Niu}},\ }\bibfield
  {title} {\bibinfo {title} {Berry phase effects on electronic properties},\
  }\href {https://doi.org/10.1103/RevModPhys.82.1959} {\bibfield  {journal}
  {\bibinfo  {journal} {Rev. Mod. Phys.}\ }\textbf {\bibinfo {volume} {82}},\
  \bibinfo {pages} {1959} (\bibinfo {year} {2010})}\BibitemShut {NoStop}%
\bibitem [{\citenamefont {Klinkhamer}\ and\ \citenamefont
  {Volovik}(2005)}]{Klinkhamer_2005}%
  \BibitemOpen
  \bibfield  {author} {\bibinfo {author} {\bibfnamefont {F.~R.}\ \bibnamefont
  {Klinkhamer}}\ and\ \bibinfo {author} {\bibfnamefont {G.~E.}\ \bibnamefont
  {Volovik}},\ }\bibfield  {title} {\bibinfo {title} {{Emergent CPT violation
  from the splitting of Fermi points}},\ }\href
  {https://doi.org/10.1142/S0217751X05020902} {\bibfield  {journal} {\bibinfo
  {journal} {International Journal of Modern Physics A}\ }\textbf {\bibinfo
  {volume} {20}},\ \bibinfo {pages} {2795} (\bibinfo {year}
  {2005})}\BibitemShut {NoStop}%
\bibitem [{\citenamefont {Burkov}\ and\ \citenamefont
  {Balents}(2011)}]{Burkov_2011}%
  \BibitemOpen
  \bibfield  {author} {\bibinfo {author} {\bibfnamefont {A.~A.}\ \bibnamefont
  {Burkov}}\ and\ \bibinfo {author} {\bibfnamefont {L.}~\bibnamefont
  {Balents}},\ }\bibfield  {title} {\bibinfo {title} {{Weyl Semimetal in a
  Topological Insulator Multilayer}},\ }\href
  {https://doi.org/10.1103/PhysRevLett.107.127205} {\bibfield  {journal}
  {\bibinfo  {journal} {Phys. Rev. Lett.}\ }\textbf {\bibinfo {volume} {107}},\
  \bibinfo {pages} {127205} (\bibinfo {year} {2011})}\BibitemShut {NoStop}%
\bibitem [{\citenamefont {Yang}\ \emph {et~al.}(2011)\citenamefont {Yang},
  \citenamefont {Lu},\ and\ \citenamefont {Ran}}]{Yang_2011}%
  \BibitemOpen
  \bibfield  {author} {\bibinfo {author} {\bibfnamefont {K.-Y.}\ \bibnamefont
  {Yang}}, \bibinfo {author} {\bibfnamefont {Y.-M.}\ \bibnamefont {Lu}},\ and\
  \bibinfo {author} {\bibfnamefont {Y.}~\bibnamefont {Ran}},\ }\bibfield
  {title} {\bibinfo {title} {{Quantum Hall effects in a Weyl semimetal:
  Possible application in pyrochlore iridates}},\ }\href
  {https://doi.org/10.1103/PhysRevB.84.075129} {\bibfield  {journal} {\bibinfo
  {journal} {Phys. Rev. B}\ }\textbf {\bibinfo {volume} {84}},\ \bibinfo
  {pages} {075129} (\bibinfo {year} {2011})}\BibitemShut {NoStop}%
\bibitem [{\citenamefont {Nielsen}\ and\ \citenamefont
  {Ninomiya}(1983)}]{Nielsen_1983}%
  \BibitemOpen
  \bibfield  {author} {\bibinfo {author} {\bibfnamefont {H.}~\bibnamefont
  {Nielsen}}\ and\ \bibinfo {author} {\bibfnamefont {M.}~\bibnamefont
  {Ninomiya}},\ }\bibfield  {title} {\bibinfo {title} {{The Adler-Bell-Jackiw
  anomaly and Weyl fermions in a crystal}},\ }\href
  {https://doi.org/https://doi.org/10.1016/0370-2693(83)91529-0} {\bibfield
  {journal} {\bibinfo  {journal} {Physics Letters B}\ }\textbf {\bibinfo
  {volume} {130}},\ \bibinfo {pages} {389} (\bibinfo {year}
  {1983})}\BibitemShut {NoStop}%
\bibitem [{\citenamefont {Zyuzin}(2017)}]{Zyuzin_2017}%
  \BibitemOpen
  \bibfield  {author} {\bibinfo {author} {\bibfnamefont {V.~A.}\ \bibnamefont
  {Zyuzin}},\ }\bibfield  {title} {\bibinfo {title} {{Magnetotransport of Weyl
  semimetals due to the chiral anomaly}},\ }\href
  {https://doi.org/10.1103/PhysRevB.95.245128} {\bibfield  {journal} {\bibinfo
  {journal} {Phys. Rev. B}\ }\textbf {\bibinfo {volume} {95}},\ \bibinfo
  {pages} {245128} (\bibinfo {year} {2017})}\BibitemShut {NoStop}%
\bibitem [{\citenamefont {Sharma}\ \emph {et~al.}(2017)\citenamefont {Sharma},
  \citenamefont {Goswami},\ and\ \citenamefont {Tewari}}]{Sharma_2017}%
  \BibitemOpen
  \bibfield  {author} {\bibinfo {author} {\bibfnamefont {G.}~\bibnamefont
  {Sharma}}, \bibinfo {author} {\bibfnamefont {P.}~\bibnamefont {Goswami}},\
  and\ \bibinfo {author} {\bibfnamefont {S.}~\bibnamefont {Tewari}},\
  }\bibfield  {title} {\bibinfo {title} {{Chiral anomaly and longitudinal
  magnetotransport in type-II Weyl semimetals}},\ }\href
  {https://doi.org/10.1103/PhysRevB.96.045112} {\bibfield  {journal} {\bibinfo
  {journal} {Phys. Rev. B}\ }\textbf {\bibinfo {volume} {96}},\ \bibinfo
  {pages} {045112} (\bibinfo {year} {2017})}\BibitemShut {NoStop}%
\bibitem [{\citenamefont {Vilenkin}(1980)}]{Vilenkin_1980}%
  \BibitemOpen
  \bibfield  {author} {\bibinfo {author} {\bibfnamefont {A.}~\bibnamefont
  {Vilenkin}},\ }\bibfield  {title} {\bibinfo {title} {Equilibrium
  parity-violating current in a magnetic field},\ }\href
  {https://doi.org/10.1103/PhysRevD.22.3080} {\bibfield  {journal} {\bibinfo
  {journal} {Phys. Rev. D}\ }\textbf {\bibinfo {volume} {22}},\ \bibinfo
  {pages} {3080} (\bibinfo {year} {1980})}\BibitemShut {NoStop}%
\bibitem [{\citenamefont {Fukushima}\ \emph {et~al.}(2008)\citenamefont
  {Fukushima}, \citenamefont {Kharzeev},\ and\ \citenamefont
  {Warringa}}]{Fukushima_2008}%
  \BibitemOpen
  \bibfield  {author} {\bibinfo {author} {\bibfnamefont {K.}~\bibnamefont
  {Fukushima}}, \bibinfo {author} {\bibfnamefont {D.~E.}\ \bibnamefont
  {Kharzeev}},\ and\ \bibinfo {author} {\bibfnamefont {H.~J.}\ \bibnamefont
  {Warringa}},\ }\bibfield  {title} {\bibinfo {title} {Chiral magnetic
  effect},\ }\href {https://doi.org/10.1103/PhysRevD.78.074033} {\bibfield
  {journal} {\bibinfo  {journal} {Phys. Rev. D}\ }\textbf {\bibinfo {volume}
  {78}},\ \bibinfo {pages} {074033} (\bibinfo {year} {2008})}\BibitemShut
  {NoStop}%
\bibitem [{\citenamefont {Zyuzin}\ \emph {et~al.}(2012)\citenamefont {Zyuzin},
  \citenamefont {Wu},\ and\ \citenamefont {Burkov}}]{Zyuzin_2012}%
  \BibitemOpen
  \bibfield  {author} {\bibinfo {author} {\bibfnamefont {A.~A.}\ \bibnamefont
  {Zyuzin}}, \bibinfo {author} {\bibfnamefont {S.}~\bibnamefont {Wu}},\ and\
  \bibinfo {author} {\bibfnamefont {A.~A.}\ \bibnamefont {Burkov}},\ }\bibfield
   {title} {\bibinfo {title} {Weyl semimetal with broken time reversal and
  inversion symmetries},\ }\href {https://doi.org/10.1103/PhysRevB.85.165110}
  {\bibfield  {journal} {\bibinfo  {journal} {Phys. Rev. B}\ }\textbf {\bibinfo
  {volume} {85}},\ \bibinfo {pages} {165110} (\bibinfo {year}
  {2012})}\BibitemShut {NoStop}%
\bibitem [{\citenamefont {Cortijo}(2016)}]{Cortijo_2016}%
  \BibitemOpen
  \bibfield  {author} {\bibinfo {author} {\bibfnamefont {A.}~\bibnamefont
  {Cortijo}},\ }\bibfield  {title} {\bibinfo {title} {{Linear magnetochiral
  effect in Weyl semimetals}},\ }\href
  {https://doi.org/10.1103/PhysRevB.94.241105} {\bibfield  {journal} {\bibinfo
  {journal} {Phys. Rev. B}\ }\textbf {\bibinfo {volume} {94}},\ \bibinfo
  {pages} {241105} (\bibinfo {year} {2016})}\BibitemShut {NoStop}%
\bibitem [{\citenamefont {Kundu}\ \emph {et~al.}(2020)\citenamefont {Kundu},
  \citenamefont {Siu}, \citenamefont {Yang},\ and\ \citenamefont
  {Jalil}}]{Kundu_2020}%
  \BibitemOpen
  \bibfield  {author} {\bibinfo {author} {\bibfnamefont {A.}~\bibnamefont
  {Kundu}}, \bibinfo {author} {\bibfnamefont {Z.~B.}\ \bibnamefont {Siu}},
  \bibinfo {author} {\bibfnamefont {H.}~\bibnamefont {Yang}},\ and\ \bibinfo
  {author} {\bibfnamefont {M.~B.~A.}\ \bibnamefont {Jalil}},\ }\bibfield
  {title} {\bibinfo {title} {{Magnetotransport of Weyl semimetals with tilted
  Dirac cones}},\ }\href {https://doi.org/10.1088/1367-2630/aba98d} {\bibfield
  {journal} {\bibinfo  {journal} {New Journal of Physics}\ }\textbf {\bibinfo
  {volume} {22}},\ \bibinfo {pages} {083081} (\bibinfo {year}
  {2020})}\BibitemShut {NoStop}%
\bibitem [{\citenamefont {Castro~Neto}\ \emph {et~al.}(2009)\citenamefont
  {Castro~Neto}, \citenamefont {Guinea}, \citenamefont {Peres}, \citenamefont
  {Novoselov},\ and\ \citenamefont {Geim}}]{Castro_2009}%
  \BibitemOpen
  \bibfield  {author} {\bibinfo {author} {\bibfnamefont {A.~H.}\ \bibnamefont
  {Castro~Neto}}, \bibinfo {author} {\bibfnamefont {F.}~\bibnamefont {Guinea}},
  \bibinfo {author} {\bibfnamefont {N.~M.~R.}\ \bibnamefont {Peres}}, \bibinfo
  {author} {\bibfnamefont {K.~S.}\ \bibnamefont {Novoselov}},\ and\ \bibinfo
  {author} {\bibfnamefont {A.~K.}\ \bibnamefont {Geim}},\ }\bibfield  {title}
  {\bibinfo {title} {The electronic properties of graphene},\ }\href
  {https://doi.org/10.1103/revmodphys.81.109} {\bibfield  {journal} {\bibinfo
  {journal} {Reviews of Modern Physics}\ }\textbf {\bibinfo {volume} {81}},\
  \bibinfo {pages} {109–162} (\bibinfo {year} {2009})}\BibitemShut {NoStop}%
\bibitem [{Note3()}]{Note3}%
  \BibitemOpen
  \bibinfo {note} {Interesting results along this direction were obtained in
  Ref. \cite {Habe_2022} after submission of this manuscript.}\BibitemShut
  {Stop}%
\bibitem [{\citenamefont {Moore}\ \emph {et~al.}(2008)\citenamefont {Moore},
  \citenamefont {Ran},\ and\ \citenamefont {Wen}}]{Moore_2008}%
  \BibitemOpen
  \bibfield  {author} {\bibinfo {author} {\bibfnamefont {J.~E.}\ \bibnamefont
  {Moore}}, \bibinfo {author} {\bibfnamefont {Y.}~\bibnamefont {Ran}},\ and\
  \bibinfo {author} {\bibfnamefont {X.-G.}\ \bibnamefont {Wen}},\ }\bibfield
  {title} {\bibinfo {title} {{Topological Surface States in Three-Dimensional
  Magnetic Insulators}},\ }\href
  {https://doi.org/10.1103/PhysRevLett.101.186805} {\bibfield  {journal}
  {\bibinfo  {journal} {Phys. Rev. Lett.}\ }\textbf {\bibinfo {volume} {101}},\
  \bibinfo {pages} {186805} (\bibinfo {year} {2008})}\BibitemShut {NoStop}%
\bibitem [{\citenamefont {Deng}\ \emph {et~al.}(2013)\citenamefont {Deng},
  \citenamefont {Wang}, \citenamefont {Shen},\ and\ \citenamefont
  {Duan}}]{Deng_2013}%
  \BibitemOpen
  \bibfield  {author} {\bibinfo {author} {\bibfnamefont {D.-L.}\ \bibnamefont
  {Deng}}, \bibinfo {author} {\bibfnamefont {S.-T.}\ \bibnamefont {Wang}},
  \bibinfo {author} {\bibfnamefont {C.}~\bibnamefont {Shen}},\ and\ \bibinfo
  {author} {\bibfnamefont {L.-M.}\ \bibnamefont {Duan}},\ }\bibfield  {title}
  {\bibinfo {title} {Hopf insulators and their topologically protected surface
  states},\ }\href {https://doi.org/10.1103/PhysRevB.88.201105} {\bibfield
  {journal} {\bibinfo  {journal} {Phys. Rev. B}\ }\textbf {\bibinfo {volume}
  {88}},\ \bibinfo {pages} {201105} (\bibinfo {year} {2013})}\BibitemShut
  {NoStop}%
\bibitem [{\citenamefont {Liu}\ \emph {et~al.}(2017)\citenamefont {Liu},
  \citenamefont {Vafa},\ and\ \citenamefont {Xu}}]{Liu_2017}%
  \BibitemOpen
  \bibfield  {author} {\bibinfo {author} {\bibfnamefont {C.}~\bibnamefont
  {Liu}}, \bibinfo {author} {\bibfnamefont {F.}~\bibnamefont {Vafa}},\ and\
  \bibinfo {author} {\bibfnamefont {C.}~\bibnamefont {Xu}},\ }\bibfield
  {title} {\bibinfo {title} {{Symmetry-protected topological Hopf insulator and
  its generalizations}},\ }\href {https://doi.org/10.1103/PhysRevB.95.161116}
  {\bibfield  {journal} {\bibinfo  {journal} {Phys. Rev. B}\ }\textbf {\bibinfo
  {volume} {95}},\ \bibinfo {pages} {161116} (\bibinfo {year}
  {2017})}\BibitemShut {NoStop}%
\bibitem [{\citenamefont {Deng}\ \emph {et~al.}(2017)\citenamefont {Deng},
  \citenamefont {Wang}, \citenamefont {Sun},\ and\ \citenamefont
  {Duan}}]{Deng_2018}%
  \BibitemOpen
  \bibfield  {author} {\bibinfo {author} {\bibfnamefont {D.-L.}\ \bibnamefont
  {Deng}}, \bibinfo {author} {\bibfnamefont {S.-T.}\ \bibnamefont {Wang}},
  \bibinfo {author} {\bibfnamefont {K.}~\bibnamefont {Sun}},\ and\ \bibinfo
  {author} {\bibfnamefont {L.-M.}\ \bibnamefont {Duan}},\ }\bibfield  {title}
  {\bibinfo {title} {{Probe Knots and Hopf Insulators with Ultracold Atoms}},\
  }\href {https://doi.org/10.1088/0256-307x/35/1/013701} {\bibfield  {journal}
  {\bibinfo  {journal} {Chinese Physics Letters}\ }\textbf {\bibinfo {volume}
  {35}},\ \bibinfo {pages} {013701} (\bibinfo {year} {2017})}\BibitemShut
  {NoStop}%
\bibitem [{\citenamefont {\"Unal}\ \emph {et~al.}(2019)\citenamefont {\"Unal},
  \citenamefont {Eckardt},\ and\ \citenamefont {Slager}}]{Unal_2019}%
  \BibitemOpen
  \bibfield  {author} {\bibinfo {author} {\bibfnamefont {F.~N.}\ \bibnamefont
  {\"Unal}}, \bibinfo {author} {\bibfnamefont {A.}~\bibnamefont {Eckardt}},\
  and\ \bibinfo {author} {\bibfnamefont {R.-J.}\ \bibnamefont {Slager}},\
  }\bibfield  {title} {\bibinfo {title} {{Hopf characterization of
  two-dimensional Floquet topological insulators}},\ }\href
  {https://doi.org/10.1103/PhysRevResearch.1.022003} {\bibfield  {journal}
  {\bibinfo  {journal} {Phys. Rev. Research}\ }\textbf {\bibinfo {volume}
  {1}},\ \bibinfo {pages} {022003} (\bibinfo {year} {2019})}\BibitemShut
  {NoStop}%
\bibitem [{\citenamefont {Schuster}\ \emph
  {et~al.}(2021{\natexlab{a}})\citenamefont {Schuster}, \citenamefont
  {Flicker}, \citenamefont {Li}, \citenamefont {Kotochigova}, \citenamefont
  {Moore}, \citenamefont {Ye},\ and\ \citenamefont {Yao}}]{Schuster_2021}%
  \BibitemOpen
  \bibfield  {author} {\bibinfo {author} {\bibfnamefont {T.}~\bibnamefont
  {Schuster}}, \bibinfo {author} {\bibfnamefont {F.}~\bibnamefont {Flicker}},
  \bibinfo {author} {\bibfnamefont {M.}~\bibnamefont {Li}}, \bibinfo {author}
  {\bibfnamefont {S.}~\bibnamefont {Kotochigova}}, \bibinfo {author}
  {\bibfnamefont {J.~E.}\ \bibnamefont {Moore}}, \bibinfo {author}
  {\bibfnamefont {J.}~\bibnamefont {Ye}},\ and\ \bibinfo {author}
  {\bibfnamefont {N.~Y.}\ \bibnamefont {Yao}},\ }\bibfield  {title} {\bibinfo
  {title} {{Realizing Hopf Insulators in Dipolar Spin Systems}},\ }\href
  {https://doi.org/10.1103/PhysRevLett.127.015301} {\bibfield  {journal}
  {\bibinfo  {journal} {Phys. Rev. Lett.}\ }\textbf {\bibinfo {volume} {127}},\
  \bibinfo {pages} {015301} (\bibinfo {year} {2021}{\natexlab{a}})}\BibitemShut
  {NoStop}%
\bibitem [{\citenamefont {Schuster}\ \emph
  {et~al.}(2021{\natexlab{b}})\citenamefont {Schuster}, \citenamefont
  {Flicker}, \citenamefont {Li}, \citenamefont {Kotochigova}, \citenamefont
  {Moore}, \citenamefont {Ye},\ and\ \citenamefont {Yao}}]{Schuster_2021a}%
  \BibitemOpen
  \bibfield  {author} {\bibinfo {author} {\bibfnamefont {T.}~\bibnamefont
  {Schuster}}, \bibinfo {author} {\bibfnamefont {F.}~\bibnamefont {Flicker}},
  \bibinfo {author} {\bibfnamefont {M.}~\bibnamefont {Li}}, \bibinfo {author}
  {\bibfnamefont {S.}~\bibnamefont {Kotochigova}}, \bibinfo {author}
  {\bibfnamefont {J.~E.}\ \bibnamefont {Moore}}, \bibinfo {author}
  {\bibfnamefont {J.}~\bibnamefont {Ye}},\ and\ \bibinfo {author}
  {\bibfnamefont {N.~Y.}\ \bibnamefont {Yao}},\ }\bibfield  {title} {\bibinfo
  {title} {Floquet engineering ultracold polar molecules to simulate
  topological insulators},\ }\href
  {https://doi.org/10.1103/PhysRevA.103.063322} {\bibfield  {journal} {\bibinfo
   {journal} {Phys. Rev. A}\ }\textbf {\bibinfo {volume} {103}},\ \bibinfo
  {pages} {063322} (\bibinfo {year} {2021}{\natexlab{b}})}\BibitemShut
  {NoStop}%
\bibitem [{\citenamefont {Wang}\ \emph {et~al.}(2014)\citenamefont {Wang},
  \citenamefont {Deng},\ and\ \citenamefont {Duan}}]{Wang_2014}%
  \BibitemOpen
  \bibfield  {author} {\bibinfo {author} {\bibfnamefont {S.-T.}\ \bibnamefont
  {Wang}}, \bibinfo {author} {\bibfnamefont {D.-L.}\ \bibnamefont {Deng}},\
  and\ \bibinfo {author} {\bibfnamefont {L.-M.}\ \bibnamefont {Duan}},\
  }\bibfield  {title} {\bibinfo {title} {{Probe of Three-Dimensional Chiral
  Topological Insulators in an Optical Lattice}},\ }\href
  {https://doi.org/10.1103/PhysRevLett.113.033002} {\bibfield  {journal}
  {\bibinfo  {journal} {Phys. Rev. Lett.}\ }\textbf {\bibinfo {volume} {113}},\
  \bibinfo {pages} {033002} (\bibinfo {year} {2014})}\BibitemShut {NoStop}%
\bibitem [{\citenamefont {Lian}\ \emph {et~al.}(2019)\citenamefont {Lian},
  \citenamefont {Wang}, \citenamefont {Lu}, \citenamefont {Huang},
  \citenamefont {Wang}, \citenamefont {Yuan}, \citenamefont {Zhang},
  \citenamefont {Ouyang}, \citenamefont {Wang}, \citenamefont {Huang},
  \citenamefont {He}, \citenamefont {Chang}, \citenamefont {Deng},\ and\
  \citenamefont {Duan}}]{Lian_2019}%
  \BibitemOpen
  \bibfield  {author} {\bibinfo {author} {\bibfnamefont {W.}~\bibnamefont
  {Lian}}, \bibinfo {author} {\bibfnamefont {S.-T.}\ \bibnamefont {Wang}},
  \bibinfo {author} {\bibfnamefont {S.}~\bibnamefont {Lu}}, \bibinfo {author}
  {\bibfnamefont {Y.}~\bibnamefont {Huang}}, \bibinfo {author} {\bibfnamefont
  {F.}~\bibnamefont {Wang}}, \bibinfo {author} {\bibfnamefont {X.}~\bibnamefont
  {Yuan}}, \bibinfo {author} {\bibfnamefont {W.}~\bibnamefont {Zhang}},
  \bibinfo {author} {\bibfnamefont {X.}~\bibnamefont {Ouyang}}, \bibinfo
  {author} {\bibfnamefont {X.}~\bibnamefont {Wang}}, \bibinfo {author}
  {\bibfnamefont {X.}~\bibnamefont {Huang}}, \bibinfo {author} {\bibfnamefont
  {L.}~\bibnamefont {He}}, \bibinfo {author} {\bibfnamefont {X.}~\bibnamefont
  {Chang}}, \bibinfo {author} {\bibfnamefont {D.-L.}\ \bibnamefont {Deng}},\
  and\ \bibinfo {author} {\bibfnamefont {L.}~\bibnamefont {Duan}},\ }\bibfield
  {title} {\bibinfo {title} {{Machine Learning Topological Phases with a
  Solid-State Quantum Simulator}},\ }\href
  {https://doi.org/10.1103/PhysRevLett.122.210503} {\bibfield  {journal}
  {\bibinfo  {journal} {Phys. Rev. Lett.}\ }\textbf {\bibinfo {volume} {122}},\
  \bibinfo {pages} {210503} (\bibinfo {year} {2019})}\BibitemShut {NoStop}%
\bibitem [{\citenamefont {Neupert}\ \emph {et~al.}(2012)\citenamefont
  {Neupert}, \citenamefont {Santos}, \citenamefont {Ryu}, \citenamefont
  {Chamon},\ and\ \citenamefont {Mudry}}]{Neupert_2012}%
  \BibitemOpen
  \bibfield  {author} {\bibinfo {author} {\bibfnamefont {T.}~\bibnamefont
  {Neupert}}, \bibinfo {author} {\bibfnamefont {L.}~\bibnamefont {Santos}},
  \bibinfo {author} {\bibfnamefont {S.}~\bibnamefont {Ryu}}, \bibinfo {author}
  {\bibfnamefont {C.}~\bibnamefont {Chamon}},\ and\ \bibinfo {author}
  {\bibfnamefont {C.}~\bibnamefont {Mudry}},\ }\bibfield  {title} {\bibinfo
  {title} {Noncommutative geometry for three-dimensional topological
  insulators},\ }\href {https://doi.org/10.1103/PhysRevB.86.035125} {\bibfield
  {journal} {\bibinfo  {journal} {Phys. Rev. B}\ }\textbf {\bibinfo {volume}
  {86}},\ \bibinfo {pages} {035125} (\bibinfo {year} {2012})}\BibitemShut
  {NoStop}%
\bibitem [{\citenamefont {Habe}(2022)}]{Habe_2022}%
  \BibitemOpen
  \bibfield  {author} {\bibinfo {author} {\bibfnamefont {T.}~\bibnamefont
  {Habe}},\ }\bibfield  {title} {\bibinfo {title} {{Optical conductivity of the
  threefold Hopf semimetal}},\ }\href
  {https://doi.org/10.1103/PhysRevB.106.205204} {\bibfield  {journal} {\bibinfo
   {journal} {Phys. Rev. B}\ }\textbf {\bibinfo {volume} {106}},\ \bibinfo
  {pages} {205204} (\bibinfo {year} {2022})}\BibitemShut {NoStop}%
\bibitem [{\citenamefont {Lu}\ \emph {et~al.}(2015)\citenamefont {Lu},
  \citenamefont {Wang}, \citenamefont {Ye}, \citenamefont {Ran}, \citenamefont
  {Fu}, \citenamefont {Joannopoulos},\ and\ \citenamefont
  {Soljačić}}]{Lu_2015}%
  \BibitemOpen
  \bibfield  {author} {\bibinfo {author} {\bibfnamefont {L.}~\bibnamefont
  {Lu}}, \bibinfo {author} {\bibfnamefont {Z.}~\bibnamefont {Wang}}, \bibinfo
  {author} {\bibfnamefont {D.}~\bibnamefont {Ye}}, \bibinfo {author}
  {\bibfnamefont {L.}~\bibnamefont {Ran}}, \bibinfo {author} {\bibfnamefont
  {L.}~\bibnamefont {Fu}}, \bibinfo {author} {\bibfnamefont {J.~D.}\
  \bibnamefont {Joannopoulos}},\ and\ \bibinfo {author} {\bibfnamefont
  {M.}~\bibnamefont {Soljačić}},\ }\bibfield  {title} {\bibinfo {title}
  {{Experimental observation of Weyl points}},\ }\href
  {https://doi.org/10.1126/science.aaa9273} {\bibfield  {journal} {\bibinfo
  {journal} {Science}\ }\textbf {\bibinfo {volume} {349}},\ \bibinfo {pages}
  {622} (\bibinfo {year} {2015})}\BibitemShut {NoStop}%
\bibitem [{\citenamefont {Chen}\ \emph {et~al.}(2016)\citenamefont {Chen},
  \citenamefont {Xiao},\ and\ \citenamefont {Chan}}]{Chen_2016}%
  \BibitemOpen
  \bibfield  {author} {\bibinfo {author} {\bibfnamefont {W.-J.}\ \bibnamefont
  {Chen}}, \bibinfo {author} {\bibfnamefont {M.}~\bibnamefont {Xiao}},\ and\
  \bibinfo {author} {\bibfnamefont {C.~T.}\ \bibnamefont {Chan}},\ }\bibfield
  {title} {\bibinfo {title} {{Photonic crystals possessing multiple Weyl points
  and the experimental observation of robust surface states}},\ }\href
  {https://doi.org/10.1038/ncomms13038} {\bibfield  {journal} {\bibinfo
  {journal} {Nature Communications}\ }\textbf {\bibinfo {volume} {7}},\
  \bibinfo {pages} {13038} (\bibinfo {year} {2016})}\BibitemShut {NoStop}%
\bibitem [{\citenamefont {Riwar}\ \emph {et~al.}(2016)\citenamefont {Riwar},
  \citenamefont {Houzet}, \citenamefont {Meyer},\ and\ \citenamefont
  {Nazarov}}]{Riwar_2016}%
  \BibitemOpen
  \bibfield  {author} {\bibinfo {author} {\bibfnamefont {R.-P.}\ \bibnamefont
  {Riwar}}, \bibinfo {author} {\bibfnamefont {M.}~\bibnamefont {Houzet}},
  \bibinfo {author} {\bibfnamefont {J.~S.}\ \bibnamefont {Meyer}},\ and\
  \bibinfo {author} {\bibfnamefont {Y.~V.}\ \bibnamefont {Nazarov}},\
  }\bibfield  {title} {\bibinfo {title} {{Multi-terminal Josephson junctions as
  topological matter}},\ }\href {https://doi.org/10.1038/ncomms11167}
  {\bibfield  {journal} {\bibinfo  {journal} {Nature Communications}\ }\textbf
  {\bibinfo {volume} {7}},\ \bibinfo {pages} {11167} (\bibinfo {year}
  {2016})}\BibitemShut {NoStop}%
\bibitem [{\citenamefont {Wang}\ \emph {et~al.}(2017)\citenamefont {Wang},
  \citenamefont {Xiao}, \citenamefont {Liu}, \citenamefont {Zhu},\ and\
  \citenamefont {Chan}}]{Wang_2017}%
  \BibitemOpen
  \bibfield  {author} {\bibinfo {author} {\bibfnamefont {Q.}~\bibnamefont
  {Wang}}, \bibinfo {author} {\bibfnamefont {M.}~\bibnamefont {Xiao}}, \bibinfo
  {author} {\bibfnamefont {H.}~\bibnamefont {Liu}}, \bibinfo {author}
  {\bibfnamefont {S.}~\bibnamefont {Zhu}},\ and\ \bibinfo {author}
  {\bibfnamefont {C.~T.}\ \bibnamefont {Chan}},\ }\bibfield  {title} {\bibinfo
  {title} {{Optical Interface States Protected by Synthetic Weyl Points}},\
  }\href {https://doi.org/10.1103/PhysRevX.7.031032} {\bibfield  {journal}
  {\bibinfo  {journal} {Phys. Rev. X}\ }\textbf {\bibinfo {volume} {7}},\
  \bibinfo {pages} {031032} (\bibinfo {year} {2017})}\BibitemShut {NoStop}%
\bibitem [{\citenamefont {Zhu}\ \emph {et~al.}(2017)\citenamefont {Zhu},
  \citenamefont {Zhang}, \citenamefont {Yan}, \citenamefont {Xing},\ and\
  \citenamefont {Zhu}}]{Zhu_2017}%
  \BibitemOpen
  \bibfield  {author} {\bibinfo {author} {\bibfnamefont {Y.-Q.}\ \bibnamefont
  {Zhu}}, \bibinfo {author} {\bibfnamefont {D.-W.}\ \bibnamefont {Zhang}},
  \bibinfo {author} {\bibfnamefont {H.}~\bibnamefont {Yan}}, \bibinfo {author}
  {\bibfnamefont {D.-Y.}\ \bibnamefont {Xing}},\ and\ \bibinfo {author}
  {\bibfnamefont {S.-L.}\ \bibnamefont {Zhu}},\ }\bibfield  {title} {\bibinfo
  {title} {{Emergent pseudospin-1 Maxwell fermions with a threefold degeneracy
  in optical lattices}},\ }\href {https://doi.org/10.1103/PhysRevA.96.033634}
  {\bibfield  {journal} {\bibinfo  {journal} {Phys. Rev. A}\ }\textbf {\bibinfo
  {volume} {96}},\ \bibinfo {pages} {033634} (\bibinfo {year}
  {2017})}\BibitemShut {NoStop}%
\bibitem [{\citenamefont {Zhang}\ \emph {et~al.}(2018)\citenamefont {Zhang},
  \citenamefont {Zhu}, \citenamefont {Zhao}, \citenamefont {Yan},\ and\
  \citenamefont {Zhu}}]{Zhang_2018a}%
  \BibitemOpen
  \bibfield  {author} {\bibinfo {author} {\bibfnamefont {D.-W.}\ \bibnamefont
  {Zhang}}, \bibinfo {author} {\bibfnamefont {Y.-Q.}\ \bibnamefont {Zhu}},
  \bibinfo {author} {\bibfnamefont {Y.~X.}\ \bibnamefont {Zhao}}, \bibinfo
  {author} {\bibfnamefont {H.}~\bibnamefont {Yan}},\ and\ \bibinfo {author}
  {\bibfnamefont {S.-L.}\ \bibnamefont {Zhu}},\ }\bibfield  {title} {\bibinfo
  {title} {Topological quantum matter with cold atoms},\ }\href
  {https://doi.org/10.1080/00018732.2019.1594094} {\bibfield  {journal}
  {\bibinfo  {journal} {Advances in Physics}\ }\textbf {\bibinfo {volume}
  {67}},\ \bibinfo {pages} {253} (\bibinfo {year} {2018})}\BibitemShut
  {NoStop}%
\bibitem [{\citenamefont {Tan}\ \emph {et~al.}(2018)\citenamefont {Tan},
  \citenamefont {Zhang}, \citenamefont {Liu}, \citenamefont {Xue},
  \citenamefont {Yu}, \citenamefont {Zhu}, \citenamefont {Yan}, \citenamefont
  {Zhu},\ and\ \citenamefont {Yu}}]{Tan_2018}%
  \BibitemOpen
  \bibfield  {author} {\bibinfo {author} {\bibfnamefont {X.}~\bibnamefont
  {Tan}}, \bibinfo {author} {\bibfnamefont {D.-W.}\ \bibnamefont {Zhang}},
  \bibinfo {author} {\bibfnamefont {Q.}~\bibnamefont {Liu}}, \bibinfo {author}
  {\bibfnamefont {G.}~\bibnamefont {Xue}}, \bibinfo {author} {\bibfnamefont
  {H.-F.}\ \bibnamefont {Yu}}, \bibinfo {author} {\bibfnamefont {Y.-Q.}\
  \bibnamefont {Zhu}}, \bibinfo {author} {\bibfnamefont {H.}~\bibnamefont
  {Yan}}, \bibinfo {author} {\bibfnamefont {S.-L.}\ \bibnamefont {Zhu}},\ and\
  \bibinfo {author} {\bibfnamefont {Y.}~\bibnamefont {Yu}},\ }\bibfield
  {title} {\bibinfo {title} {{Topological Maxwell Metal Bands in a
  Superconducting Qutrit}},\ }\href
  {https://doi.org/10.1103/PhysRevLett.120.130503} {\bibfield  {journal}
  {\bibinfo  {journal} {Phys. Rev. Lett.}\ }\textbf {\bibinfo {volume} {120}},\
  \bibinfo {pages} {130503} (\bibinfo {year} {2018})}\BibitemShut {NoStop}%
\bibitem [{\citenamefont {Fulga}\ \emph {et~al.}(2018)\citenamefont {Fulga},
  \citenamefont {Fallani},\ and\ \citenamefont {Burrello}}]{Fulga_2018}%
  \BibitemOpen
  \bibfield  {author} {\bibinfo {author} {\bibfnamefont {I.~C.}\ \bibnamefont
  {Fulga}}, \bibinfo {author} {\bibfnamefont {L.}~\bibnamefont {Fallani}},\
  and\ \bibinfo {author} {\bibfnamefont {M.}~\bibnamefont {Burrello}},\
  }\bibfield  {title} {\bibinfo {title} {Geometrically protected triple-point
  crossings in an optical lattice},\ }\href
  {https://doi.org/10.1103/PhysRevB.97.121402} {\bibfield  {journal} {\bibinfo
  {journal} {Phys. Rev. B}\ }\textbf {\bibinfo {volume} {97}},\ \bibinfo
  {pages} {121402} (\bibinfo {year} {2018})}\BibitemShut {NoStop}%
\bibitem [{\citenamefont {Hu}\ \emph {et~al.}(2018)\citenamefont {Hu},
  \citenamefont {Hou}, \citenamefont {Zhang},\ and\ \citenamefont
  {Zhang}}]{Hu_2018b}%
  \BibitemOpen
  \bibfield  {author} {\bibinfo {author} {\bibfnamefont {H.}~\bibnamefont
  {Hu}}, \bibinfo {author} {\bibfnamefont {J.}~\bibnamefont {Hou}}, \bibinfo
  {author} {\bibfnamefont {F.}~\bibnamefont {Zhang}},\ and\ \bibinfo {author}
  {\bibfnamefont {C.}~\bibnamefont {Zhang}},\ }\bibfield  {title} {\bibinfo
  {title} {{Topological Triply Degenerate Points Induced by
  Spin-Tensor-Momentum Couplings}},\ }\href
  {https://doi.org/10.1103/PhysRevLett.120.240401} {\bibfield  {journal}
  {\bibinfo  {journal} {Phys. Rev. Lett.}\ }\textbf {\bibinfo {volume} {120}},\
  \bibinfo {pages} {240401} (\bibinfo {year} {2018})}\BibitemShut {NoStop}%
\bibitem [{\citenamefont {Tan}\ \emph {et~al.}(2021)\citenamefont {Tan},
  \citenamefont {Zhang}, \citenamefont {Zheng}, \citenamefont {Yang},
  \citenamefont {Song}, \citenamefont {Han}, \citenamefont {Dong},
  \citenamefont {Wang}, \citenamefont {Lan}, \citenamefont {Yan}, \citenamefont
  {Zhu},\ and\ \citenamefont {Yu}}]{Tan_2021}%
  \BibitemOpen
  \bibfield  {author} {\bibinfo {author} {\bibfnamefont {X.}~\bibnamefont
  {Tan}}, \bibinfo {author} {\bibfnamefont {D.-W.}\ \bibnamefont {Zhang}},
  \bibinfo {author} {\bibfnamefont {W.}~\bibnamefont {Zheng}}, \bibinfo
  {author} {\bibfnamefont {X.}~\bibnamefont {Yang}}, \bibinfo {author}
  {\bibfnamefont {S.}~\bibnamefont {Song}}, \bibinfo {author} {\bibfnamefont
  {Z.}~\bibnamefont {Han}}, \bibinfo {author} {\bibfnamefont {Y.}~\bibnamefont
  {Dong}}, \bibinfo {author} {\bibfnamefont {Z.}~\bibnamefont {Wang}}, \bibinfo
  {author} {\bibfnamefont {D.}~\bibnamefont {Lan}}, \bibinfo {author}
  {\bibfnamefont {H.}~\bibnamefont {Yan}}, \bibinfo {author} {\bibfnamefont
  {S.-L.}\ \bibnamefont {Zhu}},\ and\ \bibinfo {author} {\bibfnamefont
  {Y.}~\bibnamefont {Yu}},\ }\bibfield  {title} {\bibinfo {title}
  {{Experimental Observation of Tensor Monopoles with a Superconducting
  Qudit}},\ }\href {https://doi.org/10.1103/PhysRevLett.126.017702} {\bibfield
  {journal} {\bibinfo  {journal} {Phys. Rev. Lett.}\ }\textbf {\bibinfo
  {volume} {126}},\ \bibinfo {pages} {017702} (\bibinfo {year}
  {2021})}\BibitemShut {NoStop}%
\bibitem [{\citenamefont {Haldane}(1988)}]{Haldane_1988}%
  \BibitemOpen
  \bibfield  {author} {\bibinfo {author} {\bibfnamefont {F.~D.~M.}\
  \bibnamefont {Haldane}},\ }\bibfield  {title} {\bibinfo {title} {{Model for a
  Quantum Hall Effect without Landau Levels: Condensed-Matter Realization of
  the "Parity Anomaly"}},\ }\href {https://doi.org/10.1103/PhysRevLett.61.2015}
  {\bibfield  {journal} {\bibinfo  {journal} {Phys. Rev. Lett.}\ }\textbf
  {\bibinfo {volume} {61}},\ \bibinfo {pages} {2015} (\bibinfo {year}
  {1988})}\BibitemShut {NoStop}%
\bibitem [{\citenamefont {Sun}\ \emph {et~al.}(2018)\citenamefont {Sun},
  \citenamefont {Zhang},\ and\ \citenamefont {Bzdu\ifmmode~\check{s}\else
  \v{s}\fi{}ek}}]{Sun_2018}%
  \BibitemOpen
  \bibfield  {author} {\bibinfo {author} {\bibfnamefont {X.-Q.}\ \bibnamefont
  {Sun}}, \bibinfo {author} {\bibfnamefont {S.-C.}\ \bibnamefont {Zhang}},\
  and\ \bibinfo {author} {\bibfnamefont {T.~c.~v.}\ \bibnamefont
  {Bzdu\ifmmode~\check{s}\else \v{s}\fi{}ek}},\ }\bibfield  {title} {\bibinfo
  {title} {{Conversion Rules for Weyl Points and Nodal Lines in Topological
  Media}},\ }\href {https://doi.org/10.1103/PhysRevLett.121.106402} {\bibfield
  {journal} {\bibinfo  {journal} {Phys. Rev. Lett.}\ }\textbf {\bibinfo
  {volume} {121}},\ \bibinfo {pages} {106402} (\bibinfo {year}
  {2018})}\BibitemShut {NoStop}%
\bibitem [{\citenamefont {Bouhon}\ \emph {et~al.}(2020)\citenamefont {Bouhon},
  \citenamefont {Wu}, \citenamefont {Slager}, \citenamefont {Weng},
  \citenamefont {Yazyev},\ and\ \citenamefont {Bzdušek}}]{Bouhon_2020}%
  \BibitemOpen
  \bibfield  {author} {\bibinfo {author} {\bibfnamefont {A.}~\bibnamefont
  {Bouhon}}, \bibinfo {author} {\bibfnamefont {Q.}~\bibnamefont {Wu}}, \bibinfo
  {author} {\bibfnamefont {R.-J.}\ \bibnamefont {Slager}}, \bibinfo {author}
  {\bibfnamefont {H.}~\bibnamefont {Weng}}, \bibinfo {author} {\bibfnamefont
  {O.~V.}\ \bibnamefont {Yazyev}},\ and\ \bibinfo {author} {\bibfnamefont
  {T.}~\bibnamefont {Bzdušek}},\ }\bibfield  {title} {\bibinfo {title}
  {{Non-Abelian reciprocal braiding of Weyl points and its manifestation in
  ZrTe}},\ }\href {https://doi.org/10.1038/s41567-020-0967-9} {\bibfield
  {journal} {\bibinfo  {journal} {Nature Physics}\ }\textbf {\bibinfo {volume}
  {16}},\ \bibinfo {pages} {1137} (\bibinfo {year} {2020})}\BibitemShut
  {NoStop}%
\bibitem [{\citenamefont {Palumbo}\ and\ \citenamefont
  {Goldman}(2018)}]{Palumbo_2018}%
  \BibitemOpen
  \bibfield  {author} {\bibinfo {author} {\bibfnamefont {G.}~\bibnamefont
  {Palumbo}}\ and\ \bibinfo {author} {\bibfnamefont {N.}~\bibnamefont
  {Goldman}},\ }\bibfield  {title} {\bibinfo {title} {{Revealing Tensor
  Monopoles through Quantum-Metric Measurements}},\ }\href
  {https://doi.org/10.1103/PhysRevLett.121.170401} {\bibfield  {journal}
  {\bibinfo  {journal} {Phys. Rev. Lett.}\ }\textbf {\bibinfo {volume} {121}},\
  \bibinfo {pages} {170401} (\bibinfo {year} {2018})}\BibitemShut {NoStop}%
\bibitem [{\citenamefont {Zhu}\ \emph {et~al.}(2020)\citenamefont {Zhu},
  \citenamefont {Goldman},\ and\ \citenamefont {Palumbo}}]{Zhu_2020}%
  \BibitemOpen
  \bibfield  {author} {\bibinfo {author} {\bibfnamefont {Y.-Q.}\ \bibnamefont
  {Zhu}}, \bibinfo {author} {\bibfnamefont {N.}~\bibnamefont {Goldman}},\ and\
  \bibinfo {author} {\bibfnamefont {G.}~\bibnamefont {Palumbo}},\ }\bibfield
  {title} {\bibinfo {title} {Four-dimensional semimetals with tensor monopoles:
  From surface states to topological responses},\ }\href
  {https://doi.org/10.1103/PhysRevB.102.081109} {\bibfield  {journal} {\bibinfo
   {journal} {Phys. Rev. B}\ }\textbf {\bibinfo {volume} {102}},\ \bibinfo
  {pages} {081109} (\bibinfo {year} {2020})}\BibitemShut {NoStop}%
\bibitem [{\citenamefont {Raoux}\ \emph {et~al.}(2014)\citenamefont {Raoux},
  \citenamefont {Morigi}, \citenamefont {Fuchs}, \citenamefont {Pi\'echon},\
  and\ \citenamefont {Montambaux}}]{Raoux_2014}%
  \BibitemOpen
  \bibfield  {author} {\bibinfo {author} {\bibfnamefont {A.}~\bibnamefont
  {Raoux}}, \bibinfo {author} {\bibfnamefont {M.}~\bibnamefont {Morigi}},
  \bibinfo {author} {\bibfnamefont {J.-N.}\ \bibnamefont {Fuchs}}, \bibinfo
  {author} {\bibfnamefont {F.}~\bibnamefont {Pi\'echon}},\ and\ \bibinfo
  {author} {\bibfnamefont {G.}~\bibnamefont {Montambaux}},\ }\bibfield  {title}
  {\bibinfo {title} {{From Dia- to Paramagnetic Orbital Susceptibility of
  Massless Fermions}},\ }\href {https://doi.org/10.1103/PhysRevLett.112.026402}
  {\bibfield  {journal} {\bibinfo  {journal} {Phys. Rev. Lett.}\ }\textbf
  {\bibinfo {volume} {112}},\ \bibinfo {pages} {026402} (\bibinfo {year}
  {2014})}\BibitemShut {NoStop}%
\bibitem [{\citenamefont {Jackson}(2012)}]{Jackson_2012}%
  \BibitemOpen
  \bibfield  {author} {\bibinfo {author} {\bibfnamefont {J.}~\bibnamefont
  {Jackson}},\ }\href {https://books.google.fr/books?id=8qHCZjJHRUgC} {\emph
  {\bibinfo {title} {Classical Electrodynamics}}}\ (\bibinfo  {publisher}
  {Wiley},\ \bibinfo {year} {2012})\BibitemShut {NoStop}%
\bibitem [{\citenamefont {Li}\ \emph {et~al.}(2016)\citenamefont {Li},
  \citenamefont {Roy},\ and\ \citenamefont {Das~Sarma}}]{Li_2016}%
  \BibitemOpen
  \bibfield  {author} {\bibinfo {author} {\bibfnamefont {X.}~\bibnamefont
  {Li}}, \bibinfo {author} {\bibfnamefont {B.}~\bibnamefont {Roy}},\ and\
  \bibinfo {author} {\bibfnamefont {S.}~\bibnamefont {Das~Sarma}},\ }\bibfield
  {title} {\bibinfo {title} {Weyl fermions with arbitrary monopoles in magnetic
  fields: Landau levels, longitudinal magnetotransport, and density-wave
  ordering},\ }\href {https://doi.org/10.1103/PhysRevB.94.195144} {\bibfield
  {journal} {\bibinfo  {journal} {Phys. Rev. B}\ }\textbf {\bibinfo {volume}
  {94}},\ \bibinfo {pages} {195144} (\bibinfo {year} {2016})}\BibitemShut
  {NoStop}%
\bibitem [{\citenamefont {Wang}\ \emph {et~al.}(2019)\citenamefont {Wang},
  \citenamefont {Duan}, \citenamefont {Glazman},\ and\ \citenamefont
  {Alexandradinata}}]{Wang_2019}%
  \BibitemOpen
  \bibfield  {author} {\bibinfo {author} {\bibfnamefont {C.}~\bibnamefont
  {Wang}}, \bibinfo {author} {\bibfnamefont {W.}~\bibnamefont {Duan}}, \bibinfo
  {author} {\bibfnamefont {L.}~\bibnamefont {Glazman}},\ and\ \bibinfo {author}
  {\bibfnamefont {A.}~\bibnamefont {Alexandradinata}},\ }\bibfield  {title}
  {\bibinfo {title} {{Landau quantization of nearly degenerate bands and full
  symmetry classification of Landau level crossings}},\ }\href
  {https://doi.org/10.1103/PhysRevB.100.014442} {\bibfield  {journal} {\bibinfo
   {journal} {Phys. Rev. B}\ }\textbf {\bibinfo {volume} {100}},\ \bibinfo
  {pages} {014442} (\bibinfo {year} {2019})}\BibitemShut {NoStop}%
\bibitem [{\citenamefont {K\"onye}\ and\ \citenamefont
  {Ogata}(2021)}]{Konye_2021}%
  \BibitemOpen
  \bibfield  {author} {\bibinfo {author} {\bibfnamefont {V.}~\bibnamefont
  {K\"onye}}\ and\ \bibinfo {author} {\bibfnamefont {M.}~\bibnamefont
  {Ogata}},\ }\bibfield  {title} {\bibinfo {title} {{Microscopic theory of
  magnetoconductivity at low magnetic fields in terms of Berry curvature and
  orbital magnetic moment}},\ }\href
  {https://doi.org/10.1103/PhysRevResearch.3.033076} {\bibfield  {journal}
  {\bibinfo  {journal} {Phys. Rev. Research}\ }\textbf {\bibinfo {volume}
  {3}},\ \bibinfo {pages} {033076} (\bibinfo {year} {2021})}\BibitemShut
  {NoStop}%
\bibitem [{\citenamefont {Cayssol}\ and\ \citenamefont
  {Fuchs}(2021)}]{Cayssol_2021}%
  \BibitemOpen
  \bibfield  {author} {\bibinfo {author} {\bibfnamefont {J.}~\bibnamefont
  {Cayssol}}\ and\ \bibinfo {author} {\bibfnamefont {J.~N.}\ \bibnamefont
  {Fuchs}},\ }\bibfield  {title} {\bibinfo {title} {Topological and geometrical
  aspects of band theory},\ }\href {https://doi.org/10.1088/2515-7639/abf0b5}
  {\bibfield  {journal} {\bibinfo  {journal} {Journal of Physics: Materials}\
  }\textbf {\bibinfo {volume} {4}},\ \bibinfo {pages} {034007} (\bibinfo {year}
  {2021})}\BibitemShut {NoStop}%
\bibitem [{\citenamefont {Thonhauser}\ and\ \citenamefont
  {Vanderbilt}(2006)}]{Thonhauser_2006}%
  \BibitemOpen
  \bibfield  {author} {\bibinfo {author} {\bibfnamefont {T.}~\bibnamefont
  {Thonhauser}}\ and\ \bibinfo {author} {\bibfnamefont {D.}~\bibnamefont
  {Vanderbilt}},\ }\bibfield  {title} {\bibinfo {title}
  {{Insulator/Chern-insulator transition in the Haldane model}},\ }\href
  {https://doi.org/10.1103/PhysRevB.74.235111} {\bibfield  {journal} {\bibinfo
  {journal} {Phys. Rev. B}\ }\textbf {\bibinfo {volume} {74}},\ \bibinfo
  {pages} {235111} (\bibinfo {year} {2006})}\BibitemShut {NoStop}%
\bibitem [{\citenamefont {Semenoff}(1984)}]{Semenoff_1984}%
  \BibitemOpen
  \bibfield  {author} {\bibinfo {author} {\bibfnamefont {G.~W.}\ \bibnamefont
  {Semenoff}},\ }\bibfield  {title} {\bibinfo {title} {{Condensed-Matter
  Simulation of a Three-Dimensional Anomaly}},\ }\href
  {https://doi.org/10.1103/PhysRevLett.53.2449} {\bibfield  {journal} {\bibinfo
   {journal} {Phys. Rev. Lett.}\ }\textbf {\bibinfo {volume} {53}},\ \bibinfo
  {pages} {2449} (\bibinfo {year} {1984})}\BibitemShut {NoStop}%
\bibitem [{\citenamefont {Nelson}\ \emph {et~al.}(2021)\citenamefont {Nelson},
  \citenamefont {Neupert}, \citenamefont {Bzdu\v{s}ek},\ and\ \citenamefont
  {Alexandradinata}}]{Nelson_2021a}%
  \BibitemOpen
  \bibfield  {author} {\bibinfo {author} {\bibfnamefont {A.}~\bibnamefont
  {Nelson}}, \bibinfo {author} {\bibfnamefont {T.}~\bibnamefont {Neupert}},
  \bibinfo {author} {\bibfnamefont {T.}~\bibnamefont {Bzdu\v{s}ek}},\ and\
  \bibinfo {author} {\bibfnamefont {A.}~\bibnamefont {Alexandradinata}},\
  }\bibfield  {title} {\bibinfo {title} {{Multicellularity of Delicate
  Topological Insulators}},\ }\href
  {https://doi.org/10.1103/PhysRevLett.126.216404} {\bibfield  {journal}
  {\bibinfo  {journal} {Phys. Rev. Lett.}\ }\textbf {\bibinfo {volume} {126}},\
  \bibinfo {pages} {216404} (\bibinfo {year} {2021})}\BibitemShut {NoStop}%
\end{thebibliography}%

\end{document}